# Calculation of the electrostatic energy of two charged helices on rods in a generalized braid geometry

Dominic Lee

*Department of Chemistry, Imperial College London, SW7 2AZ, London, UK*

**Abstract**

This is a technical document that outlines a calculation of an electrostatic interaction energy between two rods, with charge helices on them, forming a braid. We deal here with screened electrostatics. A general braid geometry is considered, though to obtain local expressions for the energy the curvature of the rods is considered to be small. Further approximations are made for small tilt angle. This is a generalization of the calculations given in the supplemental material of [R. Cortini et Al, Biophys. J. **101** 875 (2011)] for a straight symmetric braid structure, using a different method of calculation where the braid geometry does not need to be supposed a priori.

## 0. Introduction

The goal of these calculations is to develop a local screened electrostatic energy functional for a braid of two rods with charge helices on them. This can be done if one considers the braid curvature to be small compared to the inverse electrostatic screening length (Debye length). In this document, we develop a perturbation expansion for the curvature of the rods that relies on this. We consider braid geometries where the separation between the centre lines of the two rods is considered fixed. However, all other geometric parameters are allowed to vary. The other parameters that characterize the braid are: $\eta(s)$, the angle between the two tangent vectors of the rod centre lines, the orientation of the tangent vector of the braid axis, $\hat{\mathbf{t}}_A(s)$ and shape of the helices. This goes beyond the work of [1], describing DNA, where we considered the braid axis to be straight and $\eta(s)$ to be constant.

Such an electrostatic energy functional may be combined with the elastic rod theory of a braid [2]. The local geometric parameters of a braid formed of such rods can then be obtained through minimization of the energy functional subject to mechanical work by external forces [2]. This may be of general applied mathematical interest, but may be also extended to the biophysics of helically charged biopolymers [3] with the inclusion of statistical mechanics. Indeed, there are experiments where charged DNA is forced into a braided configuration [4,5,6]. There are theories that include electrostatics [7,8,9] that describe such experiments, but braid geometry is assumed a priori and the molecules are assumed to be uniformly charged. It would be interesting to develop a mathematical theory that relaxes these assumptions.

The paper is divided into following sections. In section 1 we start by characterizing the local geometry of our braided system and build up an analytical geometry formalism to handle the electrostatic calculation. In section 2, as a warm up, we look at the situation where the dielectric constant is uniform throughout space and there is electrostatic screening both outside and inside the rods. Next, in section 3, we describe how this gets modified when we have rods of low dielectric constant where there is no longer electrostatic screening inside them and the image charge expansion. To calculate an expansion of the image charges that we need for the electrostatic calculation, in section 4 we consider the boundary value problem of one point charge on one of the rods of low dielectric. It is here we start to develop a perturbation expansion that combines the electrostatic image charge expansion with that of the local curvature of the rods. In sections 5 and 6, we utilize results of section 4 to develop energy functionals that describe the electrostatic interaction between the two rods. In section 7, we make an approximation valid when the charged helices have the same average pitch and are close to being ideal (having a constant pitch) and commensurate with each other. In sections 8, 9 and 10, we expand our results out for small $\eta(s)$, obtaining simplified energy functionals. In sections 1-10, we consider one helical line of charge, in section 11 we show how this can be extended to more general charge distributions with helical symmetry. Finally, in section 12, we discuss the results and the further development of this theory.

## 1. Defining the local geometry of the braid

### *1.1 The geometry of the centre lines*

We consider centrelines of two charged rods forming a braid. We label these 1 and 2. We first define position vectors $\mathbf{r}_1(s)$ and $\mathbf{r}_2(s)$ in a lab frame that determine the orientations of molecular centrelines 1 and 2 respectively

$$\mathbf{r}_1(s) = \int_0^s \hat{\mathbf{t}}_1(\tau)\sigma_1(\tau)d\tau + \mathbf{r}_{1,0}, \qquad \mathbf{r}_2(s') = \int_0^{s'} \hat{\mathbf{t}}_2(\tau)\sigma_2(\tau)d\tau' + \mathbf{r}_{2,0}, \qquad (1.1)$$

where $\sigma_1(s) = \dfrac{ds_1}{ds}$ and $\sigma_1(s) = \dfrac{ds_2}{ds}$, $s_1$ and $s_2$ are the arc-lengths of the centre lines of rods 1 and 2, respectively, whereas $s$ and $\tau$ are arc-lengths of the braid axis, which we define below. The vectors $\mathbf{r}_{1,0}$ and $\mathbf{r}_{2,0}$ define the positions of the ends of rods 1 and 2 at $s=0$, respectively.

We will deal with braided structures where

$$\mathbf{r}_1(s) - \mathbf{r}_2(s) = -R\hat{\mathbf{d}}(s), \qquad (1.2)$$

and the magnitude $R$ remains constant. The vector $\hat{\mathbf{d}}(s)$ is chosen to be perpendicular to both $\hat{\mathbf{t}}_1(s)$ and $\hat{\mathbf{t}}_2(s)$, the tangent vectors. We define the braid axis as the midpoint between the two centre line curves, so that

$$\mathbf{r}_A(s) = \frac{\mathbf{r}_1(s) + \mathbf{r}_2(s)}{2} = \mathbf{r}_2(s) - \frac{R}{2}\hat{\mathbf{d}}(s) = \mathbf{r}_1(s) + \frac{R}{2}\hat{\mathbf{d}}(s). \qquad (1.3)$$

The coordinate $s$ is defined so that it is indeed measured in arc-length, namely

$$\left|\frac{d\mathbf{r}_A}{ds}\right| = 1, \qquad \frac{d\mathbf{r}_A}{ds} = \hat{\mathbf{t}}_A(s). \qquad (1.4)$$

This implies that from Eqs. (1.1) and (1.3)

$$\hat{\mathbf{t}}_A(s) = \frac{\sigma_2(s)\hat{\mathbf{t}}_2(s) + \sigma_1(s)\hat{\mathbf{t}}_1(s)}{2} = \sigma_2(s)\hat{\mathbf{t}}_2(s) - \frac{R}{2}\frac{d\hat{\mathbf{d}}(s)}{ds} = \sigma_1(s)\hat{\mathbf{t}}_1(s) + \frac{R}{2}\frac{d\hat{\mathbf{d}}(s)}{ds}. \quad (1.5)$$

It is useful to define the tilt angles of the molecular centre lines with respect to the braid axis through the dot products

$$\hat{\mathbf{t}}_1(s).\hat{\mathbf{t}}_A(s) = \cos\eta_1(s), \quad \hat{\mathbf{t}}_2(s).\hat{\mathbf{t}}_A(s) = \cos\eta_2(s). \quad (1.6)$$

This allows us to write from Eq. (1.5)

$$2 = \sigma_1(s)\cos\eta_1(s) + \sigma_2(s)\cos\eta_2(s). \quad (1.7)$$

The tilt angle between the two rod centre lines is given by $\eta(s) = \eta_1(s) + \eta_2(s)$. Now, let us parameterize the tangent of the braid axis and $\hat{\mathbf{d}}(s)$. We have two angles $\alpha(s)$ and $\beta(s)$ that define the orientation of the tangent vector of the braid axis at the point $s$ along the braid. We also have the angle $\phi_0(s)$ that defines the relative orientation of $\hat{\mathbf{d}}(s)$ around the tangent vectors. Through these angles we may write

$$\hat{\mathbf{d}}(s) = \mathbf{T}_\beta(s)\mathbf{T}_\alpha(s)\mathbf{T}_{\phi_0}(s)\begin{pmatrix}1\\0\\0\end{pmatrix}, \quad \hat{\mathbf{t}}_A(s) = \mathbf{T}_\beta(s)\mathbf{T}_\alpha(s)\mathbf{T}_{\phi_0}(s)\begin{pmatrix}0\\0\\1\end{pmatrix}, \quad (1.8)$$

where the rotation matrices through angles $\alpha(s)$, $\beta(s)$ and $\phi_0(s)$ are given by

$$\mathbf{T}_\alpha(s) = \begin{pmatrix}1 & 0 & 0\\ 0 & \cos\alpha(s) & -\sin\alpha(s)\\ 0 & \sin\alpha(s) & \cos\alpha(s)\end{pmatrix}, \quad \mathbf{T}_\beta(s) = \begin{pmatrix}\cos\beta(s) & 0 & -\sin\beta(s)\\ 0 & 1 & 0\\ \sin\beta(s) & 0 & \cos\beta(s)\end{pmatrix}, \quad (1.9)$$

$$\mathbf{T}_{\phi_0}(s) = \begin{pmatrix}\cos\phi_0(s) & -\sin\phi_0(s) & 0\\ \sin\phi_0(s) & \cos\phi_0(s) & 0\\ 0 & 0 & 1\end{pmatrix}.$$

Then, through Eq. (1.6) we are able to write

$$\hat{\mathbf{t}}_1(s) = \mathbf{T}_\beta(s)\mathbf{T}_\alpha(s)\mathbf{T}_{\phi_0}(s)\begin{pmatrix}0\\ \sin(\eta_1(s))\\ \cos(\eta_1(s))\end{pmatrix}, \quad \hat{\mathbf{t}}_2(s) = \mathbf{T}_\beta(s)\mathbf{T}_\alpha(s)\mathbf{T}_{\phi_0}(s)\begin{pmatrix}0\\ -\sin(\eta_2(s))\\ \cos(\eta_2(s))\end{pmatrix}. \quad (1.10)$$

Inspection of Eq. (1.10) gives us the additional requirement that

$$0 = \sigma_1(s)\sin\eta_1(s) - \sigma_2(s)\sin\eta_2(s). \quad (1.11)$$

Solving both Eqs. (1.7) and (1.11) for $\sigma_1(s)$ and $\sigma_2(s)$ allows us to write:

$$\frac{2\sin\eta_2(s)}{\sin(\eta_1(s)+\eta_2(s))}=\sigma_1(s) \quad \frac{2\sin\eta_1(s)}{\sin(\eta_1(s)+\eta_2(s))}=\sigma_2(s). \tag{1.12}$$

To complete the set of vectors that define 'Braid frames' [2] we define $\hat{\mathbf{n}}_1(s)=\hat{\mathbf{t}}_1(s)\times\hat{\mathbf{d}}(s)$ and $\hat{\mathbf{n}}_2(s)=\hat{\mathbf{t}}_2(s)\times\hat{\mathbf{d}}(s)$ (also we can define $\hat{\mathbf{n}}_A(s)=\hat{\mathbf{t}}_A(s)\times\hat{\mathbf{d}}(s)$). This gives us two orthogonal sets of basis vectors, $\{\hat{\mathbf{d}},\hat{\mathbf{n}}_1,\hat{\mathbf{t}}_1\}$ and $\{\hat{\mathbf{d}},\hat{\mathbf{n}}_2,\hat{\mathbf{t}}_2\}$. To describe how the basis sets change with $s$, instead of the angles $\eta(s)$, $\alpha(s)$, $\beta(s)$ and $\phi_0(s)$, we can define the generalized braid rotational frequency vectors ($\mu=1,2,A$)

$$\mathbf{\Omega}_\mu(s)=\omega_{\mu,1}(s)\hat{\mathbf{t}}_\mu(s)+\omega_{\mu,2}(s)\hat{\mathbf{d}}(s)+\omega_{\mu,3}(s)\hat{\mathbf{n}}_\mu(s), \tag{1.13}$$

Here, $\omega_{\mu,1}(s)$ is the rate of precession of $\hat{\mathbf{d}}(s)$ and $\hat{\mathbf{n}}_\mu(s)$ about $\hat{\mathbf{t}}_\mu(s)$, $\omega_{\mu,2}(s)$ is the rate of precession of $\hat{\mathbf{t}}_\mu(s)$ and $\hat{\mathbf{n}}_\mu(s)$ about $\hat{\mathbf{d}}(s)$ and $\omega_{\mu,3}(s)$ is the rate of precession of $\hat{\mathbf{t}}_\mu(s)$ and $\hat{\mathbf{d}}(s)$ about $\hat{\mathbf{n}}_\mu(s)$. This allows us to write (note that $\sigma_A(s)=1$):

$$\frac{d\hat{\mathbf{d}}(s)}{ds}=\sigma_\mu(s)\left(\mathbf{\Omega}_\mu(s)\times\hat{\mathbf{d}}(s)\right)=\sigma_\mu(s)\left(\omega_{\mu,1}(s)\hat{\mathbf{n}}_\mu(s)-\omega_{\mu,3}(s)\hat{\mathbf{t}}_\mu(s)\right), \tag{1.14}$$

$$\frac{d\hat{\mathbf{n}}_\mu(s)}{ds}=\sigma_\mu(s)\left(\mathbf{\Omega}_\mu(s)\times\hat{\mathbf{n}}_\mu(s)\right)=\sigma_\mu(s)\left(\omega_{\mu,2}(s)\hat{\mathbf{t}}_\mu(s)-\omega_{\mu,1}(s)\hat{\mathbf{d}}(s)\right), \tag{1.15}$$

$$\frac{d\hat{\mathbf{t}}_\mu(s)}{ds}=\sigma_\mu(s)\left(\mathbf{\Omega}_\mu(s)\times\hat{\mathbf{t}}_\mu(s)\right)=\sigma_\mu(s)\left(\omega_{\mu,3}(s)\hat{\mathbf{d}}(s)-\omega_{\mu,2}(s)\hat{\mathbf{n}}_\mu(s)\right). \tag{1.16}$$

Now, we want to express these angular frequencies in terms of the derivatives of $\alpha(s)$, $\beta(s)$, $\phi_0(s)$ and $\eta(s)$ with respect to $s$. To do this it is useful to calculate the following matrices from Eq. (1.9)

$$\mathbf{T}_\alpha^{-1}(s)\frac{d\mathbf{T}_\alpha(s)}{ds}=\frac{d\alpha(s)}{ds}\begin{pmatrix}0&0&0\\0&0&-1\\0&1&0\end{pmatrix}, \quad T_{\phi_0}^{-1}(s)\frac{dT_{\phi_0}(s)}{ds}=\frac{d\phi_0(s)}{ds}\begin{pmatrix}0&-1&0\\1&0&0\\0&0&0\end{pmatrix}, \tag{1.17}$$

$$T_\beta^{-1}(s)\frac{dT_\beta(s)}{ds}=\frac{d\beta(s)}{ds}\begin{pmatrix}0&0&-1\\0&0&0\\1&0&0\end{pmatrix}.$$

Then, Eqs. (1.9), (1.14)-(1.17) allow us to express the angular frequencies as

$$\omega_{\mu,1}(s) = \frac{1}{\sigma_\mu(s)} \left( \cos(\delta_\mu \eta_\mu(s)) \frac{d\phi_0(s)}{ds} - \sin \phi_0(s) \sin(\delta_\mu \eta_\mu(s)) \frac{d\alpha(s)}{ds} + \right.$$

$$\left. + \left[ \sin \alpha(s) \cos(\delta_\mu \eta_\mu(s)) - \sin(\delta_\mu \eta_\mu(s)) \cos \alpha(s) \cos \phi_0(s) \right] \frac{d\beta(s)}{ds} \right),$$

(1.18)

$$\omega_{\mu,3}(s) = -\frac{1}{\sigma_\mu(s)} \left( \sin(\delta_\mu \eta_\mu(s)) \frac{d\phi_0(s)}{ds} + \sin \phi_0(s) \cos(\delta_\mu \eta_\mu(s)) \frac{d\alpha(s)}{ds} + \right.$$

$$\left. \left[ \sin \alpha(s) \sin(\delta_\mu \eta_\mu(s)) + \cos(\delta_\mu \eta_\mu(s)) \cos \alpha(s) \cos \phi_0(s) \right] \frac{d\beta(s)}{ds} \right),$$

(1.19)

$$\omega_{\mu,2}(s) = \frac{1}{\sigma_\mu(s)} \left( \cos \phi_0(s) \frac{d\alpha(s)}{ds} - \cos \alpha(s) \sin \phi_0(s) \frac{d\beta(s)}{ds} - \delta_\mu \frac{d\eta_\mu(s)}{ds} \right),$$

(1.20)

where $\delta_1 = 1, \delta_A = 0$ and $\delta_2 = -1$. It is useful to rewrite Eqs. (1.18)-(1.20) in the following manner ( $\mu = 1, 2$ )

$$\omega_{\mu,1}(s) = \frac{\sin(\eta_1(s) + \eta_2(s))}{2 \sin \eta_\nu(s)} \left( \cos \eta_\mu(s) \omega_{A,1}(s) + \delta_\mu \sin \eta_\mu(s) \omega_{A,3}(s) \right),$$

(1.21)

$$\omega_{\mu,3}(s) = \frac{\sin(\eta_1(s) + \eta_2(s))}{2 \sin \eta_\nu(s)} \left( \cos \eta_\nu(s) \omega_{A,3}(s) - \delta_\mu \sin \eta_\nu(s) \omega_{A,1}(s) \right),$$

(1.22)

$$\omega_{\mu,2}(s) = \frac{\sin(\eta_1(s) + \eta_2(s))}{2 \sin \eta_\nu(s)} \left( \omega_{A,2}(s) - \delta_\mu \frac{d\eta_\mu(s)}{ds} \right).$$

(1.23)

It is also useful to obtain expressions for both $\sigma_1(s)$ and $\sigma_2(s)$ in terms of the angular frequencies. Substituting Eq. (1.14) into (1.5), we find

$$\hat{\mathbf{t}}_A(s) = \sigma_1(s) \left( R\omega_{1,1}(s)/2 \hat{\mathbf{n}}_1(s) + (1 - R\omega_{1,3}(s)/2) \hat{\mathbf{t}}_1(s) \right)$$
$$= \sigma_2(s) \left( -R\omega_{2,1}(s)/2 \hat{\mathbf{n}}_2(s) + (1 + R\omega_{2,3}(s)/2) \hat{\mathbf{t}}_2(s) \right).$$

(1.24)

Taking the modulus of Eq. (1.24) then allows us to express

$$\sigma_1(s) = \frac{1}{\sqrt{\left(\frac{R\omega_{1,1}(s)}{2}\right)^2 + \left(1 - \frac{R\omega_{1,3}(s)}{2}\right)^2}}, \quad \text{and} \quad \sigma_2(s) = \frac{1}{\sqrt{\left(\frac{R\omega_{2,1}(s)}{2}\right)^2 + \left(1 + \frac{R\omega_{2,3}(s)}{2}\right)^2}}.$$

(1.25)

Using both Eqs. (1.12) and (1.25), we can express the ratio between $\sin \eta_1(s)$ and $\sin \eta_2(s)$ as

$$\frac{\sin \eta_1(s)}{\sin \eta_2(s)} = \frac{\sqrt{\left(\frac{R\omega_{2,1}}{2}\right)^2 + \left(1 + \frac{R\omega_{2,3}}{2}\right)^2}}{\sqrt{\left(\frac{R\omega_{1,1}}{2}\right)^2 + \left(1 - \frac{R\omega_{1,3}}{2}\right)^2}}. \tag{1.26}$$

Also, using Eqs. (1.21) and (1.22) and (1.25), it is possible to get equations on $\Delta\eta(s) = \eta_1(s) - \eta_2(s)$ and $\omega_{A,1}(s)$ which determine them as functions of $\sin\eta(s) = \sin(\eta_1(s) + \eta_2(s))$ and $\omega_{A,3}(s)$ which are

$$4\sin^2\left(\frac{\eta(s) - \Delta\eta(s)}{2}\right) + 2R\sin\eta(s)\sin\left(\frac{\eta(s) + \Delta\eta(s)}{2}\right)\sin\left(\frac{\eta(s) - \Delta\eta(s)}{2}\right)\omega_{A,1}(s)$$

$$-2R\sin\eta(s)\cos\left(\frac{\eta(s) + \Delta\eta(s)}{2}\right)\sin\left(\frac{\eta(s) - \Delta\eta(s)}{2}\right)\omega_{A,3}(s) = \sin^2\eta(s)\left(1 - \frac{R^2\left(\omega_{A,1}(s)^2 + \omega_{A,3}(s)^2\right)}{4}\right),$$
$$\tag{1.27}$$

$$4\sin^2\left(\frac{\eta(s) + \Delta\eta(s)}{2}\right) + 2R\sin\eta(s)\sin\left(\frac{\eta(s) + \Delta\eta(s)}{2}\right)\sin\left(\frac{\eta(s) - \Delta\eta(s)}{2}\right)\omega_{A,1}(s)$$

$$+2R\sin\eta(s)\cos\left(\frac{\eta(s) - \Delta\eta(s)}{2}\right)\sin\left(\frac{\eta(s) + \Delta\eta(s)}{2}\right)\omega_{A,3}(s) = \sin^2\eta(s)\left(1 - \frac{R^2\left(\omega_{A,1}(s)^2 + \omega_{A,3}(s)^2\right)}{4}\right).$$
$$\tag{1.28}$$

We can subtract Eq. (1.27) from (1.28) obtaining

$$\sin\Delta\eta(s) = -\frac{R\omega_{A,3}(s)\sin\eta(s)}{2}. \tag{1.29}$$

We can also substitute Eq. (1.29) into either Eq (1.27) or (1.28), so obtaining a relationship between $\omega_{A,1}(s)$ and $\omega_{A,3}(s)$,

$$\omega_{A,1}(s) = -\frac{2}{R\sin\eta(s)}\left(\sqrt{1 - \frac{R^2\omega_{A,3}(s)^2\sin^2\eta(s)}{4}} - \cos\eta(s)\right). \tag{1.30}$$

We can substitute into Eqs (1.21)-and (1.22) obtaining

$$\omega_{\mu,1}(s) = \frac{\sin(\eta(s))}{2\sin\left(\frac{\eta(s)}{2}\right)C\left(R\sin\eta(s)\omega_{A,3}(s)\right) - 2\delta_\mu \cos\left(\frac{\eta(s)}{2}\right)S\left(R\sin\eta(s)\omega_{A,3}(s)\right)}$$

$$\left[\left(C\left(R\sin\eta(s)\omega_{A,3}(s)\right)\cos\left(\frac{\eta(s)}{2}\right) - \delta_\mu S\left(R\sin\eta(s)\omega_{A,3}(s)\right)\sin\left(\frac{\eta(s)}{2}\right)\right)\omega_{A,1}(s) \right. \quad (1.31)$$

$$\left. + \left(\delta_\mu \sin\left(\frac{\eta(s)}{2}\right)C\left(R\sin\eta(s)\omega_{A,3}(s)\right) + S\left(R\sin\eta(s)\omega_{A,3}(s)\right)\cos\left(\frac{\eta(s)}{2}\right)\right)\omega_{A,3}(s)\right],$$

$$\omega_{\mu,3}(s) = \frac{\sin(\eta(s))}{2\sin\left(\frac{\eta(s)}{2}\right)C\left(R\sin\eta(s)\omega_{A,3}(s)\right) - 2\delta_\mu \cos\left(\frac{\eta(s)}{2}\right)S\left(R\sin\eta(s)\omega_{A,3}(s)\right)}$$

$$\left[\left(C\left(R\sin\eta(s)\omega_{A,3}(s)\right)\cos\left(\frac{\eta(s)}{2}\right) - \delta_\mu S\left(R\sin\eta(s)\omega_{A,3}(s)\right)\sin\left(\frac{\eta(s)}{2}\right)\right)\omega_{A,3}(s) \right. \quad (1.32)$$

$$\left. - \left(\sin\left(\frac{\eta(s)}{2}\right)\delta_\mu C\left(R\sin\eta(s)\omega_{A,3}(s)\right) + S\left(R\sin\eta(s)\omega_{A,3}(s)\right)\cos\left(\frac{\eta(s)}{2}\right)\right)\omega_{A,1}(s)\right],$$

where

$$C(x) = \left(\frac{\sqrt{1-x^2/4}+1}{2}\right)^{1/2} \text{ and } S(x) = -\left(\frac{1-\sqrt{1-x^2/4}}{2}\right)^{1/2}. \quad (1.33)$$

To determine $\omega_{\mu,2}(s)$ as a function of $\omega_{A,2}(s)$ $\sin\eta(s)$ $\omega_{A,1}(s)$ and $\omega_{A,3}(s)$ we need equations for both $\eta_1'(s)$ and $\eta_2'(s)$ (the prime here means differentiation with respect to argument) which are got through differentiating Eq. (1.29) with respect to s and substituting the result for $\Delta\eta(s)'$ into Eq (1.23) ($\mu = 1, 2$)

$$\omega_{\mu,2}(s) = \frac{\sin(\eta(s))}{2\sin\left(\frac{\eta(s)}{2}\right)C\left(R\omega_{A,3}(s)\sin\eta(s)\right) - 2\delta_\mu \cos\left(\frac{\eta(s)}{2}\right)S\left(R\omega_{A,3}(s)\sin\eta(s)\right)}$$

$$\left(\omega_{A,2}(s) - \frac{\delta_\mu}{2}\frac{d\eta(s)}{ds} + \frac{R\left(\omega_{A,3}(s)\frac{d\eta(s)}{ds}\cos\eta(s) + \frac{d\omega_{A,3}(s)}{ds}\sin\eta(s)\right)}{2\sqrt{1-R^2\omega_{A,3}(s)^2\sin^2\eta(s)}}\right). \quad (1.34)$$

### 1.2 Definitions of terminology for specific cases of braid

We define a *symmetric* braid as one for which $\sigma_1(s) = \sigma_2(s)$, here Eqs. (1.12) and (1.29) must be satisfied and $\omega_{A,3}(s) = 0$.

We define a *regular* braid as one for which

$$\frac{d\eta(s)}{ds} = 0. \tag{1.35}$$

A *straight* braid is braid with a straight axis, which means that

$$\frac{d\alpha(s)}{ds} = 0 \text{ and } \frac{d\beta(s)}{ds} = 0, \tag{1.36}$$

and therefore is a *symmetric* braid for which $\omega_{A,2}(s) = 0$.

### *1.3 Describing the trajectories of the helices*

We will now characterize the trajectories of helices. The helices will be taken lie on the surface of the rods, which we suppose have the same radius. The circular rod cross sections, of radius $a$, lie in the planes spanned by: $\hat{\mathbf{d}}(s)$ and $\hat{\mathbf{n}}_1(s)$ for rod 1, as well as $\hat{\mathbf{d}}(s)$ and $\hat{\mathbf{n}}_2(s)$ for rod 2. Therefore, the position vectors that trace out the helices are given by

$$\mathbf{r}_{H,1}(s) = \mathbf{r}_1(s) + a\hat{\mathbf{v}}_1(s), \quad \mathbf{r}_{H,2}(s) = \mathbf{r}_2(s) + a\hat{\mathbf{v}}_2(s). \tag{1.37}$$

The vectors $\hat{\mathbf{v}}_1(s)$ and $\hat{\mathbf{v}}_2(s)$ describe the orientations of the helices at point $s$ relative to $\hat{\mathbf{d}}(s)$. We can define the angle that $\hat{\mathbf{v}}_\mu(s)$ makes with $\hat{\mathbf{d}}(s)$, which we call $\xi_\mu(s)$ ($\mu = 1, 2$). Therefore, we can write

$$\hat{\mathbf{v}}_\mu(s) \cdot \hat{\mathbf{d}}(s) = \cos \xi_\mu(s), \tag{1.38}$$

and $\hat{\mathbf{v}}_\mu(s)$ can be parameterized as

$$\hat{\mathbf{v}}_\mu(s) = \mathbf{T}_\beta(s)\mathbf{T}_\alpha(s)\mathbf{T}_{\phi_0}(s)\begin{pmatrix} \cos \xi_\mu(s) \\ \sin \xi_\mu(s) \cos \eta_\mu(s) \\ -\delta_\mu \sin \xi_\mu(s) \sin \eta_\mu(s) \end{pmatrix}. \tag{1.39}$$

We can also define the vectors $\hat{\mathbf{u}}_\mu(s) = \hat{\mathbf{t}}_\mu(s) \times \hat{\mathbf{v}}_\mu(s)$, which are

$$\hat{\mathbf{u}}_\mu(s) = \mathbf{T}_\beta(s)\mathbf{T}_\alpha(s)\mathbf{T}_{\phi_0}(s)\begin{pmatrix} -\sin \xi_\mu(s) \\ \cos \xi_\mu(s) \cos \eta_1(s) \\ -\delta_\mu \cos \xi_\mu(s) \sin \eta_\mu(s) \end{pmatrix}. \tag{1.40}$$

For the rotating frames, $\{\hat{\mathbf{v}}_1, \hat{\mathbf{u}}_1, \hat{\mathbf{t}}_1\}$ and $\{\hat{\mathbf{v}}_2, \hat{\mathbf{u}}_2, \hat{\mathbf{t}}_2\}$ that describe the orientation of the helices about the centre lines, we may define the angular frequency vectors ($\mu = 1, 2$)

$$\tilde{\mathbf{\Omega}}_\mu(s) = \tilde{\omega}_{\mu,1}(s)\hat{\mathbf{t}}_\mu(s) + \tilde{\omega}_{\mu,2}(s)\hat{\mathbf{v}}_\mu(s) + \tilde{\omega}_{\mu,3}(s)\hat{\mathbf{u}}_\mu(s). \tag{1.41}$$

Here, $\tilde{\omega}_{\mu,1}(s)$ is the rate of precession of $\hat{\mathbf{v}}_\mu(s)$ and $\hat{\mathbf{u}}_\mu(s)$ about $\hat{\mathbf{t}}_\mu(s)$, $\omega_{\mu,2}(s)$ is the rate of precession of $\hat{\mathbf{t}}_\mu(s)$ and $\hat{\mathbf{u}}_\mu(s)$ about $\hat{\mathbf{v}}_\mu(s)$ and $\omega_{\mu,3}(s)$ is the rate of precession of $\hat{\mathbf{t}}_\mu(s)$ and $\hat{\mathbf{v}}_\mu(s)$ about $\hat{\mathbf{u}}_\mu(s)$. This allows us to write ($\mu = 1,2$):

$$\frac{d\hat{\mathbf{v}}_\mu(s)}{ds} = \sigma_\mu(s)\left(\mathbf{\Omega}_\mu(s) \times \hat{\mathbf{v}}_\mu(s)\right) = \sigma_\mu(s)\left(\tilde{\omega}_{\mu,1}(s)\hat{\mathbf{u}}_\mu(s) - \tilde{\omega}_{\mu,3}(s)\hat{\mathbf{t}}_\mu(s)\right), \tag{1.42}$$

$$\frac{d\hat{\mathbf{u}}_\mu(s)}{ds} = \sigma_\mu(s)\left(\mathbf{\Omega}_\mu(s) \times \hat{\mathbf{u}}_\mu(s)\right) = \sigma_1(s)\left(\tilde{\omega}_{\mu,2}(s)\hat{\mathbf{t}}_\mu(s) - \tilde{\omega}_{\mu,1}(s)\hat{\mathbf{v}}_\mu(s)\right), \tag{1.43}$$

$$\frac{d\hat{\mathbf{t}}_\mu(s)}{ds} = \sigma_\mu(s)\left(\mathbf{\Omega}_\mu(s) \times \hat{\mathbf{t}}_\mu(s)\right) = \sigma_\mu(s)\left(\tilde{\omega}_{\mu,3}(s)\hat{\mathbf{v}}_1(s) - \tilde{\omega}_{\mu,2}(s)\hat{\mathbf{u}}_1(s)\right). \tag{1.44}$$

Using Eqs. (1.40) and (1.42)-(1.44) allows us to express

$$\tilde{\omega}_{\mu,1}(s) = \omega_{\mu,1}(s) + \frac{1}{\sigma_\mu(s)}\frac{d\xi_\mu(s)}{ds}, \tag{1.45}$$

$$\tilde{\omega}_{\mu,2}(s) = \omega_{\mu,2}(s)\cos\xi_\mu(s) + \omega_{\mu,3}(s)\sin\xi_\mu(s), \tag{1.46}$$

$$\tilde{\omega}_{\mu,3}(s) = \omega_{\mu,3}(s)\cos\xi_2(s) - \omega_{\mu,2}(s)\sin\xi_2(s). \tag{1.47}$$

### 1.4 Describing the surface of the rods and introducing zero torsion frames

The surfaces of the rods forming the braid are described through vector equations

$$\mathbf{S}_1(s,t) = \mathbf{r}_1(s) + a\hat{\mathbf{N}}_1(s,t), \quad \mathbf{S}_2(s,t) = \mathbf{r}_2(s) + a\hat{\mathbf{N}}_2(s,t). \tag{1.48}$$

As well as the arc-length of the braid axis coordinate $s$ we have coordinate $t$, running from $0$ to $\pi$ that give the position on the circumference of the rod cross sections at fixed $s$. The unit normal vectors $\hat{\mathbf{N}}_\mu(s,t)$ ($\mu = 1,2$) to the surface obey the condition

$$\hat{\mathbf{N}}_\mu(s,t).\hat{\mathbf{t}}(s) = 0. \tag{1.49}$$

We can satisfy these constraints by writing:

$$\hat{\mathbf{N}}_\mu(s,t) = \cos t\, \hat{\mathbf{U}}_\mu(s) + \sin t\, \hat{\mathbf{V}}_\mu(s). \tag{1.50}$$

In what follows it will be convenient to choose basis vectors $\hat{\mathbf{U}}_\mu$ and $\hat{\mathbf{V}}_\mu$ so that they **do not precess** around the tangent vector. It is also useful to define a vector set $\hat{\mathbf{U}}_A, \hat{\mathbf{V}}_A, \hat{\mathbf{t}}_A$ for which ($\mu = 1, 2$)

$$\hat{\mathbf{V}}_\mu(s) = \cos\eta_\mu(s)\hat{\mathbf{V}}_A(s) - \delta_\mu \sin\eta_\mu(s)\hat{\mathbf{t}}_A(s) \tag{1.51}$$

This means that we define new angular frequency vectors (for $\mu = 1, 2, A$):

$$\bar{\mathbf{\Omega}}_\mu(s) = \bar{\omega}_{\mu,2}(s)\hat{\mathbf{U}}_1(s) + \bar{\omega}_{\mu,3}(s)\hat{\mathbf{V}}_1(s), \tag{1.52}$$

Here, $\bar{\omega}_{\mu,2}(s)$ is the rate of precession of $\hat{\mathbf{t}}_\mu(s)$ and $\hat{\mathbf{V}}_\mu(s)$ about $\hat{\mathbf{U}}_\mu(s)$ and $\bar{\omega}_{\mu,3}(s)$ is the rate of precession of $\hat{\mathbf{t}}_\mu(s)$ and $\hat{\mathbf{U}}_\mu(s)$ about $\hat{\mathbf{V}}_\mu(s)$. This leads to the following relations for (note that $\sigma_A(s) = 1$)

$$\frac{d\hat{\mathbf{U}}_\mu(s)}{ds} = \sigma_\mu(s)\left(\bar{\mathbf{\Omega}}_\mu(s) \times \hat{\mathbf{U}}_\mu(s)\right) = -\sigma_\mu(s)\bar{\omega}_{\mu,3}(s)\hat{\mathbf{t}}_\mu(s), \tag{1.53}$$

$$\frac{d\hat{\mathbf{V}}_\mu(s)}{ds} = \sigma_\mu(s)\left(\bar{\mathbf{\Omega}}_\mu(s) \times \hat{\mathbf{V}}_\mu(s)\right) = \sigma_\mu(s)\bar{\omega}_{\mu,2}(s)\hat{\mathbf{t}}_\mu(s), \tag{1.54}$$

$$\frac{d\hat{\mathbf{t}}_\mu(s)}{ds} = \sigma_\mu(s)\left(\bar{\mathbf{\Omega}}_\mu(s) \times \hat{\mathbf{t}}_\mu(s)\right) = \sigma_\mu(s)\left(\bar{\omega}_{\mu,3}\hat{\mathbf{U}}_\mu(s) - \bar{\omega}_{\mu,2}\hat{\mathbf{V}}_\mu(s)\right). \tag{1.55}$$

Note that omega bars are related to the braid frequencies through the following relations

$$\bar{\omega}_{\mu,2}(s) = \omega_{\mu,2}(s)\cos\gamma_\mu(s;s_0) - \omega_{\mu,3}(s)\sin\gamma_\mu(s;s_0), \tag{1.56}$$

$$\bar{\omega}_{\mu,3}(s) = \omega_{\mu,3}(s)\cos\gamma_\mu(s;s_0) + \omega_{\mu,2}(s)\sin\gamma_\mu(s;s_0), \tag{1.57}$$

where the phases $\gamma_\mu(s;s_0)$ can be defined as

$$\gamma_\mu(s;s_0) = \int_{s_0}^{s} \sigma_\mu(s')\omega_{\mu,1}(s')ds', \tag{1.58}$$

We can choose the point $s = s_0$ so that

$$\hat{\mathbf{U}}_\mu(s_0) = \hat{\mathbf{d}}(s_0), \quad \hat{\mathbf{V}}_\mu(s_0) = \hat{\mathbf{n}}_\mu(s_0). \tag{1.59}$$

### 1.5 Some geometric formulas for small $R\omega_{A,3}(s)$

For small $R\omega_{A,3}(s)$ we may write

$$\omega_{1,A}(s) \approx -\frac{2(1-\cos\eta(s))}{R\sin\eta(s)} + O(R^2\omega_{3,A}(s)^2). \tag{1.60}$$

Expanding out in powers of $R\omega_{A,3}(s)$ we may write from Eqs. (1.12), (1.31)-(1.34) the following

$$\sin\eta_\mu(s) = \sin\left(\frac{\eta(s)}{2}\right) - \frac{\delta_\mu R \omega_{3,A}(s)}{4}\cos\left(\frac{\eta(s)}{2}\right)\sin(\eta(s)), \tag{1.61}$$

$$\sigma_\mu(s) \approx \frac{1}{\cos\left(\frac{\eta(s)}{2}\right)} + \frac{\delta_\mu R \omega_{A,3}(s)}{2}\cos\left(\frac{\eta(s)}{2}\right), \tag{1.62}$$

$$\omega_{\mu,1}(s) \approx \cos\left(\frac{\eta(s)}{2}\right)\left(\delta_\mu \sin\left(\frac{\eta(s)}{2}\right)\omega_{A,3}(s)\right.$$
$$\left. - \frac{2(1-\cos\eta(s))}{R\sin\eta(s)}\left(\left(1 - \frac{\delta_\mu R\omega_{A,3}(s)\cos^2\left(\frac{\eta(s)}{2}\right)}{2}\right)\cos\left(\frac{\eta(s)}{2}\right) + \frac{\delta_\mu R\sin\eta(s)\omega_{A,3}(s)}{4}\sin\left(\frac{\eta(s)}{2}\right)\right)\right), \tag{1.63}$$

$$\omega_{\mu,3}(s) = \cos\left(\frac{\eta(s)}{2}\right)\left(\cos\left(\frac{\eta(s)}{2}\right)\omega_{A,3}(s)\right.$$
$$\left. + \frac{2(1-\cos\eta(s))}{R\sin\eta(s)}\left(\left(1 - \frac{\delta_\mu R\omega_{A,3}(s)\cos^2\left(\frac{\eta(s)}{2}\right)}{2}\right)\sin\left(\frac{\eta(s)}{2}\right)\delta_\mu - \frac{\delta_\mu R\sin\eta(s)\omega_{A,3}(s)}{4}\cos\left(\frac{\eta(s)}{2}\right)\right)\right), \tag{1.64}$$

$$\omega_{\mu,2}(s) \approx \cos\left(\frac{\eta(s)}{2}\right)$$
$$\left(\left(\omega_{A,2}(s) - \frac{\delta_\mu}{2}\frac{d\eta(s)}{ds}\right)\left(1 - \frac{\delta_\mu R\omega_{A,3}(s)\cos^2\left(\frac{\eta(s)}{2}\right)}{2}\right) + \frac{R\left(\omega_{A,3}(s)\frac{d\eta(s)}{ds}\cos\eta(s) + \frac{d\omega_{A,3}(s)}{ds}\sin\eta(s)\right)}{2}\right). \tag{1.65}$$

It is also useful to write Eqs. (1.60)-(1.65) for small $\sin\eta(s)$ (where we have neglected $O(R\omega_{A,3}\sin\eta(s))$)

$$\sin\eta_\mu(s) = \frac{\sin(\eta(s))}{2}\left(1 - \frac{\delta_\mu R\omega_{3,A}(s)}{2}\right) \qquad \omega_{1,A}(s) \approx -\frac{\sin\eta(s)}{R} \tag{1.66}$$

$$\sigma_\mu(s) \approx 1 + \frac{\delta_\mu R}{2}\omega_{A,3}(s) \approx 1 + \frac{\delta_\mu R}{2}\omega_{\mu,3}(s) \quad \omega_{\mu,3}(s) \approx \omega_{A,3}(s) \quad \omega_{\mu,1}(s) \approx \omega_{A,1}(s) \tag{1.67}$$

## 2. Interaction energy without dielectric cores

Now, let us consider the helices to be uniformly charged lines with linear charge density $e/l_c$ per unit arc length of the centre lines of rods 1 and 2. We suppose that both helices lie in a medium which has dielectric constant $\varepsilon_w$ and also that the electrostatic interaction is Debye screened with inverse screening length $\kappa_D$, so that the electrostatic potential, at the point $\mathbf{r}$, of a point charge of charge $e$ at $\mathbf{r}'$ is given by (in Gaussian units)

$$\phi_{\text{point}}(\mathbf{r}-\mathbf{r}') = \frac{4\pi e}{\varepsilon_w} \frac{1}{(2\pi)^3} \int d^3k \, \frac{\exp(i\mathbf{k}.(\mathbf{r}-\mathbf{r}'))}{\mathbf{k}^2 + \kappa_D^2}. \tag{2.1}$$

By summing up all the point charge contributions along helices (using Eq. (2.1)) we can write the interaction between two charged single helices in a braided configuration as

$$E = \frac{4\pi e^2}{\varepsilon_w (2\pi)^3 l_c^2} \int_0^{L_B} ds \int_0^{L_B} ds' \sigma_2(s') \sigma_1(s) \int d^3k \, \frac{\exp(i\mathbf{k}.(\mathbf{r}_1(s) - \mathbf{r}_2(s')))}{\mathbf{k}^2 + \kappa_D^2} \exp(ia\mathbf{k}.(\hat{\mathbf{v}}_1(s) - \hat{\mathbf{v}}_2(s'))),$$

$$\tag{2.2}$$

where we have also used Eq. (1.37). Now we will attempt to evaluate Eq. (2.2) for a braid where the angular frequencies described by Eq. (1.13) are smaller than the inverse Debye screening length, which admits a perturbation expansion. First, we change variables to $\tau = \frac{s+s'}{2}$ and $\tau' = \frac{s-s'}{2}$ which gives

$$E = \frac{8\pi e^2}{\varepsilon_w (2\pi)^3 l_c^2} \int_0^{L_B} d\tau \int_{-\infty}^{\infty} d\tau' \sigma_2(\tau - \tau') \sigma_1(\tau + \tau') \times$$
$$\int d^3k \, \frac{\exp(i\mathbf{k}.(\mathbf{r}_1(\tau + \tau') - \mathbf{r}_2(\tau - \tau')))}{\mathbf{k}^2 + \kappa_D^2} \exp(ia\mathbf{k}.(\hat{\mathbf{v}}_1(\tau + \tau') - \hat{\mathbf{v}}_2(\tau - \tau'))). \tag{2.3}$$

Here, we have approximated the limits of integration on the $\tau'$ integral with plus and minus infinity, which is fine provided that the length of the braid exceeds all other length scales. Next, we Taylor expand the position vectors of the centre lines in powers of $\tau'$ about the point $\tau$

$$\mathbf{r}_1(\tau + \tau') - \mathbf{r}_2(\tau - \tau') \approx -R\hat{\mathbf{d}}(\tau) + \tau'(\sigma_1(\tau)\hat{\mathbf{t}}_1(\tau) + \sigma_2(\tau)\hat{\mathbf{t}}_2(\tau)) = -R\hat{\mathbf{d}}(\tau) + 2\tau'\hat{\mathbf{t}}_A(\tau), \tag{2.4}$$

where we have used Eqs. (1.1), (1.2) and (1.5).

Now we use the results of Section 1.4 (where we have chosen our points coincidence Eq. (1.59) at the points where the charges lie). Note that the super script $(\mu)$ on $\hat{\mathbf{U}}_A^{(\mu)}(\tau + \tau')$, implies that the chosen coincidence is for the charges on rod $\mu$.

$$\hat{\mathbf{v}}_1(\tau+\tau') = \cos\xi_1(\tau+\tau')\hat{\mathbf{d}}(\tau+\tau') + \sin\xi_1(\tau+\tau')\hat{\mathbf{n}}_1(\tau+\tau') = \hat{\mathbf{N}}_1(\tau+\tau',\xi_1(\tau+\tau'))$$
$$= \cos\xi_1(\tau+\tau')\hat{\mathbf{U}}_A^{(1)}(\tau+\tau') + \sin\xi_1(\tau+\tau')\cos\eta_1(\tau+\tau')\hat{\mathbf{V}}_A^{(1)}(\tau+\tau') - \sin\xi_1(\tau+\tau')\sin\eta_1(\tau+\tau')\hat{\mathbf{t}}_A(\tau+\tau'),$$
(2.5)

$$\hat{\mathbf{v}}_2(\tau-\tau') = \cos\xi_2(\tau-\tau')\hat{\mathbf{d}}(\tau-\tau') + \sin\xi_2(\tau-\tau')\hat{\mathbf{n}}_2(\tau-\tau') = \hat{\mathbf{N}}_2(\tau-\tau',\xi_1(\tau-\tau'))$$
$$= \cos\xi_2(\tau-\tau')\hat{\mathbf{U}}_A^{(2)}(\tau-\tau') + \sin\xi_2(\tau-\tau')\cos\eta_2(\tau-\tau')\hat{\mathbf{V}}_A^{(2)}(\tau-\tau') + \sin\xi_2(\tau-\tau')\sin\eta_2(\tau-\tau')\hat{\mathbf{t}}_A(\tau-\tau').$$
(2.6)

Now, we neglect braid axis curvature by making the approximation:

$$\hat{\mathbf{v}}_1(\tau+\tau') \approx \cos\xi_1(\tau+\tau')\hat{\mathbf{U}}_A^{(1)}(\tau) + \sin\xi_1(\tau+\tau')\cos\eta_1(\tau+\tau')\hat{\mathbf{V}}_A^{(1)}(\tau) - \sin\xi_1(\tau+\tau')\sin\eta_1(\tau+\tau')\hat{\mathbf{t}}_A(\tau),$$
(2.7)

$$\hat{\mathbf{v}}_2(\tau-\tau') \approx \cos\xi_2(\tau-\tau')\hat{\mathbf{U}}_A^{(2)}(\tau) + \sin\xi_2(\tau-\tau')\cos\eta_2(\tau-\tau')\hat{\mathbf{V}}_A^{(2)}(\tau) + \sin\xi_2(\tau-\tau')\sin\eta_2(\tau-\tau')\hat{\mathbf{t}}_A(\tau).$$
(2.8)

We can see that the next to leading order expansion of $\hat{\mathbf{U}}_A^{(\mu)}(\tau)$ and $\hat{\mathbf{V}}_A^{(\mu)}(\tau)$ will generate up linear terms in both $\bar{\omega}_{A,2}(s)$ and $\bar{\omega}_{A,3}(s)$ (see Eqs. (1.53) and (1.54)).

Then, it is useful to perform local transformations to frames, at each value of $\tau$, where $\hat{\mathbf{t}}_A(\tau).\mathbf{k} = k_z$ $\hat{\mathbf{d}}(\tau).\mathbf{k} = k_x$. Under these transformations the Jacobian is 1, and $\mathbf{k}^2$ is an invariant and each transformation can be done independently. Therefore we are able to write

$$E \approx \frac{8\pi e^2}{\varepsilon_w (2\pi)^3 l_c^2} \int_0^{L_B} d\tau \int_{-\infty}^{\infty} d\tau' \sigma_2(\tau-\tau')\sigma_1(\tau+\tau') \int d^3k \frac{\exp(-ik_x R)\exp(2ik_z \tau)}{\mathbf{k}^2 + \kappa_D^2} \exp(ia\mathbf{k}.(\hat{\mathbf{v}}_1'(\tau,\tau') - \hat{\mathbf{v}}_2'(\tau,\tau'))),$$
(2.9)

where now, in these transformed frames, we have

$$\hat{\mathbf{v}}_1'(\tau,\tau') \approx \tilde{\mathbf{T}}(-\gamma_A(\tau;\tau'+\tau)) \begin{pmatrix} \cos\xi_1(\tau+\tau') \\ \sin\xi_1(\tau+\tau')\cos\eta_1(\tau+\tau') \\ -\sin\xi_1(\tau+\tau')\sin\eta_1(\tau+\tau') \end{pmatrix},$$
(2.10)

and

$$\hat{\mathbf{v}}_2'(\tau,\tau') \approx \tilde{\mathbf{T}}(-\gamma_A(\tau;\tau-\tau')) \begin{pmatrix} \cos\xi_2(\tau-\tau') \\ \sin\xi_2(\tau-\tau')\cos\eta_2(\tau-\tau') \\ \sin\xi_2(\tau-\tau')\sin\eta_2(\tau-\tau') \end{pmatrix},$$
(2.11)

.

where $\tilde{\mathbf{T}}(\phi_0(\tau)) = \mathbf{T}_{\phi_0}(\tau)$. (2.12)

In finding Eqs. (2.10) and (2.11), we have used the relationships Eq. (1.56) and (1.57). Direct substitution of Eqs. (2.10) and (2.11) into Eq. (2.9) allows us to rewrite the interaction energy

$$E \approx \frac{8\pi e^2}{\varepsilon_w (2\pi)^3 l_c^2} \int_0^{L_B} d\tau \int_{-\infty}^{\infty} d\tau' \sigma_2(\tau-\tau')\sigma_1(\tau+\tau') \int_{-\infty}^{\infty} dk_z \int_0^{2\pi} d\phi_K \int_0^{\infty} KdK \frac{\exp(-iRK\cos(\phi_K))}{\mathbf{k}^2 + \kappa_D^2}$$
$$\exp(iaKR_{H1}(\tau+\tau')\cos(\tilde{\xi}_1(\tau+\tau')+\gamma_A(\tau+\tau';\tau)-\phi_K))$$
$$\exp(-iaKR_{H2}(\tau-\tau')\cos(\tilde{\xi}_2(\tau-\tau')+\gamma_A(\tau+\tau';\tau)-\phi_K))$$
$$\exp(ik_z(2\tau' + aZ_{H1}(\tau+\tau') - aZ_{H2}(\tau-\tau'))),$$
(2.13)

where

$$R_{H\mu}(\tau+\tau') = \sqrt{\cos(\xi_\mu(\tau+\tau'))^2 + \cos^2\eta_\mu(\tau+\tau')\sin(\xi_\mu(\tau+\tau'))^2},$$
(2.14)

$$\tilde{\xi}_\mu(\tau+\tau') = \tan^{-1}\left(\cos\eta_\mu(\tau+\tau')\tan(\xi_\mu(\tau+\tau'))\right),$$
(2.15)

$$Z_{H\mu}(\tau-\tau') = -\delta_\mu \sin\xi_\mu(\tau-\tau')\sin\eta_\mu(\tau-\tau').$$
(2.16)

We will now use the following mathematical identities:

$$\exp(-iRK\cos(\phi_K)) \equiv \sum_m i^{-m} J_m(RK) \exp(-im\phi_K),$$
(2.17)

$$\exp(iaKR_{H1}(\tau+\tau')\cos(\tilde{\xi}_1(\tau+\tau')+\gamma_A(\tau+\tau';\tau)-\phi_K))$$
$$\equiv \sum_n i^n J_n(aKR_{H1}(\tau+\tau')) \exp(in(\tilde{\xi}_1(\tau+\tau')+\gamma_A(\tau+\tau';\tau)-\phi_K)),$$
(2.18)

$$\exp(-iaKR_{H2}(\tau-\tau')\cos(\tilde{\xi}_2(\tau-\tau')+\gamma_A(\tau-\tau';\tau)-\phi_K))$$
$$\equiv \sum_n i^{-n'} J_{n'}(aKR_{H2}(\tau-\tau')) \exp(-in'(\tilde{\xi}_2(\tau-\tau')+\gamma_A(\tau-\tau';\tau)-\phi_K))).$$
(2.19)

Eqs. (2.17)-(2.19) allow us to perform the $\phi_K$ integral in Eq. (2.13) and so write

$$E = \frac{8\pi e^2}{\varepsilon_w (2\pi)^2 l_c^2} \sum_{n,n'} \int_0^{L_B} d\tau \int_{-\infty}^{\infty} d\tau' \sigma_2(\tau-\tau')\sigma_1(\tau+\tau') \int_{-\infty}^{\infty} dk_z \int_0^{\infty} KdK \frac{J_{n-n'}(RK) J_n(aKR_{H1}(\tau+\tau')) J_{n'}(aKR_{H2}(\tau-\tau'))}{K^2 + k_z^2 + \kappa_D^2}$$
$$\exp(in(\tilde{\xi}_1(\tau+\tau')+\gamma_A(\tau+\tau';\tau)))\exp(-in'(\tilde{\xi}_2(\tau-\tau')+\gamma_A(\tau+\tau';\tau)))$$
$$\exp(ik_z(2\tau' + aZ_{H1}(\tau+\tau') - aZ_{H2}(\tau-\tau'))).$$
(2.20)

The integral over $K$ may then be evaluated leaving us with

$$E = \frac{8\pi e^2}{\varepsilon_w (2\pi)^2 l_c^2} \sum_{n,n'} \int_0^{L_B} d\tau \int_{-\infty}^{\infty} d\tau' \sigma_2(\tau-\tau')\sigma_1(\tau+\tau') \int_{-\infty}^{\infty} dk_z (-1)^{n'} K_{n'-n}\left(R\sqrt{k_z^2+\kappa_D^2}\right) I_n\left(aR_{H1}(\tau+\tau')\sqrt{k_z^2+\kappa_D^2}\right)$$
$$I_{n'}\left(aR_{H2}(\tau-\tau')\sqrt{k_z^2+\kappa_D^2}\right)\exp(in(\tilde{\xi}_1(\tau+\tau')+\gamma_A(\tau+\tau';\tau)))\exp(-in'(\tilde{\xi}_2(\tau-\tau')+\gamma_A(\tau-\tau';\tau)))$$
$$\exp(ik_z(2\tau' + aZ_{H1}(\tau+\tau') - aZ_{H2}(\tau-\tau'))).$$
(2.21)

Now, we can perform a trick. We notice that we could have written

$$\exp(-iaKR_H(\tau-\tau')\cos(\tilde{\xi}_2(\tau-\tau')+\gamma_A(\tau-\tau';\tau)-\phi_k))$$
$$=\sum_{m,n} i^{-m} J_{m-n}\left(\frac{aK}{2}(1-\cos\eta_2(\tau-\tau'))\right) J_n\left(\frac{aK}{2}(1+\cos\eta_2(\tau-\tau'))\right) \qquad (2.22)$$
$$\exp(im(\phi_k-\gamma(\tau-\tau';\tau)))\exp(-2in\xi_2(\tau-\tau'))\exp(im\xi_2(\tau-\tau')).$$

This, then, suggests the following addition formula:

$$J_m(aKR_{H2}(\tau-\tau'))\exp(-im\tilde{\xi}_2(\tau-\tau'))=$$
$$=\sum_n J_{m-n}\left(\frac{aK}{2}(1-\cos\eta_2(\tau-\tau'))\right) J_n\left(\frac{aK}{2}(1+\cos\eta_2(\tau-\tau'))\right)\exp(-2in\xi_2((\tau-\tau')))\exp(im\xi_2((\tau-\tau'))),$$
$$(2.23)$$

or alternatively (analytically continuing Eq.(2.23) into the complex plane) we may write

$$I_{n'}\left(aR_{H2}(\tau-\tau')\sqrt{k_z^2+\kappa_D^2}\right)\exp(-in'\tilde{\xi}_2(\tau-\tau'))$$
$$=\sum_{m'} I_{n'-m'}\left(\frac{a\sqrt{k_z^2+\kappa_D^2}}{2}(1-\cos\eta_2(\tau-\tau'))\right) I_{m'}\left(\frac{a\sqrt{k_z^2+\kappa_D^2}}{2}(1+\cos\eta_2(\tau-\tau'))\right) \qquad (2.24)$$
$$\exp(-2im'\xi_2(\tau-\tau'))\exp(in'\xi_2(\tau-\tau')).$$

Similarly, we find the expression

$$I_n\left(aR_{H1}(\tau+\tau')\sqrt{k_z^2+\kappa_D^2}\right)\exp(in\tilde{\xi}_1(\tau+\tau'))$$
$$=\sum_m I_{n-m}\left(\frac{a\sqrt{k_z^2+\kappa_D^2}}{2}(1-\cos\eta_1(\tau+\tau'))\right) I_m\left(\frac{a\sqrt{k_z^2+\kappa_D^2}}{2}(1+\cos\eta_1(\tau+\tau'))\right) \qquad (2.25)$$
$$\exp(2im\xi_1(\tau+\tau'))\exp(-in\xi_1(\tau+\tau')).$$

We can also express

$$\exp(iak_z\delta_\mu Z_{H\mu}(\tau+\tau'))=\exp\left(-iak_z\sin\eta_\mu(\tau+\tau')\sin\left(\xi_\mu(\tau+\tau')\right)\right)$$
$$=\sum_j J_j(ak_z\sin\eta_\mu(\tau+\tau'))\exp(-ij(\xi_\mu(\tau+\tau'))). \qquad (2.26)$$

Substitution of Eqs. (2.24)-(2.26) into Eq. (2.21) yields the following result

$$E = \frac{8\pi e^2}{\varepsilon_w (2\pi)^2 l_c^2} \sum_{n,n',m,m',j,j'} \int_0^{L_B} d\tau \int_{-\infty}^{\infty} d\tau' \sigma_2(\tau-\tau')\sigma_1(\tau+\tau') \int_{-\infty}^{\infty} dk_z (-1)^{n'} K_{n'-n}\left(R\sqrt{k_z^2+\kappa_D^2}\right) I_{n'-m'}\left(\frac{a\sqrt{k_z^2+\kappa_D^2}}{2}(1-\cos\eta_2(\tau-\tau'))\right)$$

$$I_{m'}\left(\frac{a\sqrt{k_z^2+\kappa_D^2}}{2}(1+\cos\eta_2(\tau-\tau'))\right) I_{n-m}\left(\frac{a\sqrt{k_z^2+\kappa_D^2}}{2}(1-\cos\eta_1(\tau+\tau'))\right) I_m\left(\frac{a\sqrt{k_z^2+\kappa_D^2}}{2}(1+\cos\eta_1(\tau+\tau'))\right)$$

$$\exp(-in(\xi_1(\tau+\tau')-\gamma_A(\tau+\tau';\tau)))\exp(in'(\xi_2(\tau-\tau')-\gamma_A(\tau-\tau';\tau)))\exp(-2im'\xi_2(\tau-\tau'))\exp(2im\xi_1(\tau+\tau'))$$
$$\exp(2ik_z\tau')J_j(ak_z\sin\eta_1(\tau+\tau'))J_{j'}(ak_z\sin\eta_2(\tau-\tau'))\exp(-ij'(\xi_2(\tau-\tau')))\exp(-ij(\xi_1(\tau+\tau'))).$$
(2.27)

To proceed any further and make the theory local we need to perform the following additional Taylor expansions

$$\eta_1(\tau+\tau')\approx\eta_1(\tau),\quad \eta_2(\tau-\tau')\approx\eta_2(\tau),\quad \sigma_1(\tau+\tau')\approx\sigma_1(\tau),\quad \sigma_2(\tau-\tau')\approx\sigma_2(\tau). \quad (2.28)$$

$$\xi_1(\tau+\tau')\approx\xi_1(\tau)+\frac{d\xi_1(\tau)}{d\tau}\tau',\quad \xi_1(\tau-\tau')\approx\xi_1(\tau)-\frac{d\xi_1(\tau)}{d\tau}\tau', \quad (2.29)$$

$$\gamma_A(\tau+\tau';\tau)\approx\tau'\omega_{A,1}(s)\quad \gamma_A(\tau-\tau';\tau)\approx-\tau'\omega_{A,1}(\tau), \quad (2.30)$$

Substitution of these expansions, Eqs. (2.28)-(2.30), into (2.27) gives

$$E \approx \frac{8\pi e^2}{\varepsilon_w (2\pi)^2 l_c^2} \sum_{n,n',m,m',j,j'} \int_0^{L_B} d\tau \int_{-\infty}^{\infty} d\tau' \sigma_2(\tau)\sigma_1(\tau) \int_{-\infty}^{\infty} dk_z (-1)^{n'} K_{n'-n}\left(R\sqrt{k_z^2+\kappa_D^2}\right) I_{n'-m'}\left(\frac{a\sqrt{k_z^2+\kappa_D^2}}{2}(1-\cos\eta_2(\tau))\right)$$

$$I_{m'}\left(\frac{a\sqrt{k_z^2+\kappa_D^2}}{2}(1+\cos\eta_2(\tau))\right) I_{n-m}\left(\frac{a\sqrt{k_z^2+\kappa_D^2}}{2}(1-\cos\eta_1(\tau))\right) I_m\left(\frac{a\sqrt{k_z^2+\kappa_D^2}}{2}(1+\cos\eta_1(\tau))\right)\exp(2ik_z\tau')$$

$$\exp\left(-i(n\xi_1(\tau)-n'\xi_2(\tau))\right)\exp(2i(m\xi_1(\tau)-m'\xi_2(\tau)))\exp(i(-j'\xi_2(\tau)-j\xi_1(\tau)))J_j(ak_z\sin\eta_1(\tau))J_{j'}(ak_z\sin\eta_2(\tau))$$

$$\exp\left(i\tau'\left(n\left(\omega_{A,1}(\tau)-\frac{d\xi_1(\tau)}{d\tau}\right)+n'\left(\omega_{A,1}(\tau)-\frac{d\xi_2(\tau)}{d\tau}\right)\right)\right)\exp\left(i\tau'\left((2m-j)\frac{d\xi_1(\tau)}{d\tau}+(2m'+j')\frac{d\xi_2(\tau)}{d\tau}\right)\right).$$
(2.31)

On performing the $\tau'$ integration and the $k_z$ integration we obtain

$$E \approx \frac{2e^2}{\varepsilon_w l_c^2} \sum_{n,n',m,m',j,j'} \int_0^{L_B} d\tau \sigma_2(\tau)\sigma_1(\tau)(-1)^{n'} K_{n'-n}\left(R\kappa_{n,n',m,m',j,j'}\right) I_{n'-m'}\left(\frac{a\kappa_{n,n',m,m',j,j'}}{2}(1-\cos\eta_2(\tau))\right)$$

$$I_{m'}\left(\frac{a\kappa_{n,n',m,m',j,j'}}{2}(1+\cos\eta_2(\tau))\right) I_{n-m}\left(\frac{a\kappa_{n,n',m,m',j,j'}}{2}(1-\cos\eta_1(\tau))\right) I_m\left(\frac{a\kappa_{n,n',m,m',j,j'}}{2}(1+\cos\eta_1(\tau))\right)$$

$$\exp(i((2m-n-j)\xi_1(\tau)-(2m'-n'+j')\xi_2(\tau)))J_j(ak_{n,n',m,m',j,j'}\sin\eta_1(\tau))J_{j'}(ak_{n,n',m,m',j,j'}\sin\eta_2(\tau)),$$
(2.32)

where

$$k_{n,n',m,m',j,j'} = -\frac{\left(n\left(\omega_{A,1}(s)-\frac{d\xi_1(\tau)}{d\tau}\right)+n'\left(\omega_{A,1}(s)-\frac{d\xi_2(\tau)}{d\tau}\right)\right)+\left((2m-j)\frac{d\xi_1(\tau)}{d\tau}+(2m'+j')\frac{d\xi_2(\tau)}{d\tau}\right)}{2},$$
(2.33)

$$\kappa_{n,n',m,m'} = \sqrt{k^2_{n,n',m,m'} + \kappa_D^2}. \tag{2.34}$$

Eqs. (2.32)-(2.34) form one of the general results of this paper.

## 3. Interaction energy with cores of low dielectric constant

Now, we consider a more complicated case were inside the rods there is a dielectric constant $\varepsilon_c \ll \varepsilon_w$. Also, in the rods, there is no screening; the electrostatic potential satisfies Laplace's equation. The interaction energy can be computed through an image charge expansion. Distortions of the electric field due dielectric boundaries of the two rods are taken account of by image charges. The image charges can be written as a sum of image charge contributions. To compute the leading order contribution of this series for rod $\mu$, the zeroth order images, we consider the charges on it and do not consider the dielectric core of the other rod $\nu = 3 - \mu$. The next to leading order contribution on rod $\mu$ is induced by only the zeroth order charges and charges of rod $\nu$. This is obtained by satisfying the electrostatic boundary conditions at the surface of rod $\mu$, where the only other contribution to the potential, apart from next to leading order images, is from its zeroth order images and charges on rod $\nu$. The $n$-th order image charges on rod $\mu$, where $n \geq 2$, are got by considering the contribution to the electrostatic potential that comes satisfying the electrostatic boundary conditions on the surface of rod $\mu$, where the electrostatic potentials from the $n$-th image charges on rod $\mu$ and $n-1$-th images on rod $\nu$. Each term in the series smaller than the previous one due to electrostatic screening and effectively the series can truncated; we truncate it $n = 2$. The relative size of $n$-th image charges with respect $n-1$-th images decreases with increasing $R$. Substitution of the image charge contributions into electrostatic energy gives an image charge expansion for it. We retain the first two terms in the series. Therefore, we may write the electrostatic energy as two contributions

$$E \approx E_{dir} + E_{img}. \tag{3.1}$$

The term $E_{dir}$ is the leading order contribution to the interaction energy, while $E_{img}$ is the next order contribution.

The energy $E_{dir}$ comes from an effective electrostatic interaction between charges and image charges on one rod with those on the other rod [1,3,10] and can be written as

$$E_{dir} = \frac{4\pi e^2}{\varepsilon_w l_c^2} \frac{1}{(2\pi)^3} \int_0^{L_B} ds \int_0^{L_B} ds' \sigma_1(s) \sigma_2(s') \int d^3k \frac{\rho_{eff}^{(1)}(\mathbf{k};s) \rho_{eff}^{(2)}(-\mathbf{k};s')}{\mathbf{k}^2 + \kappa_D^2}. \tag{3.2}$$

Here, $\rho_{eff}^{(1)}(\mathbf{k};s)$ and $\rho_{eff}^{(2)}(-\mathbf{k};s')$ are the Fourier transforms of the sum of the charge density of a point charge and its zeroth order image charges on rods 1 and 2, respectively. For rod 1 the point charge point is at $s$, and for rod 2 the point charge is at $s'$. Again in Eq. (3.2), as in Eq. (2.2) we sum up all the point charge contributions by performing the integrations over $s$ and $s'$.

The energy $E_{img}$ can be written as

$$E_{img} = E_{img1} + E_{img2}. \tag{3.3}$$

The energy $E_{img\mu}$ arises from the effective repulsive interaction between charges and zeroth order images on rod $\mu$ with first order images induced by them on rod $\nu$. It can be written as [1,3,10]

$$E_{img\mu} = \frac{2\pi e^2}{l_c^2 \varepsilon_w} \frac{1}{(2\pi)^3} \int_0^{L_B} ds \int_0^{L_B} ds' \sigma_\mu(s) \sigma_\mu(s') \int d^3k \frac{\rho_{eff}^{(\mu)}(\mathbf{k};s) \rho_{img}^{(\nu,\mu)}(-\mathbf{k};s')}{\mathbf{k}^2 + \kappa_D^2}. \tag{3.4}$$

Here, $\rho_{img}^{(\nu,\mu)}(-\mathbf{k};s')$ is the Fourier transform of the image charge density on rod $\nu$ induced by the zeroth order images and the point charge at $s'$ on rod $\mu$. We will now calculate both the zeroth order and first order images.

## 4. Calculation of image charges

### *4.1 The boundary value problem for a point charge*

To calculate the image charges we need first to formulate the boundary value problem for a point charge on rod $\mu$ of the braid. In terms of the surfaces described by Eq. (1.48), the boundary value problem for a charge on rod $\mu = 1, 2$, positioned at $s = s_0$ and $t = t_0$ can be formulated as

$$-\nabla^2 \phi_{ext}(\mathbf{r}) + \kappa_D^2 \phi_{ext}(\mathbf{r}) = \frac{4\pi}{\varepsilon_w} \delta(\mathbf{r} - \mathbf{S}_\mu(s_0, t_0)), \tag{4.1}$$

$$-\nabla^2 \phi_{int,1}(\mathbf{r}) = 0, \quad -\nabla^2 \phi_{int,2}(\mathbf{r}) = 0, \tag{4.2}$$

$$\phi_{int,1}(\mathbf{S}_1(\tilde{s}_1,t_1)) = \phi_{ext}(\mathbf{S}_1(\tilde{s}_1,t_1)), \quad \phi_{int,2}(\mathbf{S}_2(\tilde{s}_2,t_2)) = \phi_{ext}(\mathbf{S}_2(\tilde{s}_2,t_2)), \tag{4.3}$$

$$\varepsilon_c \hat{\mathbf{N}}_1(\tilde{s}_1,t_1) \cdot \nabla \phi_{int,1}(\mathbf{r})\big|_{\mathbf{r}=\mathbf{S}_1(\tilde{s}_1,t_1)} = \varepsilon_w \hat{\mathbf{N}}_1(\tilde{s}_1,t_1) \cdot \nabla \phi_{ext}(\mathbf{r})\big|_{\mathbf{r}=\mathbf{S}_1(\tilde{s}_1,t_1)}, \tag{4.4}$$

$$\varepsilon_c \hat{\mathbf{N}}_2(\tilde{s}_2,t_2) \cdot \nabla \phi_{int,1}(\mathbf{r})\big|_{\mathbf{r}=\mathbf{S}_2(\tilde{s}_2,t_2)} = \varepsilon_w \hat{\mathbf{N}}_2(\tilde{s}_2,t_2) \cdot \nabla \phi_{ext}(\mathbf{r})\big|_{\mathbf{r}=\mathbf{S}_2(\tilde{s}_2,t_2)}, \tag{4.5}$$

$$\nabla \phi_{ext}(\mathbf{r})\big|_{\mathbf{r}\to\infty} = 0, \quad \nabla \phi_{ext}(\mathbf{r})\big|_{\mathbf{r}\to\infty} = 0. \tag{4.6}$$

Here, $\phi_{ext}(\mathbf{r})$ is the electrostatic potential outside the rod, which is screened with inverse screening length $\kappa_D$, $\phi_{int}(\mathbf{r})$ is the potential inside the rod which is not screened, and both $\tilde{s}_1(s_1)$ and $\tilde{s}_2(s_2)$ are measured in the arc length of the braid axis, whereas $s_1$ and $s_2$ are measured in the arc-length of centre lines of rods 1 and 2, respectively. In the limit where $\varepsilon_w \gg \varepsilon_c$ we don't care about the internal potentials $\phi_{int,1}(\mathbf{r})$ and $\phi_{int,2}(\mathbf{r})$. This means that we do not need to consider Eq. (4.3). The boundary conditions given by Eq. (4.5) can then be approximated by

$$0 = \hat{\mathbf{N}}_1(\tilde{s}_1, t_1).\nabla \phi_{ext}(\mathbf{r})\big|_{\mathbf{r}=\mathbf{S}_1(\tilde{s}_1,t_1)}, \qquad 0 = \hat{\mathbf{N}}_2(\tilde{s}_2, t_2).\nabla \phi_{ext}(\mathbf{r})\big|_{\mathbf{r}=\mathbf{S}_2(\tilde{s}_2,t_2)}. \tag{4.7}$$

Now, we can analytically continue the external potential into the dielectric cylinders provided that we introduce image charges. This means that we rewrite the linearized PB equation, Eq. (4.1), as

$$-\nabla^2 \phi_{ext}(\mathbf{r}) + \kappa^2 \phi_{ext}(\mathbf{r}) = \frac{4\pi}{\varepsilon_w} \left( \rho_{ind,1}(\mathbf{r}; s_0) + \rho_{ind,2}(\mathbf{r}; s_0) \right), \tag{4.8}$$

where the effective charge densities are given by

$$\rho_{ind,\mu}(\mathbf{r}; s_0) = \frac{1}{2\pi} \int_0^{L_B} ds' \int_0^{2\pi} dt' \left[ \delta(\mathbf{r} - \tilde{\mathbf{S}}_\mu^-(s',t'))\sigma_{ind}^{(\mu)}(t',s') + \delta(\mathbf{r} - \tilde{\mathbf{S}}_\mu^+(s',t'))\delta(t'-t_0, s'-s_0) \right], \tag{4.9}$$

$$\rho_{ind,\nu}(\mathbf{r}; s_0) = \frac{1}{2\pi} \int_0^{L_B} ds' \int_0^{2\pi} dt' \delta(\mathbf{r} - \tilde{\mathbf{S}}_\nu^-(s',t'))\sigma_{ind}^{(\nu)}(t',s'), \tag{4.10}$$

where $\nu = 3 - \mu$

$$\mathbf{S}_\mu^+(\tilde{s}_\mu, t_\mu) = \mathbf{r}_\mu(\tilde{s}_\mu) + (a + \delta a)\hat{\mathbf{N}}_\mu(\tilde{s}_\mu, t_\mu), \qquad \mathbf{S}_\mu^-(\tilde{s}_\mu, t_\mu) = \mathbf{r}_\mu(\tilde{s}_\mu) + (a - \delta a)\hat{\mathbf{N}}_\mu(\tilde{s}_\mu, t_\mu), \tag{4.11}$$

and $\delta a$ is taken to be infinitesimally small value from $a$. The induced or image surface charge densities $\sigma_{ind}^{(1)}(t',s')$ and $\sigma_{ind}^{(2)}(t',s')$ are chosen so that Eq. (4.7) is satisfied. The induced charges are taken as just lying inside the cylinder, whereas the real point charge lies just outside.

We can then 'formally' solve the analytically continued Eq. (4.8) by Fourier Transform. This yields

$$\phi_{ext}(\mathbf{k}) = \frac{4\pi}{\varepsilon_w} \frac{\left( \rho_{ind,1}(\mathbf{k}; s_0) + \rho_{ind,2}(\mathbf{k}; s_0) \right)}{(\mathbf{k}^2 + \kappa_D^2)}, \tag{4.12}$$

where

$$\phi_{ext}(\mathbf{r}) = \frac{1}{(2\pi)^3} \int d^3k \, \phi_{ext}(\mathbf{k}) \exp(-i\mathbf{k}.\mathbf{r}), \tag{4.13}$$

and we can write in terms of the arc-length of the braid axis

$$\rho_{ind,\mu}(\mathbf{k}; s_0) = \frac{1}{2\pi} \int_0^{L_B} ds' \int_0^{2\pi} dt' \left( \exp(i\mathbf{k}.\mathbf{S}_\mu^-(s',t'))\sigma_{ind}^{(\mu)}(t',s') + \exp(i\mathbf{k}.\mathbf{S}_\mu^+(s',t'))\delta(t'-t_0, s'-s_0) \right), \tag{4.14}$$

$$\rho_{ind,\nu}(\mathbf{k}; s_0) = \frac{1}{2\pi} \int_0^{L_B} ds' \int_0^{2\pi} dt' \exp(i\mathbf{k}.\mathbf{S}_\nu^-(s',t'))\sigma_{ind}^{(\nu)}(t',s'). \tag{4.15}$$

Then by substituting Eq. (4.12), (4.14) and (4.15) into boundary conditions given by Eq. (4.7) we obtain equations on $\sigma_{ind}^{(1)}(t',s')$ and $\sigma_{ind}^{(2)}(t',s')$

$$\sum_{\mu'=1}^{2} \int_0^{L_B} ds' \int_0^{2\pi} dt' \int d^3k \exp(i\mathbf{k}.(\mathbf{S}_{\mu'}^-(s',t') - \mathbf{S}_1(\tilde{s}_1,t_1))) \frac{\hat{\mathbf{N}}_1(\tilde{s}_1,t_1).\mathbf{k}}{(\mathbf{k}^2 + \kappa_D^2)} \sigma_{ind}^{(\mu')}(t',s')$$
$$+ \int d^3k \exp(i\mathbf{k}.(\mathbf{S}_\mu^+(s_0,t_0) - \mathbf{S}_1(\tilde{s}_1,t_1))) \frac{\hat{\mathbf{N}}_1(\tilde{s}_1,t_1).\mathbf{k}}{(\mathbf{k}^2 + \kappa_D^2)} = 0,$$
(4.16)

$$\sum_{\mu'=1}^{2} \int_0^{L_B} ds' \int_0^{2\pi} dt' \int d^3k \exp(i\mathbf{k}.(\mathbf{S}_{\mu'}^-(s',t') - \mathbf{S}_2(\tilde{s}_2,t_2))) \frac{\hat{\mathbf{N}}_2(\tilde{s}_2,t_2).\mathbf{k}}{(\mathbf{k}^2 + \kappa_D^2)} \sigma_{ind}^{(\mu')}(t',s')$$
$$+ \int d^3k \exp(i\mathbf{k}.(\mathbf{S}_\mu^+(s_0,t_0) - \mathbf{S}_2(\tilde{s}_2,t_2))) \frac{\hat{\mathbf{N}}_2(\tilde{s}_2,t_2).\mathbf{k}}{(\mathbf{k}^2 + \kappa_D^2)} = 0.$$
(4.17)

In a step towards a complete solution of the boundary value problem, we note that we can evaluate the k-integrals. First of all we may write

$$\frac{1}{(2\pi)^3} \int d^3k \exp(i\mathbf{k}.(\mathbf{S}_1(s',t') - \mathbf{S}_1(\tilde{s}_1,t_1))) \frac{1}{(\mathbf{k}^2 + \kappa_D^2)} = \frac{1}{4\pi} \frac{\exp(-\kappa_D |\mathbf{S}_1(s',t') - \mathbf{S}_1(\tilde{s}_1,t_1)|)}{|\mathbf{S}_1(s',t') - \mathbf{S}_1(\tilde{s}_1,t_1)|}.$$
(4.18)

From Eq. (4.18) we obtain

$$\frac{1}{(2\pi)^3} \int d^3k \exp(i\mathbf{k}.(\mathbf{S}_1(s',t') - \mathbf{S}_1(\tilde{s}_1,t_1))) \frac{\hat{\mathbf{N}}_1(s_1,t_1).\mathbf{k}}{(\mathbf{k}^2 + \kappa_D^2)} = \frac{1}{4\pi} \hat{\mathbf{N}}_1(\tilde{s}_1,t_1).\nabla \left[ \frac{\exp(-\kappa_D |\mathbf{r}|)}{|\mathbf{r}|} \right]_{\mathbf{r} = \mathbf{S}_1(s',t') - \mathbf{S}_1(s_1,t_1)}$$
$$= -\frac{1}{4\pi} \frac{\hat{\mathbf{N}}_1(s_1,t_1).(\mathbf{S}_1(s',t') - \mathbf{S}_1(\tilde{s}_1,t_1)) \exp(-\kappa_D |\mathbf{S}_1(s',t') - \mathbf{S}_1(\tilde{s}_1,t_1)|)}{|\mathbf{S}_1(s',t') - \mathbf{S}_1(\tilde{s}_1,t_1)|^3}$$
$$- \frac{1}{4\pi} \frac{\kappa_D \hat{\mathbf{N}}_1(s_1,t_1).(\mathbf{S}_1(s',t') - \mathbf{S}_1(\tilde{s}_1,t_1)) \exp(-\kappa_D |\mathbf{S}_1(s',t') - \mathbf{S}_1(\tilde{s}_1,t_1)|)}{|\mathbf{S}_1(s',t') - \mathbf{S}_1(\tilde{s}_1,t_1)|^2}.$$
(4.19)

This mathematical result, Eq. (4.19) allows us to rewrite Eqs. (4.16) and (4.17) as

$$\sum_{\mu'=1}^{2} \int_0^{L_B} ds' \int_0^{2\pi} dt' \sigma_{ind}^{(\mu')}(t',s') \hat{\mathbf{N}}_1(\tilde{s}_1,t_1).(\mathbf{S}_{\mu'}^-(s',t') - \mathbf{S}_1(\tilde{s}_1,t_1)) \exp(-\kappa_D |\mathbf{S}_{\mu'}^-(s',t') - \mathbf{S}_1(\tilde{s}_1,t_1)|)$$
$$\left\{ \frac{1}{|\mathbf{S}_{\mu'}^-(s',t') - \mathbf{S}_1(\tilde{s}_1,t_1)|^3} + \frac{\kappa_D}{|\mathbf{S}_{\mu'}^-(s',t') - \mathbf{S}_1(\tilde{s}_1,t_1)|^2} \right\} + \hat{\mathbf{N}}_1(\tilde{s}_1,t_1).(\mathbf{S}_\mu^+(s_0,t_0) - \mathbf{S}_1(\tilde{s}_1,t_1))$$
$$\exp(-\kappa_D |\mathbf{S}_1(s',t') - \mathbf{S}_1(\tilde{s}_1,t_1)|) \left\{ \frac{1}{|\mathbf{S}_\mu^+(s_0,t_0) - \mathbf{S}_1(\tilde{s}_1,t_1)|^3} + \frac{\kappa_D}{|\mathbf{S}_\mu^+(s_0,t_0) - \mathbf{S}_1(\tilde{s}_1,t_1)|^2} \right\} = 0,$$
(4.20)

$$\sum_{\mu'=1}^{2} \int_{0}^{L_B} ds' \int_{0}^{2\pi} dt' \sigma_{ind}^{(\mu')}(t',s') \hat{\mathbf{N}}_2(\tilde{s}_2,t_2) \cdot (\mathbf{S}_{\mu'}^{-}(s',t') - \mathbf{S}_2(\tilde{s}_2,t_2)) \exp\left(-\kappa_D \left|\mathbf{S}_{\mu'}^{-}(s',t') - \mathbf{S}_2(\tilde{s}_2,t_2)\right|\right)$$

$$\left\{ \frac{1}{\left|\mathbf{S}_{\mu'}^{-}(s',t') - \mathbf{S}_2(s_2,t_2)\right|^3} + \frac{\kappa_D}{\left|\mathbf{S}_{\mu'}^{-}(s',t') - \mathbf{S}_2(\tilde{s}_2,t_2)\right|^2} \right\} \quad (4.21)$$

$$+ \hat{\mathbf{N}}_2(\tilde{s}_2,t_2) \cdot (\mathbf{S}_{\mu}^{+}(s_0,t_0) - \mathbf{S}_2(\tilde{s}_2,t_2)) \exp\left(-\kappa_D \left|\mathbf{S}_{\mu}^{+}(s_0,t_0) - \mathbf{S}_2(\tilde{s}_2,t_2)\right|\right)$$

$$\left\{ \frac{1}{\left|\mathbf{S}_{1}^{+}(s_0,t_0) - \mathbf{S}_2(\tilde{s}_2,t_2)\right|^3} + \frac{\kappa_D}{\left|\mathbf{S}_{\mu}^{+}(s_0,t_0) - \mathbf{S}_2(\tilde{s}_2,t_2)\right|^2} \right\} = 0.$$

### *4.3 The Image charge expansion*

Now we utilize the image charge expansion discussed in the previous section. This expansion is a perturbation expansion that relies on the fact that for the braid $R\kappa_D \gg 1$. We first note that we can also write Eq. (1.48) through Eq. (1.2) as

$$\mathbf{S}_1(\tilde{s}_1,t_1) = \mathbf{r}_2(\tilde{s}_1) - R\hat{\mathbf{d}}(\tilde{s}_1) + a\hat{\mathbf{N}}_1(\tilde{s}_1,t_1), \quad \mathbf{S}_2(\tilde{s}_2,t_2) = \mathbf{r}_1(\tilde{s}_2) + R\hat{\mathbf{d}}(\tilde{s}_2) + a\hat{\mathbf{N}}_2(\tilde{s}_2,t_2). \quad (4.22)$$

Eqs. (1.48) and (4.22) imply that

$$\left|\mathbf{S}_1(s',t') - \mathbf{S}_2(\tilde{s}_2,t_2)\right| \geq R, \quad \left|\mathbf{S}_2(s',t') - \mathbf{S}_1(\tilde{s}_1,t_1)\right| \geq R, \qquad q \quad (4.23)$$

and therefore, provided that $\kappa_D R \gg 1$, we can treat terms that include $\left|\mathbf{S}_1(s',t') - \mathbf{S}_2(\tilde{s}_2,t_2)\right|$ and $\left|\mathbf{S}_2(s',t') - \mathbf{S}_1(\tilde{s}_1,t_1)\right|$ as perturbations. If the charge lies on rod $\mu$, we can write the perturbation expansion

$$\sigma_{ind}^{(\mu)}(t',s') = \sigma_{ind,0}^{(\mu)}(t',s') + \sigma_{ind,2}^{(\mu)}(t',s') + \ldots, \quad (4.24)$$

$$\sigma_{ind}^{(\nu)}(t',s') = \sigma_{ind,1}^{(\nu)}(t',s') + \sigma_{ind,3}^{(\nu)}(t',s') + \ldots \quad (4.25)$$

In this expansion, we only require $\sigma_{ind,0}^{(\mu)}(t',s')$ and $\sigma_{ind,1}^{(\nu)}(t',s')$ to calculate $E_{dir}$ and $E_{img}$. We can then derive from Eqs.(4.20) and (4.21) that the leading and next to leading order images $\sigma_{ind,0}^{(\mu)}(t',s')$ and $\sigma_{ind,1}^{(\nu)}(t',s')$, respectively, satisfy

$$\int_0^{L_B} ds' \int_0^{2\pi} dt' \sigma_{ind,0}^{(\mu)}(t',s') \hat{\mathbf{N}}_\mu(\tilde{s}_\mu, t_\mu) \cdot (\mathbf{S}_\mu^-(s',t') - \mathbf{S}_\mu(\tilde{s}_\mu, t_\mu)) \exp\left(-\kappa_D \left|\mathbf{S}_\mu^-(s',t') - \mathbf{S}_\mu(\tilde{s}_\mu, t_\mu)\right|\right)$$

$$\left\{ \frac{1}{\left|\mathbf{S}_\mu^-(s',t') - \mathbf{S}_\mu(\tilde{s}_\mu, t_\mu)\right|^3} + \frac{\kappa_D}{\left|\mathbf{S}_\mu^-(s',t') - \mathbf{S}_\mu(\tilde{s}_\mu, t_\mu)\right|^2} \right\}$$

$$+ \hat{\mathbf{N}}_\mu(\tilde{s}_\mu, t_\mu) \cdot (\mathbf{S}_\mu^+(s_0,t_0) - \mathbf{S}_\mu(\tilde{s}_\mu, t_\mu)) \exp\left(-\kappa_D \left|\mathbf{S}_\mu^+(s_0,t_0) - \mathbf{S}_\mu(\tilde{s}_\mu, t_\mu)\right|\right)$$

$$\left\{ \frac{1}{\left|\mathbf{S}_\mu^+(s_0,t_0) - \mathbf{S}_\mu(\tilde{s}_\mu, t_\mu)\right|^3} + \frac{\kappa_D}{\left|\mathbf{S}_\mu^+(s_0,t_0) - \mathbf{S}_\mu(\tilde{s}_\mu, t_\mu)\right|^2} \right\} = 0, \quad (4.26)$$

$$\int_0^{L_B} ds' \int_0^{2\pi} dt' \sigma_{ind,0}^{(\mu)}(t',s') \hat{\mathbf{N}}_\nu(\tilde{s}_2, t_2) \cdot (\mathbf{S}_\mu^-(s',t') - \mathbf{S}_\nu(\tilde{s}_\nu, t_\nu)) \exp\left(-\kappa_D \left|\mathbf{S}_\mu(s',t') - \mathbf{S}_\nu(\tilde{s}_\nu, t_\nu)\right|\right)$$

$$\left\{ \frac{1}{\left|\mathbf{S}_\mu^-(s',t') - \mathbf{S}_\nu(\tilde{s}_2, t_2)\right|^3} + \frac{\kappa_D}{\left|\mathbf{S}_\mu^-(s',t') - \mathbf{S}_\nu(\tilde{s}_\nu, t_\nu)\right|^2} \right\}$$

$$+ \hat{\mathbf{N}}_\nu(\tilde{s}_\nu, t_\nu) \cdot (\mathbf{S}_\mu^+(s_0,t_0) - \mathbf{S}_\nu(\tilde{s}_\nu, t_\nu)) \exp\left(-\kappa_D \left|\mathbf{S}_\mu^+(s_0,t_0) - \mathbf{S}_\nu(\tilde{s}_\nu, t_\nu)\right|\right)$$

$$\left\{ \frac{1}{\left|\mathbf{S}_\mu^+(s_0,t_0) - \mathbf{S}_\nu(\tilde{s}_\nu, t_\nu)\right|^3} + \frac{\kappa_D}{\left|\mathbf{S}_\mu^+(s_0,t_0) - \mathbf{S}_\nu(\tilde{s}_\nu, t_\nu)\right|^2} \right\} \quad (4.27)$$

$$+ \int_0^{L_B} ds' \int_0^{2\pi} dt' \sigma_{ind,1}^{(\nu)}(t',s') \hat{\mathbf{N}}_\nu(\tilde{s}_\nu, t_\nu) \cdot (\mathbf{S}_\nu^-(s',t') - \mathbf{S}_\nu(\tilde{s}_\nu, t_\nu)) \exp\left(-\kappa_D \left|\mathbf{S}_\nu^-(s',t') - \mathbf{S}_\nu(\tilde{s}_\nu, t_\nu)\right|\right)$$

$$\left\{ \frac{1}{\left|\mathbf{S}_\nu^-(s',t') - \mathbf{S}_\nu(\tilde{s}_\nu, t_\nu)\right|^3} + \frac{\kappa_D}{\left|\mathbf{S}_\nu^-(s',t') - \mathbf{S}_\nu(\tilde{s}_\nu, t_\nu)\right|^2} \right\} = 0.$$

Also using Eqs. (4.14), (4.15) we can write

$$\rho_{eff}^{(\mu)}(\mathbf{k}; s_0) = \frac{1}{2\pi} \int_0^{L_B} ds' \int_0^{2\pi} dt' \left( \exp(i\mathbf{k}.\mathbf{S}_\mu^-(s',t')) \sigma_{ind,0}^{(\mu)}(t',s') + \exp(i\mathbf{k}.\mathbf{S}_\mu^+(s',t')) \delta(t'-t_0, s'-s_0) \right),$$

(4.28)

$$\rho_{img}^{(\nu,\mu)}(\mathbf{k}; s_0) = \frac{1}{2\pi} \int_0^{L_B} ds' \int_0^{2\pi} dt' \exp(i\mathbf{k}.\mathbf{S}_\nu^-(s',t')) \sigma_{ind,1}^{(\nu)}(t',s'). \quad (4.29)$$

### *4.4 Curvature expansion*

Given a specified geometry we could try and solve Eqs. (4.26) and (4.27), numerically; but as we have not specified the geometry a priori and we want to obtain analytical expressions for the image charges where the braid curvature is small. We start by expanding the vectors $\hat{\mathbf{N}}_\mu(s_\mu, t_\mu)$ and $\hat{\mathbf{t}}_\mu(s_\mu)$ about the point $s_0$. First, we consider the point charge lying on rod 1. The Taylor expansions yield (using Eqs. (1.53)-(1.55) and the choice, Eq. (1.59)) the approximate expressions

$$\hat{\mathbf{N}}_\mu(\tilde{s}_\mu, t_\mu) \approx \cos t_\mu \hat{\mathbf{d}}(s_0) + \sin t_\mu \hat{\mathbf{n}}_\mu(s_0) + \sigma_\mu(s_0)(\omega_{\mu,2}(s_0)\sin t_\mu - \omega_{\mu,3}(s_0)\cos t_\mu)\hat{\mathbf{t}}_\mu(s_0)(\tilde{s}_\mu - s_0),$$

(4.30)

$$\hat{\mathbf{t}}_\mu(\tilde{s}_\mu) \approx \hat{\mathbf{t}}_\mu(s_0) + \left.\frac{d\hat{\mathbf{t}}_\mu(\tilde{s}_\mu)}{d\tilde{s}_1}\right|_{s_\mu = s_0}(\tilde{s}_1 - s_0) = \hat{\mathbf{t}}_\mu(s_0) + \sigma_1(s_0)\left(\omega_{\mu,3}(s_0)\hat{\mathbf{d}}(s_0) - \omega_{\mu,2}(s_0)\hat{\mathbf{n}}_\mu(s_0)\right)(\tilde{s}_\mu - s_0),$$

(4.31)

where we make the choice Eq. (1.59) at the point $s_0$ where the point charge lies. The vector equations for the surfaces can be expanded out as well

$$\mathbf{S}_\mu(\tilde{s}_\mu, t_\mu) \approx \mathbf{r}_\mu(s_0) + \sigma_\mu(s_0)\hat{\mathbf{t}}_\mu(s_0)(\tilde{s}_\mu - s_0) + \frac{\sigma_\mu(s_0)^2}{2}\left(\omega_{\mu,3}(s_0)\hat{\mathbf{d}}(s_0) - \omega_{\mu,2}(s_0)\hat{\mathbf{n}}_\mu(s_0)\right)(\tilde{s}_\mu - s_0)^2$$
$$+ a\cos t_\mu \hat{\mathbf{d}}(s_0) + a\sin t_\mu \hat{\mathbf{n}}_\mu(s_1) + a\sigma_\mu(s_0)(\omega_{\mu,2}(s_0)\sin t_\mu - \omega_{\mu,3}(s_0)\cos t_\mu)\hat{\mathbf{t}}_\mu(s_0)(\tilde{s}_\mu - s_0),$$

(4.32)

where we have neglected the derivatives of $\sigma_\mu(s_0)$. This is because, by looking at Eq. (1.25), we see that they contain derivatives of the braiding frequencies that we neglect to this order of the calculation. Furthermore, we can write from Eqs. (4.32)

$$\mathbf{S}_\mu^+(s_0, t_0) - \mathbf{S}_\mu(\tilde{s}_\mu, t_\mu) \approx \sigma_\mu(s_0)\hat{\mathbf{t}}_1(s_0)(s_0 - \tilde{s}_\mu) - \frac{\sigma_\mu(s_0)^2}{2}\left(\omega_{\mu,3}(s_0)\hat{\mathbf{d}}(s_0) - \omega_{\mu,2}(s_0)\hat{\mathbf{n}}_\mu(s_0)\right)(\tilde{s}_\mu - s_0)^2$$
$$+ \left((a + \delta a)\cos t_0 - a\cos t_\mu\right)\hat{\mathbf{d}}(s_0) + \left((a + \delta a)\sin t_0 - a\sin t_\mu\right)\hat{\mathbf{n}}_\mu(s_0)$$
$$- a\sigma_\mu(s_0)(\omega_{\mu,2}(s_0)\sin t_\mu - \omega_{\mu,3}(s_0)\cos t_\mu)\hat{\mathbf{t}}_\mu(s_0)(\tilde{s}_\mu - s_0),$$

(4.33)

$$\mathbf{S}_\mu^-(s', t') - \mathbf{S}_\mu(\tilde{s}_\mu, t_\mu) \approx \sigma_\mu(s_0)\hat{\mathbf{t}}_\mu(s_0)(s' - \tilde{s}_\mu) + \frac{\sigma_\mu(s_0)^2}{2}\left(\omega_{\mu,3}(s_0)\hat{\mathbf{d}}(s_0) - \omega_{\mu,2}(s_0)\hat{\mathbf{n}}_\mu(s_0)\right)\left((s' - s_0)^2 - (\tilde{s}_\mu - s_0)^2\right)$$
$$+ \left((a - \delta a)\cos t' - a\cos t_\mu\right)\hat{\mathbf{d}}(s_0) + \left((a - \delta a)\sin t' - a\sin t_\mu\right)\hat{\mathbf{n}}_\mu(s_0)$$
$$- a\sigma_\mu(s_0)(\omega_{\mu,2}(s_0)\sin t_\mu - \omega_{\mu,3}(s_0)\cos t_\mu)\hat{\mathbf{t}}_\mu(s_0)(\tilde{s}_\mu - s_0)$$
$$+ (a - \delta a)\sigma_\mu(s_0)(\omega_{\mu,2}(s_0)\sin t' - \omega_{\mu,3}(s_0)\cos t')\hat{\mathbf{t}}_\mu(s_0)(s' - s_0),$$

(4.34)

$$\mathbf{S}_\nu^-(s',t') - \mathbf{S}_\mu(\tilde{s}_\mu, t_\mu) = \delta_\mu R \hat{\mathbf{d}}(s_0) + \sigma_\nu(s_0)\hat{\mathbf{t}}_\nu(s_0)(s'-s_0) - \sigma_\mu(s_0)\hat{\mathbf{t}}_\mu(s_0)(\tilde{s}_\mu - s_0)$$

$$+ \frac{\sigma_\nu(s_0)^2}{2}\left(\omega_{\nu,3}(s_0)\hat{\mathbf{d}}(s_0) - \omega_{\nu,2}(s_0)\hat{\mathbf{n}}_2(s_0)\right)(s'-s_0)^2 - \frac{\sigma_\mu(s_0)^2}{2}\left(\omega_{\mu,3}(s_0)\hat{\mathbf{d}}(s_0) - \omega_{\mu,2}(s_0)\hat{\mathbf{n}}_1(s_0)\right)(\tilde{s}_\mu - s_0)^2$$

$$+ (a-\delta a)\cos t' \hat{\mathbf{d}}(s_0) + (a-\delta a)\sin t' \hat{\mathbf{n}}_\nu(s_0) + (a-\delta a)\sigma_\nu(s_0)(\omega_{\nu,2}(s_0)\sin t' - \omega_{\nu,3}(s_0)\cos t')\hat{\mathbf{t}}_\nu(s_0)(s'-s_0)$$

$$- a\cos t_\mu \hat{\mathbf{d}}(s_0) - a\sin t_\mu \hat{\mathbf{n}}_\mu(s_0) - a\sigma_\mu(s_0)(\omega_{\mu,2}(s_0)\sin t_\mu - \omega_{\mu,3}(s_0)\cos t_\mu)\hat{\mathbf{t}}_\mu(s_0)(\tilde{s}_\mu - s_0).$$

(4.35)

It is actually simpler to go back to Eq. (4.16) and (4.17) and perform the curvature expansion before the image charge expansion and perform the $k$-integrations later on. Therefore, we substitute Eqs. (4.33)-(4.35) into Eq. (4.16) and expanded out for small $\omega_{\mu,2}(s_0), \omega_{\mu,3}(s_0), \omega_{\nu,2}(s_0)$ and $\omega_{\nu,3}(s_0)$. Also, we are allowed to choose the following for the vector dot product

$$\mathbf{k}.\hat{\mathbf{d}} = K\cos\phi_K, \quad \mathbf{k}.\hat{\mathbf{n}}_\mu(s_0) = K\sin\phi_K, \quad \mathbf{k}.\hat{\mathbf{t}}_\mu(s_0) = k_z,$$

(4.36)

and perform the image charge expansion (Eqs. (4.24) and (4.25)). This allows us to write:

$$\int_0^{L_B} ds' \int_0^{2\pi} dt' \int d^3k \exp\left(ik_z\sigma_\mu(s_0)(s'-\tilde{s}_\mu) + i(a-\delta a)K\cos(t'-\phi_K) - iaK\cos(t_\mu - \phi_K)\right)$$

$$\left(1 + \frac{i\sigma_\mu(s_0)^2}{2}\left(\omega_{\mu,3}(s_0)K\cos\phi_K - \omega_{\mu,2}(s_0)K\sin\phi_K\right)\left((s'-s_0)^2 - (\tilde{s}_\mu - s_0)^2\right)\right.$$

$$-ia\sigma_\mu(s_0)(\omega_{\mu,2}(s_0)\sin t_\mu - \omega_{\mu,3}(s_0)\cos t_\mu)k_z(\tilde{s}_\mu - s_0) + i(a-\delta a)\sigma_\mu(s_0)(\omega_{\mu,2}(s_0)\sin t' - \omega_{\mu,3}(s_0)\cos t')k_z(s'-s_0)\right)$$

$$\left(K\cos(t_\mu - \phi_K) + \sigma_\mu(s_0)(\omega_{\mu,2}(s_0)\sin t_\mu - \omega_{\mu,3}(s_0)\cos t_\mu)k_z(\tilde{s}_\mu - s_0)\right)\frac{1}{(\mathbf{k}^2 + \kappa_D^2)}\sigma_{ind,0}^{(\mu)}(t',s')$$

$$+ \int d^3k \exp\left(ik_z\sigma_\mu(s_0)(s_0 - \tilde{s}_\mu) + i(a+\delta a)K\cos(t_0 - \phi_K) - iaK\cos(t_\mu - \phi_K)\right)$$

$$\left(1 - i\frac{\sigma_\mu(s_0)^2}{2}\left(\omega_{\mu,3}(s_0)K\cos\phi_K - \omega_{\mu,2}(s_0)K\sin\phi_K\right)(\tilde{s}_\mu - s_0)^2\right.$$

$$-ia\sigma_\mu(s_0)(\omega_{\mu,2}(s_0)\sin t_\mu - \omega_{\mu,3}(s_0)\cos t_\mu)k_z(\tilde{s}_\mu - s_0)\right)$$

$$\left(K\cos(t_\mu - \phi_K) + \sigma_\mu(s_0)(\omega_{\mu,2}(s_0)\sin t_1 - \omega_{\mu,3}(s_0)\cos t_\mu)k_z(\tilde{s}_\mu - s_0)\right)\frac{1}{(\mathbf{k}^2 + \kappa_D^2)} = 0.$$

(4.37)

Now let's consider Eq. (4.17). Similarly, we can write

$$\mathbf{S}_\mu^+(s_0, t_0) - \mathbf{S}_\nu(\tilde{s}_2, t_2) \approx -\delta_\mu R\hat{\mathbf{d}}(s_0) - \sigma_\nu(s_0)\hat{\mathbf{t}}_\nu(s_0)(\tilde{s}_\nu - s_0)$$

$$- \frac{\sigma_\nu(s_0)^2}{2}\left(\omega_{2,3}(s_0)\hat{\mathbf{d}}(s_0) - \omega_{2,2}(s_0)\hat{\mathbf{n}}_\nu(s_0)\right)(\tilde{s}_\nu - s_0)^2$$

$$+ \left((a+\delta a)\cos t_0 - a\cos t_\nu\right)\hat{\mathbf{d}}(s_0) + (a+\delta a)\sin t_0 \hat{\mathbf{n}}_\mu(s_0) - a\sin t_\nu \hat{\mathbf{n}}_\nu(s_0)$$

$$- a\sigma_\nu(s_0)(\omega_{\nu,2}(s_0)\sin t_\nu - \omega_{\nu,3}(s_0)\cos t_\nu)\hat{\mathbf{t}}_\nu(s_0)(\tilde{s}_\nu - s_0),$$

(4.38)

$$\mathbf{S}_\mu^-(s',t') - \mathbf{S}_\nu(\tilde{s}_\nu, t_\nu) \approx -\delta_\mu R\hat{\mathbf{d}} + \sigma_\mu(s_0)\hat{\mathbf{t}}_\mu(s_0)(s'-s_0) - \sigma_\mu(s_0)\hat{\mathbf{t}}_\nu(s_0)(\tilde{s}_\nu - s_0)$$

$$+\frac{\sigma_\mu(s_0)^2}{2}\left(\omega_{\mu,3}(s_0)\hat{\mathbf{d}}(s_0) - \omega_{\mu,2}(s_0)\hat{\mathbf{n}}_1(s_0)\right)(s'-s_0)^2$$

$$-\frac{\sigma_\nu(s_0)^2}{2}\left(\omega_{\nu,3}(s_0)\hat{\mathbf{d}}(s_0) - \omega_{\nu,2}(s_0)\hat{\mathbf{n}}_2(s_0)\right)(\tilde{s}_\nu - s_0)^2 + \left((a-\delta a)\cos t' - a\cos t_\nu\right)\hat{\mathbf{d}}(s_0)$$

$$+(a-\delta a)\sin t'\hat{\mathbf{n}}_\mu(s_0) - a\sin t_\nu \hat{\mathbf{n}}_\nu(s_0) + (a-\delta a)\sigma_\mu(s_0)(\omega_{\mu,2}(s_0)\sin t' - \omega_{\mu,3}(s_0)\cos t')\hat{\mathbf{t}}_\mu(s_0)(s'-s_0)$$

$$-a\sigma_\nu(s_0)(\omega_{\nu,2}(s_0)\sin t_\nu - \omega_{\nu,3}(s_0)\cos t_\nu)\hat{\mathbf{t}}_2(s_0)(\tilde{s}_\nu - s_0),$$

(4.39)

$$\mathbf{S}_\nu^-(s',t') - \mathbf{S}_\nu(\tilde{s}_\nu, t_\nu) \approx \sigma_\nu(s_0)\hat{\mathbf{t}}_\nu(s_0)(s' - \tilde{s}_\nu)$$

$$+\frac{\sigma_\nu(s_0)^2}{2}\left(\omega_{\nu,3}(s_0)\hat{\mathbf{d}}(s_0) - \omega_{\nu,2}(s_0)\hat{\mathbf{n}}_\nu(s_0)\right)\left((s'-s_0)^2 - (\tilde{s}_\nu - s_0)^2\right)$$

$$+\left((a-\delta a)\cos t' - a\cos t_\nu\right)\hat{\mathbf{d}}(s_0) + \left((a-\delta a)\sin t' - a\sin t_\nu\right)\hat{\mathbf{n}}_\nu(s_0) \quad (4.40)$$

$$+(a-\delta a)\sigma_\nu(s_0)(\omega_{\nu,2}(s_0)\sin t' - \omega_{\nu,3}(s_0)\cos t')\hat{\mathbf{t}}_\nu(s_0)(s'-s_0)$$

$$-a\sigma_\nu(s_0)(\omega_{\nu,2}(s_0)\sin t_\nu - \omega_{\nu,3}(s_0)\cos t_\nu)\hat{\mathbf{t}}_\nu(s_0)(\tilde{s}_\nu - s_0).$$

Then Eqs. (4.38)-(4.40) can be substituted into Eq. (4.17) and expanded out for small $\omega_{\mu,2}(s_0)$, $\omega_{\mu,3}(s_0)$, $\omega_{\nu,2}(s_0)$ and $\omega_{\nu,3}(s_0)$, as well as the image charge expansion performed. Here, we can choose $\mathbf{k}.\hat{\mathbf{t}}_\nu(s) = k_z$ and $\mathbf{k}.\hat{\mathbf{n}}_\nu(s) = K\sin\phi_K$. However, now, through Eq. (1.10), we have the requirements that

$$\mathbf{k}.\hat{\mathbf{t}}_\mu(s) = k_z \cos\eta(s_0) + \delta_\mu K\sin\phi_K \sin\eta(s_0), \quad \mathbf{k}.\hat{\mathbf{n}}_\nu(s) = K\sin\phi_K \cos\eta(s_0) - \delta_\mu k_z \sin\eta(s_0).$$

(4.41)

This all gives us (taking the $\delta a \to 0$ limit where appropriate)

$$\int_0^{L_B} ds' \int_0^{2\pi} dt' \int d^3k \exp\left(-i\delta_\mu RK\cos\phi_K + \sigma_\mu(s_0)\left(ik_z \cos\eta(s_0) + i\delta_\mu K\sin\phi_K \sin\eta(s_0)\right)(s'-s_0) - ik_z\sigma_\nu(s_0)(\tilde{s}_\nu - s_0)\right)$$

$$\exp\left(iK\left(a\cos t' - a\cos t_\mu\right)\cos\phi_K + ia\sin t'\left(K\sin\phi_K \cos\eta(s_0) - \delta_\mu k_z \sin\eta(s_0)\right)\right)$$

$$\exp(-iaK\sin t_\nu \sin\phi_K)\left(1 + \frac{i\sigma_\mu(s_0)^2}{2}\left(-\omega_{\mu,2}(s_0)\left(K\sin\phi_K \cos\eta(s_0) - k_z \sin\eta(s_0)\right)\sin\phi_K\right)(s'-s_0)^2\right.$$

$$+\omega_{\mu,3}(s_0)K\cos\phi_K\bigg) - \frac{i\sigma_\nu(s_0)^2}{2}\left(\omega_{\nu,3}(s_0)K\cos\phi_K - \omega_{\nu,2}(s_0)K\sin\phi_K\right)(\tilde{s}_\nu - s_0)^2$$

$$+ia\sigma_\mu(s_0)(\omega_{\mu,2}(s_0)\sin t' - \omega_{\mu,3}(s_0)\cos t')\left(k_z \cos\eta(s_0) + K\sin\phi_K \sin\eta(s_0)\right)(s'-s_0)$$

$$-ia\sigma_\nu(s_0)(\omega_{\nu,2}(s_0)\sin t_\nu - \omega_{\nu,3}(s_0)\cos t_\nu)k_z(\tilde{s}_\nu - s_0)\bigg)$$

$$\left(K\cos(\phi_K - t_\nu) + k_z\sigma_\nu(s_0)(\omega_{\nu,2}(s_0)\sin t_\nu - \omega_{\nu,3}(s_0)\cos t_\nu)(\tilde{s}_\nu - s_0)\right)\frac{(\sigma_{ind,0}^{(1)}(t',s') + \delta(t'-t_0)\delta(s'-s_0))}{(\mathbf{k}^2 + \kappa_D^2)}$$

$$+\int_0^{L_B} ds' \int_0^{2\pi} dt' \int d^3k \exp\left(ik_z \sigma_\nu(s_0)(s'-s_2) + K\left((a-\delta a)\cos(t'-\phi_K) - a\cos(t_2-\phi_K)\right)\right)$$

$$\left(1 + \frac{i\sigma_\nu(s_0)^2}{2}\left(\omega_{\nu,3}(s_0)K\cos\phi_K - \omega_{\nu,2}(s_0)K\sin\phi_K\right)\left((s'-s_0)^2 - (\tilde{s}_\nu - s_0)^2\right)\right.$$

$$+i(a-\delta a)\sigma_\nu(s_0)(\omega_{\nu,2}(s_0)\sin t' - \omega_{\nu,3}(s_0)\cos t')k_z(s'-s_0)$$

$$-ia\sigma_\nu(s_0)(\omega_{\nu,2}(s_0)\sin t_\nu - \omega_{\nu,3}(s_0)\cos t_\nu)k_z(\tilde{s}_\nu - s_0)$$

$$\left(K\cos(t_\nu - \phi_K) + \sigma_\nu(s_0)(\omega_{\nu,2}(s_0)\sin t_\nu - \omega_{\nu,3}(s_0)\cos t_\nu)k_z(\tilde{s}_\nu - s_0)\right)\frac{\sigma_{ind,1}^{(\nu)}(t',s')}{(\mathbf{k}^2 + \kappa_D^2)} = 0.$$

(4.42)

We can then expand out the image charges in powers of $\omega_{\mu,2}(s_0)$ and $\omega_{\mu,3}(s_0)$

$$\sigma_{ind,0}^{(\mu)}(t',s') = \sigma_{ind,0,0}^{(\mu)}(t',s') + \omega_{\mu,2}(s_0)\sigma_{ind,0,1}^{(\mu)}(t',s') + \omega_{\mu,3}(s_0)\sigma_{ind,0,2}^{(\mu)}(t',s') + ..., \quad (4.43)$$

$$\sigma_{ind,1}^{(\nu)}(t',s') = \sigma_{ind,1,0}^{(\nu)}(t',s') + \omega_{\mu,2}(s_0)\sigma_{ind,1,1}^{(\nu)}(t',s') + \omega_{\mu,3}(s_0)\sigma_{ind,1,2}^{(\nu)}(t',s')$$
$$+\omega_{\nu,2}(s_0)\sigma_{ind,1,3}^{(\nu)}(t',s') + \omega_{\nu,3}(s_0)\sigma_{ind,1,4}^{(\nu)}(t',s')....$$

(4.44)

For the leading order images, we obtain the following expressions for the terms in the expansion

$$\int_0^{L_B} ds' \int_0^{2\pi} dt' \int d^3k \exp\left(ik_z\sigma_\mu(s_0)(s'-\tilde{s}_\mu) + i(a-\delta a)K\cos(t'-\phi_K) - iaK\cos(t_\mu - \phi_K)\right)\frac{K\cos(t_\mu - \phi_K)}{(\mathbf{k}^2 + \kappa_D^2)}\sigma_{ind,0,0}^{(1)}(t',s')$$

$$+\int d^3k \exp\left(ik_z\sigma_\mu(s_0)(s_0 - \tilde{s}_\mu) + i(a+\delta a)K\cos(t_0 - \phi_K) - iaK\cos(t_\mu - \phi_K)\right)\frac{K\cos(t_\mu - \phi_K)}{(\mathbf{k}^2 + \kappa_D^2)} = 0,$$

(4.45)

$$\int_0^{L_B} ds' \int_0^{2\pi} dt' \int d^3k \exp\left(ik_z\sigma_\mu(s_0)(s'-\tilde{s}_\mu) + i(a-\delta a)K\cos(t'-\phi_K) - iaK\cos(t_\mu - \phi_K)\right)\frac{K\cos(t_\mu - \phi_K)}{(\mathbf{k}^2 + \kappa_D^2)}$$

$$\times \sigma_{ind,0,1}^{(\mu)}(t',s') + \int_0^{L_B} ds' \int_0^{2\pi} dt' \int d^3k \exp\left(ik_z\sigma_\mu(s_0)(s'-\tilde{s}_\mu) + i(a-\delta a)K\cos(t'-\phi_K) - iaK\cos(t_\mu - \phi_K)\right)$$

$$\left(K\cos(t_\mu - \phi_K)\left(-i\frac{\sigma_\mu(s_0)^2}{2}K\sin\phi_K\left((s'-s_0)^2 - (\tilde{s}_\mu - s_0)^2\right) - iak_z\sigma_\mu(s_0)\sin t_\mu(\tilde{s}_\mu - s_0)\right.\right.$$

$$+i(a-\delta a)k_z\sigma_\mu(s_0)\sin t'(s'-s_0)\right) + ik_z\sigma_\mu(s_0)\sin t_\mu(\tilde{s}_\mu - s_0)\right)\frac{1}{(\mathbf{k}^2 + \kappa_D^2)}\sigma_{ind,0,0}^{(\mu)}(t',s')$$

$$+\int d^3k \exp\left(ik_z\sigma_\mu(s_0)(s_0 - \tilde{s}_\mu) + i(a+\delta a)K\cos(t_0 - \phi_K) - iaK\cos(t_\mu - \phi_K)\right)\frac{1}{(\mathbf{k}^2 + \kappa_D^2)}$$

$$\left(K\cos(t_\mu - \phi_K)\left(i\frac{\sigma_\mu(s_0)^2}{2}K\sin\phi_K(\tilde{s}_\mu - s_0)^2 - iak_z\sigma_\mu(s_0)\sin t_\mu(\tilde{s}_\mu - s_0)\right) + ik_z\sigma_\mu(s_0)\sin t_\mu(\tilde{s}_\mu - s_0)\right) = 0,$$

(4.46)

$$\int_0^{L_B} ds' \int_0^{2\pi} dt' \int d^3k \exp\left(ik_z \sigma_\mu(s_0)(s'-\tilde{s}_\mu) + i(a-\delta a)K\cos(t'-\phi_K) - iaK\cos(t_\mu - \phi_K)\right) \frac{K\cos(t_\mu - \phi_K)}{(\mathbf{k}^2 + \kappa_D^2)}$$

$$\times \sigma_{ind,0,2}^{(\mu)}(t',s') + \int_0^{L_B} ds' \int_0^{2\pi} dt' \int d^3k \exp\left(ik_z \sigma_\mu(s_0)(s'-\tilde{s}_\mu) + i(a-\delta a)K\cos(t'-\phi_K) - iaK\cos(t_\mu - \phi_K)\right)$$

$$\left( K\cos(t_\mu - \phi_K)\left( i\frac{\sigma_\mu(s_0)^2}{2} K\cos\phi_K \left((s'-s_0)^2 - (\tilde{s}_\mu - s_0)^2\right) + iak_z \sigma_\mu(s_0)\cos t_\mu(\tilde{s}_\mu - s_0) \right.\right.$$

$$\left.\left. -i(a-\delta a)k_z \sigma_\mu(s_0)\cos t'(s'-s_0)\right) - k_z \sigma_\mu(s_0)\cos t_\mu(\tilde{s}_\mu - s_0)\right) \frac{\sigma_{ind,0,0}^{(\mu)}(t',s')}{(\mathbf{k}^2 + \kappa_D^2)}$$

$$+ \int d^3k \exp\left(ik_z \sigma_\mu(s_0)(s_0 - \tilde{s}_\mu) + i(a+\delta a)K\cos(t_0 - \phi_K) - iaK\cos(t_\mu - \phi_K)\right) \frac{1}{(\mathbf{k}^2 + \kappa_D^2)}$$

$$\left( K\cos(t_\mu - \phi_K)\left( -i\frac{\sigma_\mu(s_0)^2}{2} K\cos\phi_K (\tilde{s}_\mu - s_0)^2 + iak_z \sigma_\mu(s_0)\cos t_\mu(\tilde{s}_\mu - s_0)\right) - \sigma_\mu(s_0)k_z \cos t_\mu(\tilde{s}_\mu - s_0)\right) = 0,$$

(4.47)

For the next to leading order images we obtain the following expressions for the terms in the expansion

$$\int_0^{L_B} ds' \int_0^{2\pi} dt' \int d^3k \exp\left(-i\delta_\mu RK\cos\phi_K + \left(ik_z \cos\eta(s_0) + i\delta_\mu K\sin\phi_K \sin\eta(s_0)\right)\sigma_\mu(s_0)(s'-s_0) - ik_z \sigma_\nu(s_0)(\tilde{s}_\nu - s_0)\right)$$

$$\times \exp\left(iK(a\cos t' - a\cos t_\nu)\cos\phi_K + ia\sin t'\left(K\sin\phi_K \cos\eta(s_0) - \delta_\mu k_z \sin\eta(s_0)\right) - iaK\sin t_\nu \sin\phi_K\right)$$

$$\times K\cos(t_\nu - \phi_K)\frac{(\sigma_{ind,0,0}^{(\mu)}(t',s') + \delta(t'-t_0)\delta(s'-s_0))}{(\mathbf{k}^2 + \kappa_D^2)}$$

$$+ \int_0^{L_B} ds' \int_0^{2\pi} dt' \int d^3k \exp\left(ik_z \sigma_\nu(s_0)(s'-\tilde{s}_\nu) + K\left((a+\delta a)\cos(t'-\phi_K) - a\cos(t_\nu - \phi_K)\right)\right)$$

$$\times \frac{K\cos(t_\nu - \phi_K)\sigma_{ind,1,0}^{(\nu)}(t',s')}{(\mathbf{k}^2 + \kappa_D^2)} = 0,$$

(4.48)

$$\int_0^{L_B} ds' \int_0^{2\pi} dt' \int d^3k \exp\left(-i\delta_\mu RK\cos\phi_K + \left(ik_z \cos\eta(s_0) + i\delta_\mu K\sin\phi_K \sin\eta(s_0)\right)\sigma_\mu(s_0)(s'-s_0) - ik_z \sigma_\nu(s_0)(\tilde{s}_\nu - s_0)\right)$$

$$\exp\left(iK(a\cos t' - a\cos t_\nu)\cos\phi_K + ia\sin t'\left(K\sin\phi_K \cos\eta(s_0) - \delta_\mu k_z \sin\eta(s_0)\right) - iaK\sin t_\nu \sin\phi_K\right)$$

$$\left( \sigma_{ind,0,1}^{(\mu)}(t',s') + \left[ -\frac{i\sigma_\mu(s_0)^2}{2}\left(K\sin\phi_K \cos\eta(s_0) - \delta_\mu k_z \sin\eta(s_0)\right)(s'-s_0)^2 + ia\sigma_\mu(s_0) \right.\right.$$

$$\left.\left. \left(k_z \cos\eta(s_0) + \delta_\mu K\sin\phi_K \sin\eta(s_0)\right)\sin t'(s'-s_0)\right](\sigma_{ind,0,0}^{(\mu)}(t',s') + \delta(s'-s_0)\delta(t'-t_0))\right) \frac{K\cos(t_\nu - \phi_K)}{(\mathbf{k}^2 + \kappa_D^2)}$$

$$+\int_0^{L_B} ds' \int_0^{2\pi} dt' \int d^3k \exp\left(ik_z \sigma_\nu(s_0)(s'-\tilde{s}_\nu) + K\left((a+\delta a)\cos(t'-\phi_K) - a\cos(t_\nu - \phi_K)\right)\right)$$

$$K\cos(t_\nu - \phi_K)\frac{\sigma^{(\nu)}_{ind,1,1}(t',s')}{(\mathbf{k}^2 + \kappa_D^2)} = 0,$$

(4.49)

$$\int_0^{L_B} ds' \int_0^{2\pi} dt' \int d^3k \exp\left(-i\delta_\mu RK\cos\phi_K + \left(ik_z\cos\eta(s_0) + i\delta_\mu K\sin\phi_K \sin\eta(s_0)\right)\sigma_\mu(s_0)(s'-s_0) - ik_z\sigma_\nu(s_0)(\tilde{s}_\nu - s_0)\right)$$

$$\exp\left(iK(a\cos t' - a\cos t_\nu)\cos\phi_K + i(a-\delta a)\sin t'\left(K\sin\phi_K \cos\eta(s_0) - k_z \sin\eta(s_0)\right) - iaK\sin t_\nu \sin\phi_K\right)$$

$$\left(\sigma^{(\mu)}_{ind,0,2}(t',s') + \left(\frac{i\sigma_\mu(s_0)^2}{2}K\cos\phi_K(s'-s_0)^2 - i(a-\delta a)\sigma_\mu(s_0)\cos t'\left(k_z\cos\eta(s_0) + \delta_\mu K\sin\phi_K \sin\eta(s_0)\right)(s'-s_0)\right)\right.$$

$$(\sigma^{(\mu)}_{ind,0,0}(t',s') + \delta(t'-t_0)\delta(s'-s_0))\Bigg)\frac{K\cos(t_\nu - \phi_K)}{(\mathbf{k}^2 + \kappa_D^2)}$$

$$+\int_0^{L_B} ds' \int_0^{2\pi} dt' \int d^3k \exp\left(ik_z\sigma_\nu(s_0)(s'-\tilde{s}_\nu) + K\left((a+\delta a)\cos(t'-\phi_K) - a\cos(t_\nu - \phi_K)\right)\right)$$

$$K\cos(t_\nu - \phi_K)\frac{\sigma^{(\nu)}_{ind,1,2}(t',s')}{(\mathbf{k}^2 + \kappa_D^2)} = 0,$$

(4.50)

$$\int_0^{L_B} ds' \int_0^{2\pi} dt' \int d^3k \exp\left(-i\delta_\mu RK\cos\phi_K + i\left(k_z\cos\eta(s_0) + \delta_\mu K\sin\phi_K \sin\eta(s_0)\right)\sigma_\nu(s_0)(s'-s_0) - i\sigma_\mu(s_0)k_z(\tilde{s}_\nu - s_0)\right)$$

$$\exp\left(iK(a\cos t' - a\cos t_\nu)\cos\phi_K + ia\sin t'\left(K\sin\phi_K \cos\eta(s_0) - \delta_\mu k_z \sin\eta(s_0)\right) - iaK\sin t_\nu \sin\phi_K\right)$$

$$\left(\left(\frac{i\sigma_2(s_0)^2}{2}K\sin\phi_K(\tilde{s}_\nu - s_0)^2 - iak_z\sigma_\nu(s_0)\sin t_\nu(\tilde{s}_\nu - s_0)\right)K\cos(t_\nu - \phi_K) + \sigma_\nu(s_0)ik_z\sin t_\nu(\tilde{s}_\nu - s_0)\right)$$

$$\frac{(\sigma^{(\mu)}_{ind,0,0}(t',s') + \delta(t'-t_0)\delta(s'-s_0))}{(\mathbf{k}^2 + \kappa_D^2)}$$

$$+\int_0^{L_B} ds' \int_0^{2\pi} dt' \int d^3k \exp\left(ik_z\sigma_\nu(s_0)(s'-s_\nu) + K\left((a-\delta a)\cos(t'-\phi_K) - a\cos(t_\nu - \phi_K)\right)\right)$$

$$\left(-\frac{i\sigma_\nu(s_0)^2}{2}K\sin\phi_K\left((s'-s_0)^2 - (\tilde{s}_\nu - s_0)^2\right) + ik_z(a-\delta a)\sigma_\nu(s_0)\sin t'(s'-s_0) - iak_z\sigma_\nu(s_0)\sin t_\nu(\tilde{s}_\nu - s_0)\right)$$

$$K\cos(t_\nu - \phi_K) + \sigma_\nu(s_0)\sin t_\nu k_z(\tilde{s}_\nu - s_0)\Bigg)\frac{\sigma^{(\nu)}_{ind,1,0}(t',s')}{(\mathbf{k}^2 + \kappa_D^2)}$$

$$+\int_0^{L_B} ds' \int_0^{2\pi} dt' \int d^3k \exp\left(ik_z\sigma_\nu(s_0)(s'-\tilde{s}_\nu) + K\left((a-\delta a)\cos(t'-\phi_K) - a\cos(t_\nu - \phi_K)\right)\right)$$

$$K\cos(t_\nu - \phi_K)\frac{\sigma^{(\nu)}_{ind,1,3}(t',s')}{(\mathbf{k}^2 + \kappa_D^2)} = 0,$$

(4.51)

$$\int_0^{L_B} ds' \int_0^{2\pi} dt' \int d^3k \exp\left(-i\delta_\mu RK \cos\phi_K + i\left(k_z \cos\eta(s_0) + \delta_\mu K \sin\phi_K \sin\eta(s_0)\right)\sigma_\mu(s_0)(s'-s_0) - ik_z\sigma_\nu(s_0)(\tilde{s}_\nu - s_0)\right)$$

$$\exp\left(iK\left(a\cos t' - a\cos t_\nu\right)\cos\phi_K + ia\sin t'\left(K\sin\phi_K \cos\eta(s_0) - \delta_\mu k_z \sin\eta(s_0)\right) - iaK\sin t_\nu \sin\phi_K\right)$$

$$\left(\left(-\frac{i\sigma_\nu(s_0)^2}{2}K\cos\phi_K(s_\nu - s_0)^2 + iak_z\sigma_\nu(s_0)\cos t_\nu(\tilde{s}_\nu - s_0)\right)K\cos(t_\nu - \phi_K) - k_z\sigma_\nu(s_0)\cos t_\nu(\tilde{s}_\nu - s_0)\right)$$

$$\frac{(\sigma_{ind,0,0}^{(\mu)}(t',s') + \delta(t'-t_0)\delta(s'-s_0))}{(\mathbf{k}^2 + \kappa_D^2)}$$

$$+\int_0^{L_B} ds' \int_0^{2\pi} dt' \int d^3k \exp\left(ik_z\sigma_\nu(s_0)(s'-\tilde{s}_\nu) + K\left(a\cos(t'-\phi_K) - a\cos(t_\nu - \phi_K)\right)\right)$$

$$\left(\left(\frac{i\sigma_\nu(s_0)^2}{2}K\cos\phi_K\left((s'-s_0)^2 - (\tilde{s}_\nu - s_0)^2\right) - i(a-\delta a)k_z\sigma_\nu(s_0)\cos t'(s'-s_0)\right.\right.$$

$$\left.\left.+iak_z\sigma_\nu(s_0)\cos t_\nu(\tilde{s}_\nu - s_0)\right)K\cos(t_\nu - \phi_K) - \sigma_\nu(s_0)k_z\cos t_\nu(\tilde{s}_\nu - s_0)\right)\frac{\sigma_{ind,1,0}^{(\nu)}(t',s')}{(\mathbf{k}^2 + \kappa_D^2)}$$

$$+\int_0^{L_B} ds' \int_0^{2\pi} dt' \int d^3k \exp\left(ik_z\sigma_\nu(s_0)(s'-\tilde{s}_\nu) + K\left(a\cos(t'-\phi_K) - a\cos(t_\nu - \phi_K)\right)\right)K\cos(t_\nu - \phi_K)$$

$$\frac{\sigma_{ind,1,4}^{(\nu)}(t',s')}{(\mathbf{k}^2 + \kappa_D^2)} = 0.$$

(4.52)

We can perform the expansion for the expressions for the charge densities (Eqs. (4.28) and (4.29))

$$\rho_{eff}^{(\mu)}(\mathbf{k};s_0) = \rho_{eff,0}^{(\mu)}(\mathbf{k};s_0) + \omega_{\mu,2}(s_0)\rho_{eff,1,1}^{(\mu)}(\mathbf{k};s_0) + \omega_{\mu,3}(s_0)\rho_{eff,1,2}^{(\mu)}(\mathbf{k};s_0) + \ldots,$$

(4.53)

$$\rho_{img}^{(\nu,\mu)}(\mathbf{k};s_0) = \rho_{img,0}^{(\nu,\mu)}(\mathbf{k};s_0) + \omega_{\mu,2}(s_0)\rho_{img,1,1}^{(\nu,\mu)}(\mathbf{k};s_0) + \omega_{\mu,3}(s_0)\rho_{img,1,2}^{(\nu,\mu)}(\mathbf{k};s_0)$$
$$+\omega_{\nu,2}(s_0)\rho_{img,1,3}^{(\nu,\mu)}(\mathbf{k};s_0) + \omega_{\nu,3}(s_0)\rho_{img,1,4}^{(\nu,\mu)}(\mathbf{k};s_0)\ldots,$$

(4.54)

which yield for $\rho_{eff}^{(\mu)}(\mathbf{k};s_0)$

$$\rho_{eff,0}^{(\mu)}(\mathbf{k};s_0) = \frac{1}{2\pi}\int_0^{L_B} ds' \int_0^{2\pi} dt' \exp\left(i\mathbf{k}.\mathbf{r}_\mu(s_0) + i\sigma_\mu(s_0)\mathbf{k}.\hat{\mathbf{t}}_\mu(s_0)(s'-s_0) + ia\cos t'\mathbf{k}.\hat{\mathbf{d}}(s_0) + ia\sin t'\mathbf{k}.\hat{\mathbf{n}}_\mu(s_0)\right)$$

$$\left(\sigma_{ind,0,0}^{(\mu)}(t',s') + \delta(t'-t_0)\delta(s'-s_0)\right),$$

(4.55)

$$\rho_{eff,1,1}^{(\mu)}(\mathbf{k};s_0) = \frac{1}{2\pi}\int_0^{L_B} ds' \int_0^{2\pi} dt' \exp\left(i\mathbf{k}.\mathbf{r}_\mu(s_0) + i\sigma_\mu(s_0)\mathbf{k}.\hat{\mathbf{t}}_\mu(s_0)(s'-s_0) + ia\cos t'\mathbf{k}.\hat{\mathbf{d}}(s_0) + ia\sin t'\mathbf{k}.\hat{\mathbf{n}}_\mu(s_0)\right)$$

$$\left(\left(-\frac{i\sigma_\mu(s_0)^2}{2}\mathbf{k}.\hat{\mathbf{n}}_\mu(s_0)(s'-s_0)^2 + ia\sigma_\mu(s_0)\sin t'\mathbf{k}.\hat{\mathbf{t}}_\mu(s_0)(s'-s_0)\right)\left(\sigma_{ind,0}^{(\mu)}(t',s') + \delta(t'-t_0)\delta(s'-s_0)\right)\right.$$

$$\left.+\sigma_{ind,0,1}^{(\mu)}(t',s')\right),$$

(4.56)

$$\rho_{eff,1,2}^{(\mu)}(\mathbf{k};s_0) = \frac{1}{2\pi}\int_0^{L_B} ds' \int_0^{2\pi} dt' \exp\left(i\mathbf{k}.\mathbf{r}_\mu(s_0) + i\sigma_\mu(s_0)\mathbf{k}.\hat{\mathbf{t}}_\mu(s_0)(s'-s_0) + ia\cos t'\mathbf{k}.\hat{\mathbf{d}}(s_0) + ia\sin t'\mathbf{k}.\hat{\mathbf{n}}_\mu(s_0)\right)$$

$$\left(\left(\frac{i\sigma_\mu(s_0)^2}{2}\mathbf{k}.\hat{\mathbf{d}}(s_0)(s'-s_0)^2 - ia\sigma_\mu(s_0)\cos t'\mathbf{k}.\hat{\mathbf{t}}_\mu(s_0)(s'-s_0)\right)\left(\sigma_{ind,0}^{(\mu)}(t',s') + \delta(t'-t_0)\delta(s'-s_0)\right)\right.$$

$$\left.+\sigma_{ind,0,2}^{(\mu)}(t',s')\right),$$

(4.57)

and for $\rho_{img}^{(2,1)}(\mathbf{k};s_0)$

$$\rho_{img,0}^{(\nu,\mu)}(\mathbf{k};s_0) = \frac{1}{2\pi}\int_0^{L_B} ds' \int_0^{2\pi} dt' \exp\left(i\mathbf{k}.\mathbf{r}_\nu(s_0) + i\sigma_1(s_0)\mathbf{k}.\hat{\mathbf{t}}_\nu(s_0)(s'-s_0) + ia\cos t'\mathbf{k}.\hat{\mathbf{d}}(s_0) + ia\sin t'\mathbf{k}.\hat{\mathbf{n}}_\nu(s_0)\right)$$

$$\sigma_{ind,2,0}^{(\nu)}(t',s'),$$

(4.58)

$$\rho_{img,1,1}^{(\nu,\mu)}(\mathbf{k};s_0) = \frac{1}{2\pi}\int_0^{L_B} ds' \int_0^{2\pi} dt' \exp\left(i\mathbf{k}.\mathbf{r}_\nu(s_0) + i\sigma_\nu(s_0)\mathbf{k}.\hat{\mathbf{t}}_\nu(s_0)(s'-s_0) + ia\cos t'\mathbf{k}.\hat{\mathbf{d}}(s_0) + ia\sin t'\mathbf{k}.\hat{\mathbf{n}}_\nu(s_0)\right)$$

$$\sigma_{ind,2,1}^{(\nu)}(t',s'),$$

(4.59)

$$\rho_{img,1,2}^{(\nu,\mu)}(\mathbf{k};s_0) = \frac{1}{2\pi}\int_0^{L_B} ds' \int_0^{2\pi} dt' \exp\left(i\mathbf{k}.\mathbf{r}_\nu(s_0) + i\sigma_\nu(s_0)\mathbf{k}.\hat{\mathbf{t}}_\nu(s_0)(s'-s_0) + ia\cos t'\mathbf{k}.\hat{\mathbf{d}}(s_0) + ia\sin t'\mathbf{k}.\hat{\mathbf{n}}_\nu(s_0)\right)$$

$$\sigma_{ind,2,2}^{(\nu)}(t',s'),$$

(4.60)

$$\rho_{img,1,3}^{(\nu,\mu)}(\mathbf{k};s_0) = \frac{1}{2\pi}\int_0^{L_B} ds' \int_0^{2\pi} dt' \exp\left(i\mathbf{k}.\mathbf{r}_\nu(s_0) + i\sigma_\nu(s_0)\mathbf{k}.\hat{\mathbf{t}}_\nu(s_0)(s'-s_0) + ia\cos t'\mathbf{k}.\hat{\mathbf{d}}(s_0) + ia\sin t'\mathbf{k}.\hat{\mathbf{n}}_\nu(s_0)\right)$$

$$\left(\sigma_{ind,2,3}^{(\nu)}(t',s') + \left(ia\sigma_\nu(s_0)\sin t'\mathbf{k}.\hat{\mathbf{t}}_\nu(s_0)(s'-s_0) - \frac{i\sigma_\nu(s_0)^2}{2}\mathbf{k}.\hat{\mathbf{n}}_\nu(s_0)(s'-s_0)^2\right)\sigma_{ind,2,0}^{(\nu)}(t',s')\right),$$

(4.61)

$$\rho_{img,1,4}^{(v,\mu)}(\mathbf{k};s_0) = \frac{1}{2\pi}\int_0^{L_B} ds' \int_0^{2\pi} dt' \exp\left(i\mathbf{k}.\mathbf{r}_v(s_0) + i\sigma_v(s_0)\mathbf{k}.\hat{\mathbf{t}}_v(s_0)(s'-s_0) + ia\cos t'\mathbf{k}.\hat{\mathbf{d}}(s_0) + ia\sin t'\mathbf{k}.\hat{\mathbf{n}}_v(s_0)\right)$$

$$\left(\sigma_{ind,2,4}^{(v)}(t',s') + \left(\frac{i\sigma_v(s_0)^2}{2}\mathbf{k}.\hat{\mathbf{n}}_v(s_0)(s'-s_0)^2 - ia\sigma_v(s_0)\cos t'\mathbf{k}.\hat{\mathbf{t}}_v(s_0)(s'-s_0)\right)\sigma_{ind,2,0}^{(v)}(t',s')\right).$$

(4.62)

### *4.6 Evaluation of the leading order terms in the curvature expansion for a charge on rod 1*

Now, we'll keep our attention on terms that are lowest order in the curvature. This is fine provided that $\kappa_D \gg \omega_{1,2}, \omega_{2,2}, \omega_{1,3}, \omega_{2,3}$. Therefore we only retain Eqs. (4.45) and (4.48). Let's start with Eq. (4.45). Using identities similar to (2.17), and the fact that $\cos x = 1/2(\exp(ix) + \exp(-ix))$, we can write Eq. (4.45) for rod 1

$$\frac{1}{2}\sum_{n,n'} \int_0^{L_B} ds' \int_0^{2\pi} dt' \int d^3k\, i^{n-n'} \exp\left(ik_z\sigma_1(s_0)(s'-\tilde{s}_1)\right) KJ_n((a-\delta a)K) J_{n'}(aK)$$

$$\left(\exp\left(in(t'-\phi_K) - i(n'+1)(t_1-\phi_K)\right) + \exp\left(in(t'-\phi_K) - i(n'-1)(t_1-\phi_K)\right)\right)\frac{1}{(\mathbf{k}^2 + \kappa_D^2)}\sigma_{ind,0,0}^{(1)}(t',s')$$

$$+\frac{1}{2}\sum_{n,n'} \int d^3k\, i^{n-n'} \exp\left(ik_z\sigma_1(s_0)(s_0-\tilde{s}_1)\right) KJ_n((a+\delta a)K) J_{n'}(aK)$$

$$\left(\exp\left(in(t_0-\phi_K) - i(n'+1)(t_1-\phi_K)\right) + \exp\left(in(t_0-\phi_K) - i(n'-1)(t_1-\phi_K)\right)\right)\frac{1}{(\mathbf{k}^2 + \kappa_D^2)} = 0.$$

(4.63)

We can then evaluate the $\phi_K$ integral and then the K-integral in Eq. (4.63), yielding

$$\frac{i}{2}\sum_n \int_0^{L_B} ds' \int_0^{2\pi} dt' \int_0^\infty dK \int_{-\infty}^\infty dk_z \exp\left(ik_z\sigma_1(s_0)(s'-\tilde{s}_1)\right) I_n\left(a\sqrt{k_z^2+\kappa_D^2}\right) K'\left(a\sqrt{k_z^2+\kappa_D^2}\right)\exp(in(t'-t_1))$$

$$\sigma_{ind,0,0}^{(1)}(t',s')$$

$$+\frac{i}{2}\sum_n \int_{-\infty}^\infty dK \int_{-\infty}^\infty dk_z \exp\left(ik_z\sigma_1(s_0)(s_0-\tilde{s}_1)\right) K_n\left(a\sqrt{k_z^2+\kappa_D^2}\right) I'\left(a\sqrt{k_z^2+\kappa_D^2}\right)\exp(in(t_0-t_1)) = 0.$$

(4.64)

It is then useful for us to write

$$\sigma_{ind,0,0}^{(1)}(t',s') = \frac{\sigma_1(s_0)}{2\pi}\sum_n \int_{-\infty}^\infty dk_z \exp(in(t_0-t'))\exp(ik_z\sigma_1(s_0)(s_0-s'))\zeta_{img0}^{(1)}(n,k_z). \qquad (4.65)$$

Eq. (4.65) allows us to express Eq.(4.64) as:

$$\frac{i}{2}\sum_n \int_0^\infty dK \int_{-\infty}^\infty dk_z \exp(i\sigma_1(s_0)k_z(s_0-\tilde{s}_1)) I_n\left(a\sqrt{k_z^2+\kappa_D^2}\right) K'\left(a\sqrt{k_z^2+\kappa_D^2}\right)\exp(in(t_0-t_1))$$
$$\zeta^{(1)}_{img\,0}(n,k_z)$$
$$+\frac{i}{2}\sum_n \int_{-\infty}^\infty dK \int_{-\infty}^\infty dk_z \exp(i\sigma_1(s_0)k_z(s_0-\tilde{s}_1)) K_n\left(a\sqrt{k_z^2+\kappa_D^2}\right) I'\left(a\sqrt{k_z^2+\kappa_D^2}\right)\exp(in(t_0-t_1))=0.$$

(4.66)

The solution to Eq. (4.66) is simply

$$\zeta^{(1)}_{img\,0}(n,k_z) = -\frac{K_n\left(a\sqrt{k_z^2+\kappa_D^2}\right) I'\left(a\sqrt{k_z^2+\kappa_D^2}\right)}{I_n\left(a\sqrt{k_z^2+\kappa_D^2}\right) K'\left(a\sqrt{k_z^2+\kappa_D^2}\right)}.$$

(4.67)

Using the fact that

$$\delta(t'-t_0)\delta(s'-s_0) = \frac{\sigma_1(s_0)}{2\pi}\sum_n \int_{-\infty}^\infty dk_z \exp(in(t_0-t'))\exp(ik_z\sigma_1(s_0)(s_0-s'))$$

(4.68)

allows us to write Eq. (4.28) for $\mu=1$ as

$$\rho^{(1)}_{eff,0}(\mathbf{k};s_0) = \frac{\sigma_1(s_0)}{(2\pi)^2}\sum_n \int_{-\infty}^\infty dq \int_0^{L_B} ds' \int_0^{2\pi} dt' \exp\left(i\mathbf{k}\cdot\mathbf{r}_1(s_0) + i\sigma_1(s_0)\mathbf{k}\cdot\hat{\mathbf{t}}_1(s_0)(s'-s_0)\right)$$
$$\exp\left(ia\cos t'\mathbf{k}\cdot\hat{\mathbf{d}}(s_0) + ia\sin t'\mathbf{k}\cdot\hat{\mathbf{n}}_1(s_0)\right)\exp(in(t_0-t'))\exp(iq\sigma(s_0)(s_0-s'))\zeta^{(1)}_{surf,0}(n,q),$$

(4.69)

where $\zeta^{(1)}_{surf,0}(n,k_z) = 1 + \zeta^{(1)}_{img\,0}(n,k_z)$. We can perform the $s'$ and $q$ integration yielding

$$\rho^{(1)}_{eff,0}(\mathbf{k};s_0) = \frac{1}{(2\pi)}\sum_n \int_0^{2\pi} d\phi \exp(-in\phi)\exp\left(i\mathbf{k}\cdot\mathbf{r}_1(s_0) + ia\cos(\phi+t_0)\mathbf{k}\cdot\hat{\mathbf{d}}(s_0) + ia\sin(\phi+t_0)\mathbf{k}\cdot\hat{\mathbf{n}}_1(s_0)\right)$$
$$\zeta^{(1)}_{surf,0}(n,\mathbf{k}\cdot\hat{\mathbf{t}}_1(s_0)),$$

(4.70)

where $\phi = t'-t_0$. Now, we concern ourselves with the next to leading order images. Therefore, we deal with Eq. (4.48). Using the definition

$$\sigma^{(2)}_{ind,1,0}(t',s') = \frac{\sigma_2(s_0)}{2\pi}\sum_l \int_{-\infty}^\infty \exp(-ilt'))\exp(ik_z\sigma_2(s_0)(s_0-s'))\zeta^{(2)}_{surf,1}(l,k_z),$$

(4.71)

we may write Eq. (4.48) (for $\mu=1$) as

$$\sum_l \int_{-\infty}^{\infty} dq \int_0^{L_B} ds' \int_0^{2\pi} dt' \int d^3k \sigma_1(s_0) \exp\left(-iRK\cos\phi_K + \left(ik_z \cos\eta(s_0) + iK\sin\phi_K \sin\eta(s_0)\right)\sigma_1(s_0)(s'-s_0)\right)$$

$$\exp\left(-ik_z\sigma_2(s_0)(\tilde{s}_2 - s_0)\right)$$

$$\exp\left(iaK(\cos t' - \cos t_2)\cos\phi_K + ia\sin t'\left(K\sin\phi_K \cos\eta(s_0) - k_z \sin\eta(s_0)\right) - iaK\sin t_2 \sin\phi_K\right)$$

$$K\cos(t_2 - \phi_K)\frac{\zeta_{surf,0}^{(1)}(l,q)}{(\mathbf{k}^2 + \kappa_D^2)}\exp(-il(t' - t_0))\exp(-iq\sigma_1(s_0)(s' - s_0))$$

$$+ \int_0^{L_B} ds' \int_0^{2\pi} dt' \int d^3k \sigma_2(s_0) \exp\left(ik_z\sigma_2(s_0)(s' - \tilde{s}_2) + K\left((a-\delta a)\cos(t' - \phi_K) - a\cos(t_2 - \phi_K)\right)\right)$$

$$\frac{K\cos(t_2 - \phi_K)\zeta_{surf,1}^{(2)}(l,q)\exp(-ilt')\exp(-iq\sigma_2(s_0)(s' - s_0))}{(\mathbf{k}^2 + \kappa_D^2)} = 0.$$

(4.72)

We can perform the $s'$ and $q$ integration in Eq. (4.72) yielding

$$\sum_l \int_0^{2\pi} dt' \int d^3k \exp\left(-iRK\cos\phi_K - ik_z\sigma_2(s_0)(\tilde{s}_2 - s_0)\right)$$

$$\exp\left(iaK(\cos t' - \cos t_2)\cos\phi_K + ia\sin t'\left(K\sin\phi_K \cos\eta(s_0) - k_z \sin\eta(s_0)\right) - iaK\sin t_2 \sin\phi_K\right)$$

$$K\cos(t_2 - \phi_K)\frac{\zeta_{surf,0}^{(1)}(l, k_z \cos\eta(s_0) + K\sin\phi_K \sin\eta(s_0))}{(\mathbf{k}^2 + \kappa_D^2)}\exp(-il(t' - t_0))$$

$$+ \sum_l \int_0^{2\pi} dt' \int d^3k \exp\left(ik_z\sigma_2(s_0)(s_0 - \tilde{s}_2) + K\left((a - \delta a)\cos(t' - \phi_K) - a\cos(t_2 - \phi_K)\right)\right)$$

$$\frac{K\cos(t_2 - \phi_K)\zeta_{surf,1}^{(2)}(l, k_z)\exp(-ilt')}{(\mathbf{k}^2 + \kappa_D^2)} = 0.$$

(4.73)

We can write the following approximation

$$\zeta_{surf,0}^{(1)}(l, k_z \cos\eta(s_0) + K\sin\phi_K \sin\eta(s_0)) \approx \zeta_{surf,0}^{(1)}(l, k_z \cos\eta(s_0)) + K\sin\phi_K \sin\eta(s_0)\zeta_{surf,0}'^{(1)}(l, k_z \cos\eta(s_0)).$$

(4.74)

This approximation may work quite well for all $\eta(s_0)$, not just for small $\sin\eta(s_0)$, provided that $\kappa_D a$ is sufficiently large. If this is so, $\zeta_{surf,0}^{(1)}(l, k_z)$ is a slowly varying function for most of $k_z$ and its derivatives are quite small. We can see this is indeed the case when we plot $\zeta_{surf,0}^{(1)}(l, k_z)$ for $l = 0, 1, 2, 3$ (see Fig 1) for $a\kappa_D = 2$.

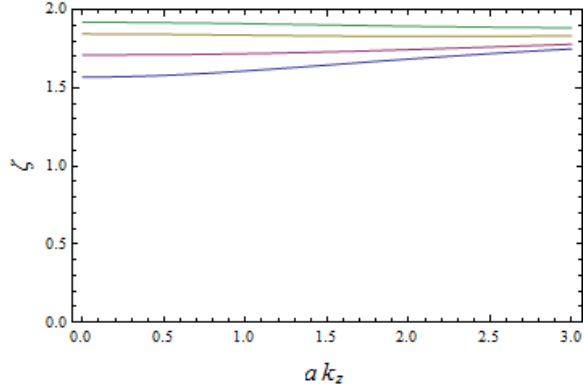

Fig 1. Showing plots of $\zeta^{(1)}_{surf,0}(l,k_z)$ for $a\kappa_D = 2$ as a function of $ak_z$ for $l=0$ (blue), $l=1$ (red), $l=2$ (yellow) and $l=3$ (green).

When we make the approximation (Eq. (4.74)) for the image charges we can further write

$$\zeta^{(2)}_{surf,1}(l,k_z) \approx \zeta^{(2)}_{surf,1,0}(l,k_z) + \sin\eta(s_0)\zeta^{(2)}_{surf,1,1}(l,k_z). \tag{4.75}$$

This allows us to express Eq. (4.73) as two separate conditions

$$\sum_l \int_0^{2\pi} dt' \int d^3k \exp(-iRK\cos\phi_K - ik_z\sigma_2(s_0)(\tilde{s}_2 - s_0))$$

$$\exp(iaK(\cos t' - \cos t_2)\cos\phi_K + ia\sin t'(K\sin\phi_K\cos\eta(s_0) - k_z\sin\eta(s_0)) - iaK\sin t_2\sin\phi_K)$$

$$K\cos(t_2 - \phi_K)\frac{\zeta^{(1)}_{surf,0}(l,k_z\cos\eta(s_0))}{(\mathbf{k}^2 + \kappa_D^2)}\exp(-il(t'-t_0))$$

$$+\sum_l \int_0^{2\pi} dt' \int d^3k \exp(ik_z\sigma_2(s_0)(s_0 - \tilde{s}_2) + K((a-\delta a)\cos(t'-\phi_K) - a\cos(t_2-\phi_K)))$$

$$\frac{K\cos(t_2-\phi_K)\zeta^{(2)}_{surf,1,0}(l,k_z)\exp(-ilt')}{(\mathbf{k}^2 + \kappa_D^2)} = 0, \tag{4.76}$$

$$\sum_l \int_0^{2\pi} dt' \int d^3k \exp(-iRK\cos\phi_K - ik_z\sigma_2(s_0)(\tilde{s}_2 - s_0))$$

$$\exp(iaK(\cos t' - \cos t_2)\cos\phi_K + ia\sin t'(K\sin\phi_K\cos\eta(s_0) - k_z\sin\eta(s_0)) - iaK\sin t_2\sin\phi_K)$$

$$K^2\sin\phi_K\cos(t_2-\phi_K)\frac{\zeta'^{(1)}_{surf,0}(l,k_z\cos\eta(s_0))}{(\mathbf{k}^2 + \kappa_D^2)}\exp(-il(t'-t_0))$$

$$+\sum_l \int_0^{2\pi} dt' \int d^3k \exp(ik_z\sigma_2(s_0)(s_0 - \tilde{s}_2) + K((a-\delta a)\cos(t'-\phi_K) - a\cos(t_2-\phi_K)))$$

$$\frac{K\cos(t_2-\phi_K)\zeta^{(2)}_{surf,1,1}(l,k_z)\exp(-ilt')}{(\mathbf{k}^2 + \kappa_D^2)} = 0. \tag{4.77}$$

Now, we can express

$$\exp\left(iaK\left(\cos t' - \cos t_2\right)\cos\phi_K + ia\sin t'\left(K\sin\phi_K\cos\eta(s_0) - k_z\sin\eta(s_0)\right) - iaK\sin t_2\sin\phi_K\right)$$
$$= iaK\tilde{R}(s_0,t')\cos\left(\tilde{\Phi}(s_0,t') - \phi_K\right) - iaK\cos(t_2 - \phi_K) - iak_z\sin t'\sin\eta(s_0),$$
(4.78)

where

$$\tilde{R}(s_0,t') = \sqrt{\cos^2 t' + \sin^2 t'\cos^2\eta(s_0)} \text{ and } \tilde{\Phi}(s_0,t') = \tan^{-1}\left(\cos\eta(s_0)\tan t'\right). \tag{4.79}$$

Eq. (4.78) allows us to write Eq. (4.76) as

$$\frac{1}{2}\sum_{l,m,n,n'}\int_0^{2\pi} dt' \int d^3k\, i^{-m+n-n'} J_m(RK) J_n(a\tilde{R}(s_0,t')K) J_{n'}(aK) \exp(-im\phi_K)$$
$$\exp\left(in\left(\tilde{\Phi}(s_0,t') - \phi_K\right)\right)\exp\left(-ik_z\sigma_2(s_0)(\tilde{s}_2 - s_0) - iak_z\sin t'\sin\eta(s_0)\right)$$
$$\left(\exp(-i(n'+1)(t_2 - \phi_K)) + \exp(-i(n'-1)(t_2 - \phi_K))\right)$$
$$\frac{K\zeta_{surf,0}^{(1)}(l,k_z\cos\eta(s_0))}{(\mathbf{k}^2 + \kappa_D^2)}\exp(-il(t' - t_0)) \tag{4.80}$$
$$+\frac{1}{2}\sum_{l,n,n'}\int_0^{2\pi} dt' \int d^3k\, i^{n-n'} J_n((a - \delta a)K) J_{n'}(aK) \exp\left(ik_z\sigma_2(s_0)(s_0 - \tilde{s}_2)\right)$$
$$\exp\left(in(t' - \phi_K)\right)\left(\exp(-i(n'+1)(t_2 - \phi_K)) + \exp(-i(n'-1)(t_2 - \phi_K))\right)$$
$$\frac{K\zeta_{surf,1,0}^{(2)}(l,k_z)\exp(-ilt')}{(\mathbf{k}^2 + \kappa_D^2)} = 0,$$

and Eq. (4.77) as

$$\frac{1}{4i}\sum_{l,m,n,n'}\int_0^{2\pi} dt' \int d^3k\, i^{-m+n-n'} J_m(RK) J_n(a\tilde{R}(s_0,t')K) J_{n'}(aK) \exp(-im\phi_K)$$
$$\exp\left(in\left(\tilde{\Phi}(s_0,t') - \phi_K\right)\right)\exp\left(-ik_z\sigma_2(s_0)(\tilde{s}_2 - s_0) - iak_z\sin t'\sin\eta(s_0)\right)$$
$$\left(\exp(-i(n'+1)(t_2 - \phi_K)) + \exp(-i(n'-1)(t_2 - \phi_K))\right)\left(\exp(i\phi_K) - \exp(-i\phi_K)\right)$$
$$\frac{K^2\zeta_{surf,0}'^{(1)}(l,k_z\cos\eta(s_0))}{(\mathbf{k}^2 + \kappa_D^2)}\exp(-il(t' - t_0)) \tag{4.81}$$
$$+\frac{1}{2}\sum_{l,n,n'}\int_0^{2\pi} dt' \int d^3k\, i^{n-n'} J_n((a - \delta a)K) J_{n'}(aK) \exp\left(ik_z\sigma_2(s_0)(s_0 - \tilde{s}_2)\right)$$
$$\exp\left(in(t' - \phi_K)\right)\left(\exp(-i(n'+1)(t_2 - \phi_K)) + \exp(-i(n'-1)(t_2 - \phi_K))\right)$$
$$\frac{K\zeta_{surf,1,1}^{(2)}(l,k_z)\exp(-ilt')}{(\mathbf{k}^2 + \kappa_D^2)} = 0.$$

In both Eqs. (4.80) and (4.81) we can perform the $\phi_K$ and $K$-integrations. This gives us

$$-\frac{i}{2}\sum_{l,n,n'}\int_0^{2\pi}dt'\int_{-\infty}^{\infty}dk_z(-1)^{n'}\zeta_{surf,0}^{(1)}(l,k_z\cos\eta(s_0))\exp(-il(t'-t_0))\sqrt{k_z^2+\kappa_D^2}$$

$$\left(K_{n'+1-n}\left(R\sqrt{k_z^2+\kappa_D^2}\right)\exp(-i(n'+1)t_2)+K_{n'-1-n}\left(R\sqrt{k_z^2+\kappa_D^2}\right)\exp(-i(n'-1)t_2)\right)$$

$$I_n\left(a\tilde{R}(s_0,t')\sqrt{k_z^2+\kappa_D^2}\right)I_{n'}\left(a\sqrt{k_z^2+\kappa_D^2}\right)\exp(in\tilde{\Phi}(s_0,t'))\exp(-ik_z\sigma_2(s_0)(s_2-s_0)-iak_z\sin t'\sin\eta(s_0))$$

$$+i\sum_{l,n}\int_0^{2\pi}dt'\int_{-\infty}^{\infty}dk_z\exp(ik_z\sigma_2(s_0)(s_0-s_2))I_n\left(a\sqrt{k_z^2+\kappa_D^2}\right)K_n'\left(a\sqrt{k_z^2+\kappa_D^2}\right)\sqrt{k_z^2+\kappa_D^2}$$

$$\exp(in(t'-t_2))\zeta_{surf,1,0}^{(2)}(l,k_z)\exp(-ilt')=0,$$

(4.82)

$$\frac{i}{2}\sum_{l,n,n'}\int_0^{2\pi}dt'\int_{-\infty}^{\infty}dk_z(-1)^{n'}I_n\left(a\tilde{R}(s_0,t')\sqrt{k_z^2+\kappa_D^2}\right)I_{n'}\left(a\sqrt{k_z^2+\kappa_D^2}\right)\frac{\sqrt{k_z^2+\kappa_D^2}\zeta_{surf,0}'^{(1)}(l,k_z\cos\eta(s_0))}{R}$$

$$\exp(-il(t'-t_0))\exp(in\tilde{\Phi}(s_0,t'))\exp(-ik_z\sigma_2(s_0)(s_2-s_0)-iak_z\sin t'\sin\eta(s_0))$$

$$\left(\exp(-i(n'+1)t_2)(n'+1-n)K_{n'+1-n}\left(R\sqrt{k_z^2+\kappa_D^2}\right)+\exp(-i(n'-1)t_2)(n'-1-n)K_{n'-1-n}\left(R\sqrt{k_z^2+\kappa_D^2}\right)\right)$$

$$+i\sum_{l,n,}\int_0^{2\pi}dt'\int_{-\infty}^{\infty}dk_z\sqrt{k_z^2+\kappa_D^2}\exp(ik_z\sigma_2(s_0)(s_0-s_2))I_n\left(a\sqrt{k_z^2+\kappa_D^2}\right)K_n'\left(a\sqrt{k_z^2+\kappa_D^2}\right)$$

$$\exp(in(t'-t_2))\zeta_{surf,1,1}^{(2)}(l,k_z)\exp(-ilt')=0.$$

(4.83)

We can shift the sums in the first terms of Eqs. (4.82) and (4.83) so that we rewrite Eq.(4.82) as

$$\int_0^{2\pi}dt'\int_{-\infty}^{\infty}dk_z\left(\sum_{l,n,n'}(-1)^{n'}\exp(-in't_2)\sqrt{k_z^2+\kappa_D^2}K_{n'-n}\left(R\sqrt{k_z^2+\kappa_D^2}\right)I_n\left(a\tilde{R}(s_0,t')\sqrt{k_z^2+\kappa_D^2}\right)I_{n'}\left(a\sqrt{k_z^2+\kappa_D^2}\right)\right.$$

$$\exp(in\tilde{\Phi}(s_0,t'))\exp(-ik_z\sigma_2(s_0)(s_2-s_0)-iak_z\sin t'\sin\eta(s_0))\zeta_{surf,0}^{(1)}(l,k_z\cos\eta(s_0))\exp(-il(t'-t_0))$$

$$+\sum_{l,n}\exp(ik_z\sigma_2(s_0)(s_0-s_2))I_n\left(a\sqrt{k_z^2+\kappa_D^2}\right)K_n'\left(a\sqrt{k_z^2+\kappa_D^2}\right)\sqrt{k_z^2+\kappa_D^2}$$

$$\left.\exp(in(t'-t_2))\zeta_{surf,1,0}^{(2)}(l,k_z)\exp(-ilt')\right)=0,$$

(4.84)

and Eq. (4.83) as

$$\int_0^{2\pi}dt'\int_0^{2\pi}dk_z\left(-\sum_{l,n,n'}I_n\left(a\tilde{R}(s_0,t')\sqrt{k_z^2+\kappa_D^2}\right)I_{n'}'\left(a\sqrt{k_z^2+\kappa_D^2}\right)K_{n'-n}\left(R\sqrt{k_z^2+\kappa_D^2}\right)(-1)^{n'}\sqrt{k_z^2+\kappa_D^2}\right.$$

$$\exp(in\tilde{\Phi}(s_0,t'))\exp(-ik_z\sigma_2(s_0)(s_2-s_0)-iak_z\sin t'\sin\eta(s_0))\exp(-in't_2)\exp(-il(t'-t_0))(n'-n)$$

$$\frac{\zeta_{surf,0}'^{(1)}(l,k_z\cos\eta(s_0))}{R}+\sum_{l,n}\exp(ik_z\sigma_2(s_0)(s_0-s_2))I_n\left(a\sqrt{k_z^2+\kappa_D^2}\right)K_n'\left(a\sqrt{k_z^2+\kappa_D^2}\right)\sqrt{k_z^2+\kappa_D^2}$$

$$\left.\exp(in(t'-t_2))\zeta_{surf,1,1}^{(2)}(l,k_z)\exp(-ilt')\right)=0.$$

(4.85)

We substitute the identities

$$I_n\left(a\sqrt{\sin^2 t' \cos^2 \eta(s_0) + \cos^2 t'}\sqrt{k_z^2 + \kappa_D^2}\right)\exp(in\tilde{\Phi}(s_0, t'))$$
$$\equiv \sum_m I_{n-m}\left(\frac{a\sqrt{k_z^2 + \kappa_D^2}}{2}(1-\cos\eta(s_0))\right) I_m\left(\frac{a\sqrt{k_z^2 + \kappa_D^2}}{2}(1+\cos\eta_1(s_0))\right)\exp(2imt')\exp(-int'),$$

(4.86)

and

$$\exp(-iak_z \sin t' \sin\eta(s_0)) \equiv \sum_{m'} J_{m'}(ak_z \sin\eta(s_0))\exp(-im't')$$
(4.87)

into Eq. (4.84) and (4.85), and perform the $t'$ integrations. The resulting equations are

$$\sum_{l,n,n',m} \int_{-\infty}^{\infty} dk_z (-1)^{n'} \exp(i(lt_0 - n't_2)) I_{n-m}\left(\frac{a\sqrt{k_z^2 + \kappa_D^2}}{2}(1-\cos\eta(s_0))\right) I_m\left(\frac{a\sqrt{k_z^2 + \kappa_D^2}}{2}(1+\cos\eta_1(s_0))\right)\sqrt{k_z^2 + \kappa_D^2}$$

$$K_{n'-n}\left(R\sqrt{k_z^2 + \kappa_D^2}\right) I'_{n'}\left(a\sqrt{k_z^2 + \kappa_D^2}\right) J_{2m-n-l}(ak_z \sin\eta(s_0))\exp(-ik_z\sigma_2(s_0)(\tilde{s}_2 - s_0))\zeta^{(1)}_{surf,0}(l, k_z \cos\eta(s_0))$$

$$+\sum_n \int_{-\infty}^{\infty} dk_z \exp(ik_z\sigma_2(s_0)(s_0 - \tilde{s}_2))\sqrt{k_z^2 + \kappa_D^2} I_n\left(a\sqrt{k_z^2 + \kappa_D^2}\right) K'_n\left(a\sqrt{k_z^2 + \kappa_D^2}\right)\exp(-int_2)\tilde{\zeta}^{(2)}_{surf,1,0}(n, k_z) = 0,$$

(4.88)

$$-\sum_{l,n,n',m} \int_{-\infty}^{\infty} dk_z (-1)^{n'} \exp(i(lt_0 - n't_2)) I_{n-m}\left(\frac{a\sqrt{k_z^2 + \kappa_D^2}}{2}(1-\cos\eta(s_0))\right) I_m\left(\frac{a\sqrt{k_z^2 + \kappa_D^2}}{2}(1+\cos\eta_1(s_0))\right)$$

$$I'_{n'}\left(a\sqrt{k_z^2 + \kappa_D^2}\right) K_{n'-n}\left(R\sqrt{k_z^2 + \kappa_D^2}\right) J_{2m-n-l}(ak_z \sin\eta(s_0))\exp(-ik_z\sigma_2(s_0)(\tilde{s}_2 - s_0))(n'-n)\sqrt{k_z^2 + \kappa_D^2}$$

$$\frac{\zeta'^{(1)}_{surf,0}(l, k_z \cos\eta(s_0))}{R} + \sum_n \int_{-\infty}^{\infty} dk_z \exp(-ik_z\sigma_2(s_0)(\tilde{s}_2 - s_0))\sqrt{k_z^2 + \kappa_D^2} I_n\left(a\sqrt{k_z^2 + \kappa_D^2}\right) K'_n\left(a\sqrt{k_z^2 + \kappa_D^2}\right)$$

$$\exp(-int_2))\zeta^{(2)}_{surf,1,1}(n, k_z) = 0.$$

(4.89)

We easily see that the solutions to Eqs. (4.88) and (4.89) are

$$\zeta^{(2)}_{surf,1,0}(n, k_z) = -\frac{I'_n\left(a\sqrt{k_z^2 + \kappa_D^2}\right)}{K'_n\left(a\sqrt{k_z^2 + \kappa_D^2}\right) I_n\left(a\sqrt{k_z^2 + \kappa_D^2}\right)} \sum_{l,n',m'} (-1)^n I_{n'-m'}\left(\frac{a\sqrt{k_z^2 + \kappa_D^2}}{2}(1-\cos\eta(s_0))\right)$$

$$I_{m'}\left(\frac{a\sqrt{k_z^2 + \kappa_D^2}}{2}(1+\cos\eta_1(s_0))\right) K_{n-n'}\left(R\sqrt{k_z^2 + \kappa_D^2}\right) J_{2m'-n'-l}(ak_z \sin\eta(s_0))$$

$$\zeta^{(1)}_{surf,0}(l, k_z \cos\eta(s_0))\exp(ilt_0) \equiv \sum_l \tilde{\zeta}^{(2)}_{surf,1,0}(n, l, k_z)\exp(ilt_0),$$

(4.90)

$$\zeta^{(2)}_{surf,1,1}(n,k_z) = \frac{I'_n\left(a\sqrt{k_z^2+\kappa_D^2}\right)}{I_n\left(a\sqrt{k_z^2+\kappa_D^2}\right)K'_n\left(a\sqrt{k_z^2+\kappa_D^2}\right)} \sum_{l,n',m'} (-1)^n I_{n'-m'}\left(\frac{a\sqrt{k_z^2+\kappa_D^2}}{2}(1-\cos\eta(s_0))\right)$$

$$I_{m'}\left(\frac{a\sqrt{k_z^2+\kappa_D^2}}{2}(1+\cos\eta_1(s_0))\right) K_{n-n'}\left(R\sqrt{k_z^2+\kappa_D^2}\right) J_{2m'-n'-l}(ak_z \sin\eta(s_0))$$

$$\frac{\zeta'^{(1)}_{surf,0}(l,k_z \cos\eta(s_0))(n-n')}{R} \exp(ilt_0) \equiv \sum_l \tilde{\zeta}^{(2)}_{surf,1,1}(n,l,k_z) \exp(ilt_0).$$

(4.91)

Using Eqs. (4.71) and (4.75) we can now write Eq. (4.58) as

$$\rho^{(2,1)}_{img,0}(\mathbf{k};s_0) = \frac{1}{2\pi} \sum_n \int_0^{L_B} ds' \int_0^{2\pi} d\phi \exp\left(i\mathbf{k}.\mathbf{r}_2(s_0) + ia\mathbf{k}.\hat{\mathbf{d}}(s_0)\cos\phi + ia\mathbf{k}.\hat{\mathbf{n}}_2(s_0)\sin\phi\right)$$

$$\exp(in\phi)(\zeta^{(2)}_{surf,1,0}(n,\mathbf{k}.\hat{\mathbf{t}}_2(s_0)) + \sin\eta(s_0)\zeta^{(2)}_{surf,1,1}(n,\mathbf{k}.\hat{\mathbf{t}}_2(s_0))).$$

(4.92)

### *4.7 Evaluation of the leading order terms in the curvature expansion for a charge on rod 2*

For the leading order term in the image charge expansion we can write

$$\sigma^{(2)}_{ind,0,0}(t',s') = \frac{\sigma_2(s_0)}{2\pi} \sum_n \int_{-\infty}^{\infty} \exp(in(t_0-t'))\exp(ik_z\sigma_2(s_0)(s_0-s'))\zeta^{(1)}_{img0}(n,k_z).$$

(4.93)

Then, when we have a charge at $s_0$ on rod 2 we find from Eq. (4.45), following similar steps to Eqs. (4.63)-(4.66) we find that

$$\zeta^{(2)}_{img0}(n,k_z) = -\frac{K_n\left(a\sqrt{k_z^2+\kappa_D^2}\right)I'_n\left(a\sqrt{k_z^2+\kappa_D^2}\right)}{I_n\left(a\sqrt{k_z^2+\kappa_D^2}\right)K'_n\left(a\sqrt{k_z^2+\kappa_D^2}\right)}.$$

(4.94)

We can then express the leading order contribution in curvature to the effective charge density $\rho^{(1)}_{eff}(\mathbf{k};s_0)$ (see Eq.(4.55))

$$\rho^{(2)}_{eff,0}(\mathbf{k};s_0) = \frac{1}{(2\pi)} \sum_n \int_0^{2\pi} d\phi \exp(-in\phi) \exp\left(i\mathbf{k}.\mathbf{r}_2(s_0) + ia\cos(\phi+t_0)\mathbf{k}.\hat{\mathbf{d}}(s_0) + ia\sin(\phi+t_0)\mathbf{k}.\hat{\mathbf{n}}_2(s_0)\right)$$

$$\zeta^{(2)}_{surf,0}(n,\mathbf{k}.\hat{\mathbf{t}}_2),$$

(4.95)

where $\zeta^{(2)}_{surf,0}(l,k_z) = 1 + \zeta^{(2)}_{img,0}(l,k_z)$.

Using the definition

$$\sigma^{(1)}_{ind,1,0}(t',s') = \frac{\sigma_1(s_0)}{2\pi} \sum_l \int_{-\infty}^{\infty} \exp(-ilt'))\exp(ik_z\sigma_1(s_0)(s_0-s'))\zeta^{(1)}_{surf,1}(l,k_z),$$

(4.96)

we can express Eq. (4.48) for the next to leading order images as

$$\sum_l \int_0^{2\pi} dt' \int d^3k \exp\left(iRK\cos\phi_K - ik_z\sigma_1(s_0)(\tilde{s}_1 - s_0)\right)$$

$$\exp\left(iaK(\cos t' - \cos t_1)\cos\phi_K + ia\sin t'(K\sin\phi_K\cos\eta(s_0) + k_z\sin\eta(s_0)) - iaK\sin t_1\sin\phi_K\right)$$

$$K\cos(t_1 - \phi_K)\frac{\zeta^{(2)}_{surf,0}(l,k_z\cos\eta(s_0) - K\sin\phi_K\sin\eta(s_0))}{(\mathbf{k}^2 + \kappa_D^2)}\exp(-il(t' - t_0))$$

$$+\sum_l \int_0^{2\pi} dt' \int d^3k \exp\left(ik_z\sigma_1(s_0)(s_0 - \tilde{s}_1) + K((a - \delta a)\cos(t' - \phi_K) - a\cos(t_1 - \phi_K))\right)$$

$$\frac{K\cos(t_1 - \phi_K)\zeta^{(1)}_{surf,1}(l,k_z)\exp(-ilt')}{(\mathbf{k}^2 + \kappa_D^2)} = 0.$$

(4.97)

We do the same approximation as Eq. (4.74) and write

$$\zeta^{(2)}_{surf,0}(l,k_z\cos\eta(s_0) - K\sin\phi_K\sin\eta(s_0)) \approx \zeta^{(2)}_{surf,0}(l,k_z\cos\eta(s_0)) - K\sin\phi_K\sin\eta(s_0)\zeta'^{(2)}_{surf,0}(l,k_z\cos\eta(s_0)),$$

(4.98)

$$\zeta^{(1)}_{surf,1}(l,k_z) \approx \zeta^{(1)}_{surf,1,0}(l,k_z) + \sin\eta(s_0)\zeta'^{(1)}_{surf,1,1}(l,k_z).$$

(4.99)

This allows us to express Eq. (4.97) as two separate conditions

$$\sum_l \int_0^{2\pi} dt' \int d^3k \exp\left(iRK\cos\phi_K - ik_z\sigma_1(s_0)(\tilde{s}_1 - s_0) - il(t' - t_0)\right) K\cos(t_1 - \phi_K)\frac{\zeta^{(2)}_{surf,0}(l,k_z\cos\eta(s_0))}{(\mathbf{k}^2 + \kappa_D^2)}$$

$$\exp\left(iaK(\cos t' - \cos t_1)\cos\phi_K + ia\sin t'(K\sin\phi_K\cos\eta(s_0) + k_z\sin\eta(s_0)) - iaK\sin t_1\sin\phi_K\right)$$

$$+\sum_l \int_0^{2\pi} dt' \int d^3k \exp\left(ik_z\sigma_1(s_0)(s_0 - \tilde{s}_1) + K((a - \delta a)\cos(t' - \phi_K) - a\cos(t_1 - \phi_K))\right)$$

$$\frac{K\cos(t_1 - \phi_K)\zeta^{(1)}_{surf,1,0}(l,k_z)\exp(-ilt')}{(\mathbf{k}^2 + \kappa_D^2)} = 0,$$

(4.100)

$$-\sum_l \int_0^{2\pi} dt' \int d^3k \exp\left(iRK\cos\phi_K - ik_z\sigma_1(s_0)(\tilde{s}_1 - s_0) - il(t' - t_0)\right) K^2\sin\phi_K\cos(t_1 - \phi_K)\frac{\zeta'^{(2)}_{surf,0}(l,k_z\cos\eta(s_0))}{(\mathbf{k}^2 + \kappa_D^2)}$$

$$\exp\left(iaK(\cos t' - \cos t_1)\cos\phi_K + ia\sin t'(K\sin\phi_K\cos\eta(s_0) + k_z\sin\eta(s_0)) - iaK\sin t_1\sin\phi_K\right)$$

$$+\sum_l \int_0^{2\pi} dt' \int d^3k \exp\left(ik_z\sigma_1(s_0)(s_0 - \tilde{s}_1) + K((a - \delta a)\cos(t' - \phi_K) - a\cos(t_1 - \phi_K))\right)$$

$$\frac{K\cos(t_1 - \phi_K)\zeta^{(1)}_{surf,1,1}(l,k_z)\exp(-ilt')}{(\mathbf{k}^2 + \kappa_D^2)} = 0.$$

(4.101)

By performing steps similar to Eqs. (4.78)-(4.89) we find for $\zeta^{(1)}_{surf,1,0}(n,k_z;t_0)$ and $\zeta^{(1)}_{surf,1,1}(n,k_z;t_0)$

$$\zeta^{(1)}_{surf,1,0}(n,k_z) = -\frac{I'_n\left(a\sqrt{k_z^2+\kappa_D^2}\right)}{I_n\left(a\sqrt{k_z^2+\kappa_D^2}\right)K'_n\left(a\sqrt{k_z^2+\kappa_D^2}\right)}\sum_{l,n',m'}(-1)^{n'}K_{n'-n}\left(R\sqrt{k_z^2+\kappa_D^2}\right)J_{n'+l-2m'}\left(ak_z\sin\eta(s_0)\right)$$

$$I_{n'-m'}\left(\frac{a\sqrt{k_z^2+\kappa_D^2}}{2}(1-\cos\eta(s_0))\right)I_{m'}\left(\frac{a\sqrt{k_z^2+\kappa_D^2}}{2}(1+\cos\eta_1(s_0))\right)\zeta^{(2)}_{surf,0}(l,k_z\cos\eta(s_0))\exp(ilt_0)$$

$$\equiv \sum_l \tilde{\zeta}^{(1)}_{surf,1,0}(n,l,k_z)\exp(ilt_0),$$

(4.102)

$$\zeta^{(1)}_{surf,1,1}(n,k_z) = \frac{I'_n\left(a\sqrt{k_z^2+\kappa_D^2}\right)}{I_n\left(a\sqrt{k_z^2+\kappa_D^2}\right)K'_n\left(a\sqrt{k_z^2+\kappa_D^2}\right)}\sum_{l,m',n'}(-1)^{n'}(n-n')I_{n'-m'}\left(\frac{a\sqrt{k_z^2+\kappa_D^2}}{2}(1-\cos\eta(s_0))\right)$$

$$I_{m'}\left(\frac{a\sqrt{k_z^2+\kappa_D^2}}{2}(1+\cos\eta(s_0))\right)J_{n'+l-2m'}(ak_z\sin\eta(s_0))K_{n'-n}\left(R\sqrt{k_z^2+\kappa_D^2}\right)\frac{\zeta'^{(2)}_{surf,0}(l,k_z\cos\eta(s_0))}{R}\exp(ilt_0)$$

$$\equiv \sum_l \tilde{\zeta}^{(1)}_{surf,1,1}(n,l,k_z)\exp(ilt_0).$$

(4.103)

Also, we have for $\rho^{(1,2)}_{img,0}(\mathbf{k};s_0)$

$$\rho^{(1,2)}_{img,0}(\mathbf{k};s_0) = \frac{1}{2\pi}\sum_n \int_0^{L_B} ds' \int_0^{2\pi} d\phi \exp\left(i\mathbf{k}.\mathbf{r}_1(s_0)+ia\mathbf{k}.\hat{\mathbf{d}}(s_0)\cos\phi+ia\mathbf{k}.\hat{\mathbf{n}}_1(s_0)\sin\phi\right)$$

$$\exp(in\phi)(\zeta^{(1)}_{surf,1,0}(n,\mathbf{k}.\hat{\mathbf{t}}_1(s_0))+\sin\eta(s_0)\zeta^{(1)}_{surf,1,1}(n,\mathbf{k}.\hat{\mathbf{t}}_1(s_0))).$$

(4.104)

Eqs. (4.90), (4.91), (4.92) (4.102), (4.103) and (4.104) form another of the general results of this paper.

## 5. Direct Electrostatic interaction for rods with low dielectric cores

We now deal with helical lines of charge which we parameterize in terms of the arc length of the braid centre line, with $s$ and $s'$ being the positions of charges on rods 1 and 2 respectively. For the helical trajectory of charges, in the image charge results of the previous section we set $t_0 = \xi_1(s)$ and $s_0 = s$ for charges on rod 1 and set $t_0 = \xi_2(s')$ and $s_0 = s'$ for charges on rod 2. Substitution of Eqs. (4.70) and (4.95) into Eq. (3.2), as well as the transform to the variables $\tau = \frac{s+s'}{2}$ and $\tau' = \frac{s-s'}{2}$, yields

$$E_{dir} = 2\int_0^{L_B} d\tau \int_{-\infty}^{\infty} d\tau' \sigma_1(\tau+\tau')\sigma_2(\tau-\tau')\tilde{E}_{pt}(\tau,\tau'),$$

(5.1)

$$\tilde{E}_{pt}(\tau,\tau') = \frac{4\pi e^2}{\varepsilon_w l_c^2} \frac{1}{(2\pi)^5} \sum_{l,l'} \int_0^{2\pi} d\phi \int_0^{2\pi} d\phi' \int d^3k \exp(-il\phi)\exp(-il'\phi') \frac{\exp(i\mathbf{k}.(\mathbf{r}_1(\tau+\tau')-\mathbf{r}_2(\tau-\tau')))}{\mathbf{k}^2+\kappa_D^2}$$
$$\exp(ia\mathbf{k}.\left(\hat{\mathbf{N}}_1^0(\tau+\tau',\phi+\zeta_1(\tau+\tau'))-\hat{\mathbf{N}}_2^0(\tau-\tau',\phi'+\zeta_2(\tau-\tau'))\right))\zeta_{surf,0}^{(1)}(l,\mathbf{k}.\hat{\mathbf{t}}_1(\tau+\tau'))\zeta_{surf,0}^{(2)}(l',-\mathbf{k}.\hat{\mathbf{t}}_2(\tau-\tau')),$$
(5.2)

where

$$\hat{\mathbf{N}}_\mu^0(s,\phi) = \cos\phi\hat{\mathbf{d}}(s)+\cos\phi\hat{\mathbf{n}}_\mu(s) = \cos\phi\hat{\mathbf{U}}_A^{(\mu)}(s)+\sin\phi\cos\eta_\mu(s)\hat{\mathbf{V}}_A^{(\mu)}(s)-\delta_\mu\sin\phi\sin\eta_\mu(s)\hat{\mathbf{t}}_A(s).$$
(5.3)

Next we Taylor expand through Eqs. (2.4), as well as in $\zeta_{surf,0}^{(1)}(l,\mathbf{k}.\hat{\mathbf{t}}_1(\tau+\tau'))$ and $\zeta_{surf,0}^{(2)}(l',-\mathbf{k}.\hat{\mathbf{t}}_2(\tau-\tau'))$

$$\hat{\mathbf{t}}_1(\tau+\tau') \approx \hat{\mathbf{t}}_1(\tau), \ \hat{\mathbf{t}}_2(\tau-\tau') \approx \hat{\mathbf{t}}_2(\tau). \tag{5.4}$$

We make approximations of $\hat{\mathbf{N}}_1^0(\tau+\tau',\phi)$ and $\hat{\mathbf{N}}_2^0(\tau-\tau',\phi')$ (in line with those of (2.7) and (2.8)), where we assume that the curvature of the braid centre line is small. As before, we choose a frame where $\hat{\mathbf{t}}_A(\tau).\mathbf{k}=k_z$ $\hat{\mathbf{d}}(\tau).\mathbf{k}=k_x=K\cos\phi_K$ so that

$$\mathbf{k}.\hat{\mathbf{t}}_\mu(\tau) = k_z \cos\eta_\mu(\tau) + K\delta_\mu \sin\phi_K \sin\eta_\mu(\tau). \tag{5.5}$$

All of the above considerations allow us to write Eq. (5.2) as

$$\tilde{E}_{pt}(\tau,\tau') = \frac{4\pi e^2}{\varepsilon_w l_c^2} \frac{1}{(2\pi)^5} \int_0^{2\pi} d\phi_K \int_{-\infty}^{\infty} dk_z \int_0^{\infty} KdK \int_0^{2\pi} d\phi \int_0^{2\pi} d\phi' \sum_{l,l'} \frac{\exp(-iKR\cos\phi_K)\exp(2i\tau'k_z)}{\mathbf{k}^2+\kappa_D^2}$$
$$\zeta_{surf}^{(1)}(l,(k_z\cos\eta_1(\tau)+K\sin\phi_K\sin\eta_1(\tau))\exp(-il\phi)\exp(ia\mathbf{k}.\hat{\mathbf{N}}_1'^0(s,\phi+\xi_1(\tau+\tau')))$$
$$\zeta_{surf}^{(2)}(l',-(k_z\cos\eta_2(\tau)-K\sin\phi_K\sin\eta_2(\tau))\exp(-il'\phi')\exp(-ia\mathbf{k}.\hat{\mathbf{N}}_2'^0(s',\phi'+\xi_2(\tau-\tau'))),$$
(5.6)

where

$$\hat{\mathbf{N}}_1'^0(\tau+\tau',\phi) = \tilde{\mathbf{T}}(\gamma_A(\tau'+\tau;\tau))\begin{pmatrix}\cos\phi\\ \sin\phi\cos\eta_1(\tau+\tau')\\ -\sin\phi\sin\eta_1(\tau+\tau')\end{pmatrix}, \tag{5.7}$$

$$\hat{\mathbf{N}}_2'^0(\tau-\tau',\phi) = \tilde{\mathbf{T}}(\gamma_A(\tau'+\tau;\tau))\begin{pmatrix}\cos\phi\\ \sin\phi\cos\eta_2(\tau-\tau')\\ \sin\phi\sin\eta_2(\tau-\tau')\end{pmatrix}. \tag{5.8}$$

The matrix $\tilde{\mathbf{T}}$ is defined by Eq. (2.12). Now in Eq. (5.6), we make similar approximations to Eq. (4.74), namely

$$\zeta_{surf}^{(1)}(l, k_z \cos\eta_1(\tau) + K\sin\phi_K \sin\eta_1(\tau))$$
$$\approx \zeta_{surf}^{(1)}(l, k_z \cos\eta_1(\tau)) + K\sin\phi_K \sin\eta_1(\tau) \zeta_{surf}'^{(1)}(l, k_z \cos\eta_1(\tau)), \qquad (5.9)$$

$$\zeta_{surf}^{(2)}(l', -(k_z \cos\eta_2(\tau) - K\sin\phi_K \sin\eta_2(\tau)))$$
$$\approx \zeta_{surf}^{(2)}(l', k_z \cos\eta_2(\tau)) - K\sin\phi_K \sin\eta_2(\tau) \zeta_{surf}'^{(2)}(l', k_z \cos\eta_2(\tau)), \qquad (5.10)$$

where we have used the property $\zeta_{surf}^{(2)}(n, -k_z) = \zeta_{surf}^{(2)}(n, k_z)$. In addition we can express

$$\mathbf{k}.\hat{\mathbf{N}}_\mu'^0(\tau+\tau', \phi + \xi_1(\tau+\tau')) = K\tilde{R}_{H\mu}(\tau+\tau', \phi)\cos\left(\tilde{\xi}_\mu(\tau+\tau', \phi) + \gamma_A(\tau+\tau'; \tau) - \phi_K\right) + k_z \tilde{Z}_{H\mu}(\tau+\tau', \phi), \qquad (5.11)$$

where

$$\tilde{R}_{H\mu}(\tau+\tau', \phi) = \sqrt{\cos\left(\xi_\mu(\tau+\tau') + \phi\right)^2 + \cos^2\eta_\mu(\tau+\tau')\sin\left(\xi_\mu(\tau+\tau') + \phi\right)^2}, \qquad (5.12)$$

$$\tilde{\xi}_\mu(\tau+\tau', \phi) = \tan^{-1}\left(\cos\eta_\mu(\tau+\tau')\tan\left(\xi_\mu(\tau+\tau') + \phi\right)\right), \qquad (5.13)$$

$$\tilde{Z}_{H\mu}(\tau+\tau', \phi) = -\delta_\mu \sin\eta_\mu(\tau+\tau')\sin\left(\xi_\mu(\tau+\tau') + \phi\right). \qquad (5.14)$$

Eqs. (5.9)-(5.14) allows us to express Eq. (5.6) as the sum of three terms

$$\tilde{E}_{pt}(\tau, \tau') \approx \tilde{E}_{pt0}(\tau, \tau') + \tilde{E}_{pt1}(\tau, \tau') + \tilde{E}_{pt2}(\tau, \tau'), \qquad (5.15)$$

$$\tilde{E}_{pt0}(\tau, \tau') = \frac{4\pi e^2}{\varepsilon_w l_c^2} \frac{1}{(2\pi)^5} \int_0^{2\pi} d\phi_K \int_{-\infty}^{\infty} dk_z \int_0^{\infty} KdK \int_0^{2\pi} d\phi \int_0^{2\pi} d\phi' \sum_{l,l'} \frac{\exp(-iKR\cos\phi_K)}{\mathbf{k}^2 + \kappa_D^2}$$
$$\exp(2i\tau' k_z) \zeta_{surf}^{(1)}(l, k_z \cos\eta_1(\tau)) \zeta_{surf}^{(2)}(l', k_z \cos\eta_2(\tau))$$
$$\exp\left(iaK\tilde{R}_{H1}(\tau+\tau', \phi)\cos\left(\tilde{\xi}_1(\tau+\tau', \phi) + \gamma_A(\tau+\tau'; \tau) - \phi_K\right)\right) \qquad (5.16)$$
$$\exp\left(-iaK\tilde{R}_{H2}(\tau-\tau', \phi')\cos\left(\tilde{\xi}_2(\tau-\tau', \phi') + \gamma_A(\tau-\tau'; \tau) - \phi_K\right)\right)$$
$$\exp\left(iak_z \tilde{Z}_{H1}(\tau+\tau', \phi)\right) \exp\left(-iak_z \tilde{Z}_{H2}(\tau-\tau', \phi')\right) \exp(-il\phi) \exp(-il'\phi'),$$

$$\tilde{E}_{pt1}(\tau,\tau') = \frac{4\pi e^2}{\varepsilon_w l_c^2} \frac{1}{(2\pi)^5} \int_0^{2\pi} d\phi_K \int_{-\infty}^{\infty} dk_z \int_0^{\infty} KdK \int_0^{2\pi} d\phi \int_0^{2\pi} d\phi' \sum_{l,l'} \frac{\exp(-iKR\cos\phi_K)}{\mathbf{k}^2 + \kappa_D^2}$$

$$\exp(2i\tau' k_z) \zeta_{surf}^{\prime(1)}(l, k_z \cos\eta_1(\tau)) \zeta_{surf}^{(2)}(l', k_z \cos\eta_2(\tau)) K \sin\phi_K \sin\eta_1(\tau)$$

$$\exp\left(iaK\tilde{R}_{H1}(\tau+\tau',\phi)\cos\left(\tilde{\xi}_1(\tau+\tau',\phi) + \gamma_A(\tau+\tau';\tau) - \phi_K\right)\right) \quad (5.17)$$

$$\exp\left(-iaK\tilde{R}_{H2}(\tau-\tau',\phi')\cos\left(\tilde{\xi}_2(\tau-\tau',\phi') + \gamma_A(\tau-\tau';\tau) - \phi_K\right)\right)$$

$$\exp\left(iak_z\tilde{Z}_{H1}(\tau+\tau',\phi)\right)\exp\left(-iak_z\tilde{Z}_{H2}(\tau-\tau',\phi')\right)\exp(-il\phi)\exp(-il'\phi'),$$

$$\tilde{E}_{pt2}(\tau,\tau') = -\frac{4\pi e^2}{\varepsilon_w l_c^2} \frac{1}{(2\pi)^5} \int_0^{2\pi} d\phi_K \int_{-\infty}^{\infty} dk_z \int_0^{\infty} KdK \int_0^{2\pi} d\phi \int_0^{2\pi} d\phi' \sum_{l,l'} \frac{\exp(-iKR\cos\phi_K)}{\mathbf{k}^2 + \kappa_D^2}$$

$$\exp(2i\tau' k_z) \zeta_{surf}^{(1)}(l, k_z \cos\eta_1(\tau)) \zeta_{surf}^{\prime(2)}(l', k_z \cos\eta_2(\tau)) K \sin\phi_K \sin\eta_2(\tau)$$

$$\exp\left(iaK\tilde{R}_{H1}(\tau+\tau',\phi)\cos\left(\tilde{\xi}_1(\tau+\tau',\phi) + \gamma_A(\tau+\tau',\tau) - \phi_K\right)\right) \quad (5.18)$$

$$\exp\left(-iaK\tilde{R}_{H2}(\tau-\tau',\phi')\cos\left(\tilde{\xi}_2(\tau-\tau',\phi') + \gamma_A(\tau-\tau',\tau) - \phi_K\right)\right)$$

$$\exp\left(iak_z\tilde{Z}_{H1}(\tau+\tau',\phi)\right)\exp\left(-iak_z\tilde{Z}_{H2}(\tau-\tau',\phi')\right)\exp(-il\phi)\exp(-il'\phi').$$

Using similar identities to Eqs. (2.17)-(2.19), Eqs. (5.16)-(5.18) can be rewritten as,

$$\tilde{E}_{pt0}(\tau,\tau') = \frac{4\pi e^2}{\varepsilon_w l_c^2} \frac{1}{(2\pi)^5} \int_0^{2\pi} d\phi_K \int_{-\infty}^{\infty} dk_z \int_0^{\infty} KdK \int_0^{2\pi} d\phi \int_0^{2\pi} d\phi' \sum_{l,l',n,n',m} (i)^{-m+n-n'}$$

$$\frac{J_m(KR) J_n(aK\tilde{R}_{H1}(\tau+\tau',\phi)) J_{n'}(aK\tilde{R}_{H2}(\tau-\tau',\phi'))}{\mathbf{k}^2 + \kappa_D^2} \zeta_{surf}^{(1)}(l, k_z \cos\eta_1(\tau)) \zeta_{surf}^{(2)}(l', k_z \cos\eta_2(\tau))$$

$$\exp\left(-im\phi_K + in\left(\tilde{\xi}_1(\tau+\tau',\phi) + \gamma_A(\tau+\tau';\tau) - \phi_K\right) - in'\left(\tilde{\xi}_2(\tau-\tau',\phi') + \gamma_A(\tau-\tau',\tau) - \phi_K\right)\right)$$

$$\exp(2i\tau' k_z) \exp\left(iak_z\tilde{Z}_{H1}(\tau+\tau',\phi)\right)\exp\left(-iak_z\tilde{Z}_{H2}(\tau-\tau',\phi')\right)\exp(-il\phi)\exp(-il'\phi'),$$

(5.19)

$$\tilde{E}_{pt1}(\tau,\tau') = \frac{2\pi e^2}{\varepsilon_w l_c^2} \frac{1}{i(2\pi)^5} \int_0^{2\pi} d\phi_K \int_{-\infty}^{\infty} dk_z \int_0^{\infty} KdK \int_0^{2\pi} d\phi \int_0^{2\pi} d\phi' \sum_{l,l',n,n',m} K\sin\eta_1(\tau)(i)^{m+n-n'}$$

$$[\exp(i(m+1)\phi_K) - \exp(i(m-1)\phi_K)] \frac{J_{-m}(KR) J_n(aK\tilde{R}_{H1}(\tau+\tau',\phi)) J_{n'}(aK\tilde{R}_{H2}(\tau-\tau',\phi'))}{\mathbf{k}^2 + \kappa_D^2}$$

$$\zeta_{surf}^{\prime(1)}(l, \sigma_1(\tau) k_z \cos\eta_1(\tau)) \zeta_{surf}^{(2)}(l', \sigma_2(\tau) k_z \cos\eta_2(\tau)) \exp(2i\tau' k_z) \exp(-il\phi) \exp(-il'\phi')$$

$$\exp\left(in\left(\tilde{\xi}_1(\tau+\tau',\phi) + \gamma_A(\tau+\tau';\tau) - \phi_K\right) - in'\left(\tilde{\xi}_2(\tau-\tau',\phi') + \gamma_A(\tau-\tau';\tau) - \phi_K\right)\right)$$

$$\exp\left(iak_z\tilde{Z}_{H1}(\tau+\tau',\phi)\right)\exp\left(-iak_z\tilde{Z}_{H2}(\tau-\tau',\phi')\right),$$

(5.20)

$$\tilde{E}_{pt2}(\tau,\tau') = -\frac{2\pi e^2}{\varepsilon_w l_c^2} \frac{1}{i(2\pi)^5} \int_0^{2\pi} d\phi_K \int_{-\infty}^{\infty} dk_z \int_0^{\infty} K dK \int_0^{2\pi} d\phi \int_0^{2\pi} d\phi' \sum_{l,l',n,n',m} K \sin\eta_2(\tau)(i)^{m+n-n'}$$

$$[\exp(i(m+1)\phi_K) - \exp(i(m-1)\phi_K)] \frac{J_{-m}(KR) J_n(aK\tilde{R}_{H1}(\tau+\tau',\phi)) J_{n'}(aK\tilde{R}_{H2}(\tau-\tau',\phi'))}{\mathbf{k}^2 + \kappa_D^2}$$

$$\zeta_{surf}^{(1)}(l, k_z \cos\eta_1(\tau)) \zeta_{surf}'^{(2)}(l', k_z \cos\eta_2(\tau)) \exp(2i\tau' k_z) \exp(-il\phi) \exp(-il'\phi')$$

$$\exp\left(in\left(\tilde{\xi}_1(\tau+\tau',\phi) + \gamma_A(\tau+\tau';\tau) - \phi_K\right) - in'\left(\tilde{\xi}_2(\tau-\tau',\phi') + \gamma_A(\tau-\tau';\tau) - \phi_K\right)\right)$$

$$\exp\left(iak_z\tilde{Z}_{H1}(\tau+\tau',\phi)\right)\exp\left(-iak_z\tilde{Z}_{H2}(\tau-\tau',\phi')\right).$$

(5.21)

On performing the $\phi_K$-integrals and then the $K$-integrals Eq. (5.19)-(5.21) become

$$\tilde{E}_{pt0}(\tau,\tau') = \frac{4\pi e^2}{\varepsilon_w l_c^2} \frac{1}{(2\pi)^4} \int_{-\infty}^{\infty} dk_z \int_0^{2\pi} d\phi \int_0^{2\pi} d\phi' \sum_{l,l',n,n'} (-1)^{n'}$$

$$K_{n-n'}\left(\sqrt{k_z^2 + \kappa_D^2} R\right) I_n\left(\sqrt{k_z^2 + \kappa_D^2} a\tilde{R}_{H1}(\tau+\tau',\phi)\right) I_{n'}\left(\sqrt{k_z^2 + \kappa_D^2} a\tilde{R}_{H2}(\tau-\tau',\phi')\right)$$

$$\zeta_{surf}^{(1)}(l, k_z \cos\eta_1(\tau)) \zeta_{surf}^{(2)}(l', k_z \cos\eta_2(\tau)) \exp(-il\phi) \exp(-il'\phi') \exp(2i\tau' k_z) \quad (5.22)$$

$$\exp\left(in\left(\tilde{\xi}_1(\tau+\tau',\phi) + \gamma_A(\tau-\tau';\tau)\right) - in'\left(\tilde{\xi}_2(\tau-\tau',\phi') + \gamma_A(\tau-\tau';\tau)\right)\right)$$

$$\exp\left(iak_z\tilde{Z}_{H1}(\tau+\tau',\phi)\right)\exp\left(-iak_z\tilde{Z}_{H2}(\tau-\tau',\phi')\right),$$

$$\tilde{E}_{pt1}(\tau,\tau') = \frac{4\pi e^2}{\varepsilon_w l_c^2} \frac{1}{(2\pi)^4} \int_{-\infty}^{\infty} dk_z \int_0^{2\pi} d\phi \int_0^{2\pi} d\phi' \sum_{l,l',n,n'} \frac{(n-n')\sin\eta_1(\tau)}{R}(-1)^{n'}$$

$$K_{n-n'}\left(R\sqrt{k_z^2 + \kappa_D^2}\right) I_n\left(a\sqrt{k_z^2 + \kappa_D^2}\tilde{R}_{H1}(\tau+\tau',\phi)\right) I_{n'}\left(a\sqrt{k_z^2 + \kappa_D^2}\tilde{R}_{H2}(\tau-\tau',\phi')\right)$$

$$\zeta_{surf}'^{(1)}(l, k_z \cos\eta_1(\tau)) \zeta_{surf}^{(2)}(l', k_z \cos\eta_2(\tau)) \exp(-il\phi) \exp(-il'\phi') \exp(2i\tau' k_z) \quad (5.23)$$

$$\exp\left(in\left(\tilde{\xi}_1(\tau+\tau',\phi) + \gamma_A(\tau+\tau';\tau)\right) - in'\left(\tilde{\xi}_2(\tau-\tau',\phi') + \gamma_A(\tau-\tau';\tau)\right)\right)$$

$$\exp\left(iak_z\tilde{Z}_{H1}(\tau+\tau',\phi)\right)\exp\left(-iak_z\tilde{Z}_{H2}(\tau-\tau',\phi')\right),$$

$$\tilde{E}_{pt2}(\tau,\tau') = -\frac{4\pi e^2}{\varepsilon_w l_c^2} \frac{1}{(2\pi)^4} \int_{-\infty}^{\infty} dk_z \int_0^{2\pi} d\phi \int_0^{2\pi} d\phi' \sum_{l,l',n,n'} \frac{(n-n')\sin\eta_2(\tau)}{R}(-1)^{n'}$$

$$K_{n-n'}\left(R\sqrt{k_z^2 + \kappa_D^2}\right) I_n\left(a\tilde{R}_{H1}(\tau+\tau',\phi)\sqrt{k_z^2 + \kappa_D^2}\right) I_{n'}\left(a\tilde{R}_{H2}(\tau-\tau',\phi')\sqrt{k_z^2 + \kappa_D^2}\right)$$

$$\zeta_{surf}^{(1)}(l, k_z \cos\eta_1(\tau)) \zeta_{surf}'^{(2)}(l', k_z \cos\eta_2(\tau)) \exp(-il\phi) \exp(-il'\phi') \exp(2i\tau' k_z) \quad (5.24)$$

$$\exp\left(in\left(\tilde{\xi}_1(\tau+\tau',\phi) + \gamma_A(\tau+\tau';\tau)\right) - in'\left(\tilde{\xi}_2(\tau-\tau',\phi') + \gamma_A(\tau-\tau';\tau)\right)\right)$$

$$\exp\left(iak_z\tilde{Z}_{H1}(\tau+\tau',\phi)\right)\exp\left(-iak_z\tilde{Z}_{H2}(\tau-\tau',\phi')\right).$$

We can use similar identities to Eqs. (2.24)-(2.26), as well as performing the $\phi$ and $\phi'$ integrations to re-express Eqs. (5.22)-(5.24)

$$\tilde{E}_{pt0}(\tau,\tau') = \frac{4\pi e^2}{\varepsilon_w l_c^2} \frac{1}{(2\pi)^2} \int_{-\infty}^{\infty} dk_z \sum_{l,l',n,n',m,m'} (-1)^{n'} \exp(2i\tau'k_z) K_{n-n'}\left(\sqrt{k_z^2 + \kappa_D^2} R\right)$$

$$I_{n-m}\left(\frac{a\sqrt{k_z^2 + \kappa_D^2}}{2}(1-\cos\eta_1(\tau+\tau'))\right) I_m\left(\frac{a\sqrt{k_z^2 + \kappa_D^2}}{2}(1+\cos\eta_1(\tau+\tau'))\right) J_{2m-n-l}(ak_z \sin\eta_1(\tau+\tau'))$$

$$I_{n'-m'}\left(\frac{a\sqrt{k_z^2 + \kappa_D^2}}{2}(1-\cos\eta_2(\tau-\tau'))\right) I_{m'}\left(\frac{a\sqrt{k_z^2 + \kappa_D^2}}{2}(1+\cos\eta_2(\tau-\tau'))\right) J_{n'-2m'-l'}(ak_z \sin\eta_2(\tau-\tau'))$$

$$\zeta_{surf}^{(1)}(l, k_z \cos\eta_1(\tau)) \zeta_{surf}^{(2)}(l', k_z \cos\eta_2(\tau)) \exp(i(l+n)\xi_1(\tau+\tau')) \exp(i(l'-n')\xi_2(\tau-\tau'))$$

$$\exp(-in(\xi_1(\tau+\tau') - \gamma_A(\tau+\tau';\tau))) \exp(in'(\xi_2(\tau-\tau') - \gamma_A(\tau-\tau';\tau))),$$

(5.25)

$$\tilde{E}_{pt1}(\tau,\tau') = \frac{4\pi e^2}{\varepsilon_w l_c^2} \frac{1}{(2\pi)^2} \int_{-\infty}^{\infty} dk_z \sum_{l,l',n,n',m,m'} (-1)^{n'} \frac{(n-n')\sin\eta_1(\tau)}{R} K_{n-n'}\left(\sqrt{k_z^2 + \kappa_D^2} R\right)$$

$$I_{n-m}\left(\frac{a\sqrt{k_z^2 + \kappa_D^2}}{2}(1-\cos\eta_1(\tau+\tau'))\right) I_m\left(\frac{a\sqrt{k_z^2 + \kappa_D^2}}{2}(1+\cos\eta_1(\tau+\tau'))\right) J_{2m-l-n}(ak_z \sin\eta_1(\tau+\tau'))$$

$$I_{n'-m'}\left(\frac{a\sqrt{k_z^2 + \kappa_D^2}}{2}(1-\cos\eta_2(\tau-\tau'))\right) I_{m'}\left(\frac{a\sqrt{k_z^2 + \kappa_D^2}}{2}(1+\cos\eta_2(\tau-\tau'))\right) J_{n'-2m'-l'}(ak_z \sin\eta_2(\tau-\tau'))$$

$$\zeta_{surf}'^{(1)}(l, k_z \cos\eta_1(\tau)) \zeta_{surf}^{(2)}(-l', k_z \cos\eta_2(\tau)) \exp(-in(\xi_1(\tau+\tau') - \gamma_A(\tau+\tau';\tau))) \exp(in'(\xi_2(\tau-\tau') - \gamma_A(\tau-\tau';\tau)))$$

$$\exp(i(l+n)\xi_1(\tau+\tau')) \exp(i(l'-n')\xi_2(\tau-\tau')) \exp(2i\tau'k_z),$$

(5.26)

$$\tilde{E}_{pt2}(\tau,\tau') = -\frac{4\pi e^2}{\varepsilon_w l_c^2} \frac{1}{(2\pi)^2} \int_{-\infty}^{\infty} dk_z \sum_{l,l',n,n',m,m'} (-1)^{n'} \frac{(n-n')\sin\eta_2(\tau)}{R} K_{n-n'}\left(\sqrt{k_z^2 + \kappa_D^2} R\right)$$

$$I_{n-m}\left(\frac{a\sqrt{k_z^2 + \kappa_D^2}}{2}(1-\cos\eta_1(\tau+\tau'))\right) I_m\left(\frac{a\sqrt{k_z^2 + \kappa_D^2}}{2}(1+\cos\eta_1(\tau+\tau'))\right) J_{2m-n-l}(ak_z \sin\eta_1(\tau+\tau'))$$

$$I_{n'-m'}\left(\frac{a\sqrt{k_z^2 + \kappa_D^2}}{2}(1-\cos\eta_2(\tau-\tau'))\right) I_{m'}\left(\frac{a\sqrt{k_z^2 + \kappa_D^2}}{2}(1+\cos\eta_2(\tau-\tau'))\right) J_{l'+2m'-n'}(ak_z \sin\eta_2(\tau-\tau'))$$

$$\zeta_{surf}^{(1)}(l, k_z \cos\eta_1(\tau)) \zeta_{surf}'^{(2)}(l', k_z \cos\eta_2(\tau)) \exp(-in(\xi_1(\tau+\tau') - \gamma_A(\tau+\tau';\tau))) \exp(in'(\xi_2(\tau-\tau') - \gamma_A(\tau-\tau';\tau)))$$

$$\exp(i(l+n)\xi_1(\tau+\tau')) \exp(i(l'-n')\xi_2(\tau-\tau')) \exp(2i\tau'k_z).$$

(5.27)

The Taylor expansions, Eqs (2.28)-(2.30) allow us the write for the $\tau'$ integrals with integrands $\tilde{E}_{pt0}(\tau,\tau')$, $\tilde{E}_{pt1}(\tau,\tau')$ and $\tilde{E}_{pt2}(\tau,\tau')$ as well as the $k_z$ integral. This gives us

$$E_{dir} = E_{dir,0} + E_{dir,1} + E_{dir,2}, \tag{5.28}$$

$$E_{dir,0} = \frac{2e^2}{\varepsilon_w l_c^2} \int_0^{L_B} d\tau \sigma_1(\tau)\sigma_2(\tau) \sum_{l,l',n,n',m,m'} (-1)^{n'} \zeta_{surf}^{(1)}(l, k_{l,l',n,n'}(\tau)\cos\eta_1(\tau)) \zeta_{surf}^{(2)}(l', k_{l,l',n,n'}(\tau)\cos\eta_2(\tau))$$

$$K_{n'-n}\left(R\kappa_{l,l',n,n'}(\tau)\right) I_{n-m}\left(\frac{a\kappa_{l,l',n,n'}(\tau)}{2}(1-\cos\eta_1(\tau))\right) I_m\left(\frac{a\kappa_{l,l',n,n'}(\tau)}{2}(1+\cos\eta_1(\tau))\right)$$

$$I_{n'-m'}\left(\frac{a\kappa_{l,l',n,n'}(\tau)}{2}(1-\cos\eta_2(\tau))\right) I_{m'}\left(\frac{a\kappa_{l,l',n,n'}(\tau)}{2}(1+\cos\eta_2(\tau))\right)$$

$$\exp(i(l\xi_1(\tau)+l'\xi_2(\tau))) J_{2m-l-n}(ak_{l,l',n,n'}(\tau)\sin\eta_1(\tau)) J_{n'-2m'-l'}(ak_{l,l',n,n'}(\tau)\sin\eta_2(\tau)), \tag{5.29}$$

$$E_{dir,1} = \frac{2e^2}{\varepsilon_w l_c^2} \int_0^{L_B} d\tau \sigma_1(\tau)\sigma_2(\tau) \sum_{l,l',n,n',m,m'} (-1)^{n'} \frac{(n-n')}{R} \sin\eta_1(\tau)$$

$$K_{n-n'}\left(R\kappa_{l,l',n,n'}(\tau)\right) I_{n-m}\left(\frac{a\kappa_{l,l',n,n'}(\tau)}{2}(1-\cos\eta_1(\tau))\right) I_m\left(\frac{a\kappa_{l,l',n,n'}(\tau)}{2}(1+\cos\eta_1(\tau))\right)$$

$$I_{n'-m'}\left(\frac{a\kappa_{l,l',n,n'}(\tau)}{2}(1-\cos\eta_2(\tau))\right) I_{m'}\left(\frac{a\kappa_{l,l',n,n'}(\tau)}{2}(1+\cos\eta_2(\tau))\right) \tag{5.30}$$

$$\zeta_{surf}'^{(1)}(l, k_{l,l',n,n'}(\tau)\cos\eta_1(\tau)) \zeta_{surf}^{(2)}(l', k_{l,l',n,n'}(\tau)\cos\eta_2(\tau)) \exp(i(l\xi_1(\tau)+l'\xi_2(\tau)))$$

$$J_{2m-l-n}(ak_{l,l',n,n'}(\tau)\sin\eta_1(\tau)) J_{n'-2m'-l'}(ak_{l,l',n,n'}(\tau)\sin\eta_2(\tau)),$$

$$E_{dir,2} = -\frac{2e^2}{\varepsilon_w l_c^2} \int_0^{L_B} d\tau \sigma_1(\tau)\sigma_2(\tau) \sum_{l,l',n,n',m,m'} (-1)^{n'} \frac{(n-n')}{R} \sin\eta_2(\tau)$$

$$K_{n'-n}\left(R\kappa_{l,l',n,n'}(\tau)\right) I_{n-m}\left(\frac{a\kappa_{l,l',n,n'}(\tau)}{2}(1-\cos\eta_1(\tau))\right) I_m\left(\frac{a\kappa_{l,l',n,n'}(\tau)}{2}(1+\cos\eta_1(\tau))\right)$$

$$I_{n'-m'}\left(\frac{a\kappa_{l,l',n,n'}(\tau)}{2}(1-\cos\eta_2(\tau))\right) I_{m'}\left(\frac{a\kappa_{l,l',n,n'}(\tau)}{2}(1+\cos\eta_2(\tau))\right) \tag{5.31}$$

$$\zeta_{surf}^{(1)}(l, k_{l,l',n,n'}(\tau)\cos\eta_1(\tau)) \zeta_{surf}'^{(2)}(l', k_{l,l',n,n'}(\tau)\cos\eta_2(\tau)) \exp(i(l\xi_1(\tau)+l'\xi_2(\tau)))$$

$$J_{2m-n-l}(ak_{l,l',n,n'}(\tau)\sin\eta_1(\tau)) J_{n'-2m'-l'}(ak_{l,l',n,n'}(\tau)\sin\eta_2(\tau)),$$

with

$$\kappa_{l,l',n,n'}(\tau) = \sqrt{k_{l,l',n,n'}(\tau)^2 + \kappa_D^2}, \tag{5.32}$$

$$k_{l,l',n,n'}(\tau) = \left(\frac{n}{2}\left(\frac{d\xi_1(\tau)}{d\tau} - \omega_{A,1}(\tau)\right) + \frac{n'}{2}\left(\frac{d\xi_2(\tau)}{d\tau} - \omega_{A,1}(\tau)\right) - \frac{(l+n)}{2}\frac{d\xi_1(\tau)}{d\tau} + \frac{(l'-n')}{2}\frac{d\xi_2(\tau)}{d\tau}\right). \tag{5.33}$$

# 6. General expressions for image charge repulsion interaction for rod with cores of low dielectric

First we deal with $E_{img1}$. We set $t_0 = \xi_1(s)$ and $s_0 = s$ in the expression for the effective charges on rod 1. In the expressions describing the image charges on rod 2 induced by rod 1 we set $t_0 = \xi_1(s')$ and $s_0 = s'$. Then, we can substitute Eqs. (4.70) and (4.92) into Eq. (3.4) and change variables $\tau = \frac{s+s'}{2}$ and $\tau' = \frac{s-s'}{2}$ so that we write

$$E_{img1} = 2\int_0^{L_B} d\tau \int_{-\infty}^{\infty} d\tau' \sigma_1(\tau+\tau')\sigma_1(\tau-\tau')\tilde{E}_{img1}(\tau,\tau'), \tag{6.1}$$

$$\tilde{E}_{img1}(\tau,\tau') = \frac{4\pi e^2}{\varepsilon_w l_c^2} \frac{1}{(2\pi)^5} \sum_{l,l'} \int_0^{2\pi} d\phi \int_0^{2\pi} d\phi' \int d^3k \exp(-il\phi)\exp(-il'\phi') \frac{\exp(i\mathbf{k}.(\mathbf{r}_1(\tau+\tau')-\mathbf{r}_2(\tau-\tau')))}{\mathbf{k}^2 + \kappa_D^2}$$
$$\exp(ia\mathbf{k}.(\hat{\mathbf{N}}_1^0(\tau+\tau',\phi+\zeta_1(s)) - \hat{\mathbf{N}}_2^0(\tau-\tau',\phi')))\zeta_{surf,0}^{(1)}(l,\mathbf{k}.\hat{\mathbf{t}}_1(\tau+\tau'))\zeta_{surf,1}^{(2)}(l,-\mathbf{k}.\hat{\mathbf{t}}_2(\tau-\tau')). \tag{6.2}$$

Next we Taylor expand through Eqs. (2.4) and (5.4). We again choose $\hat{\mathbf{t}}_A(\tau).\mathbf{k} = k_z$ $\hat{\mathbf{d}}(\tau).\mathbf{k} = k_x = K\cos\phi_K$ so that Eq. (5.5) is satisfied. Then with similar steps to those of Eq. (5.5)-(5.11) we can write Eq. (6.2) as

$$\tilde{E}_{img.1}(\tau,\tau') \approx \tilde{E}_{img,1,0}(\tau,\tau') + \tilde{E}_{img,1,1}(\tau,\tau') + \tilde{E}_{img,1,2}(\tau,\tau'), \tag{6.3}$$

$$\tilde{E}_{img1,0}(\tau,\tau') = \frac{2\pi e^2}{\varepsilon_w l_c^2} \frac{1}{(2\pi)^5} \int_0^{2\pi} d\phi_K \int_{-\infty}^{\infty} dk_z \int_0^{\infty} KdK \int_0^{2\pi} d\phi \int_0^{2\pi} d\phi' \sum_{l,l'} \frac{\exp(-iKR\cos\phi_K)\exp(2i\tau'k_z)}{\mathbf{k}^2 + \kappa_D^2}$$
$$\zeta_{surf0}^{(1)}(l,k_z\cos\eta_1(\tau))\zeta_{surf1}^{(2)}(l',-k_z\cos\eta_2(\tau);\xi_1(\tau-\tau'))\exp(-il\phi)\exp(-il'\phi')$$
$$\exp\left(iaK\tilde{R}_{H1}(\tau+\tau',\phi)\cos\left(\tilde{\xi}_1(\tau+\tau',\phi) + \gamma_A(\tau+\tau';\tau) - \phi_K\right)\right)\exp\left(iak_z\tilde{Z}_{H1}(\tau+\tau',\phi)\right)$$
$$\exp\left(-iaK\bar{R}_{H2}(\tau-\tau',\phi')\cos\left(\tilde{\Phi}_2(\tau-\tau',\phi') + \gamma_A(\tau-\tau';\tau) - \phi_K\right)\right)\exp\left(-iak_z\bar{Z}_{H2}(\tau-\tau',\phi')\right), \tag{6.4}$$

$$\tilde{E}_{img1,1}(\tau,\tau') = \frac{2\pi e^2}{\varepsilon_w l_c^2} \frac{1}{(2\pi)^5} \int_0^{2\pi} d\phi_K \int_{-\infty}^{\infty} dk_z \int_0^{\infty} KdK \int_0^{2\pi} d\phi \int_0^{2\pi} d\phi' \sum_{l,l'} \frac{\exp(-iKR\cos\phi_K)\exp(2i\tau'k_z)}{\mathbf{k}^2 + \kappa_D^2}$$
$$\zeta_{surf0}^{\prime(1)}(l,k_z\cos\eta_1(\tau))\zeta_{surf1}^{(2)}(l',-k_z\cos\eta_2(\tau);\xi_1(\tau-\tau'))K\sin\phi_K\sin\eta_1(\tau)\exp(-il\phi)\exp(-il'\phi')$$
$$\exp\left(iaK\tilde{R}_{H1}(\tau+\tau',\phi)\cos\left(\tilde{\xi}_1(\tau+\tau',\phi) + \gamma_A(\tau+\tau';\tau) - \phi_K\right)\right)\exp\left(iak_z\tilde{Z}_{H1}(\tau+\tau',\phi)\right)$$
$$\exp\left(-iaK\bar{R}_{H2}(\tau-\tau',\phi')\cos\left(\tilde{\Phi}_2(\tau-\tau',\phi') + \gamma_A(\tau-\tau';\tau) - \phi_K\right)\right)\exp\left(-iak_z\bar{Z}_{H2}(\tau-\tau',\phi')\right), \tag{6.5}$$

$$\tilde{E}_{img1,2}(\tau,\tau') = \frac{2\pi e^2}{\varepsilon_w l_c^2} \frac{1}{(2\pi)^5} \int_0^{2\pi} d\phi_K \int_{-\infty}^{\infty} dk_z \int_0^{\infty} KdK \int_0^{2\pi} d\phi \int_0^{2\pi} d\phi' \sum_{l,l'} \frac{\exp(-iKR\cos\phi_K)\exp(2i\tau'k_z)}{\mathbf{k}^2 + \kappa_D^2}$$

$$\zeta_{surf0}^{(1)}(l, k_z \cos\eta_1(\tau))\zeta_{surf1}^{\prime(2)}(l', -k_z \cos\eta_2(\tau); \xi_1(\tau-\tau'))K \sin\phi_K \sin\eta_2(\tau)\exp(-il\phi)\exp(-il'\phi')$$

$$\exp\left(iaK\tilde{R}_{H1}(\tau+\tau',\phi)\cos\left(\tilde{\xi}_1(\tau+\tau',\phi)+\gamma_A(\tau+\tau';\tau)-\phi_K\right)\right)\exp\left(iak_z\tilde{Z}_{H1}(\tau+\tau',\phi)\right)$$

$$\exp\left(-iaK\bar{R}_{H2}(\tau-\tau',\phi')\cos\left(\tilde{\Phi}_2(\tau-\tau',\phi')+\gamma_A(\tau-\tau';\tau)-\phi_K\right)\right)\exp\left(-iak_z\bar{Z}_{H2}(\tau-\tau',\phi')\right)$$

(6.6)

where we define ($\mu = 1, 2$)

$$\tilde{\Phi}_\mu(\tau-\tau',\phi') = \tan^{-1}\left(\cos\eta_\mu(\tau-\tau')\tan(\phi')\right), \tag{6.7}$$

$$\bar{R}_{H\mu}(\tau-\tau',\phi') = \sqrt{\cos(\phi')^2 + \cos^2\eta_\mu(\tau-\tau')\sin(\phi')^2}, \tag{6.8}$$

$$\bar{Z}_{H\mu}(\tau-\tau',\phi') = \sin\eta_\mu(\tau-\tau')\sin(\phi'). \tag{6.9}$$

Following similar steps as Eqs. (5.19)-(5.27) and using Eqs (4.75) (4.90), (4.91) we may write

$$\tilde{E}_{img1,0}(\tau,\tau') = \frac{2\pi e^2}{\varepsilon_w l_c^2} \frac{1}{(2\pi)^2} \int_{-\infty}^{\infty} dk_z \sum_{l,l',n,n',m,m',j'} (-1)^{n'} K_{n'-n}\left(\sqrt{k_z^2 + \kappa_D^2}R\right)\exp(2ik_z\tau')$$

$$I_{n-m}\left(\frac{a\sqrt{k_z^2 + \kappa_D^2}}{2}(1-\cos\eta_1(\tau+\tau'))\right)I_m\left(\frac{a\sqrt{k_z^2 + \kappa_D^2}}{2}(1+\cos\eta_1(\tau+\tau'))\right)J_{2m-n-l}(ak_z\sin\eta_1(\tau+\tau'))$$

$$I_{n'-m'}\left(\frac{a\sqrt{k_z^2 + \kappa_D^2}}{2}(1-\cos\eta_2(\tau-\tau'))\right)I_{m'}\left(\frac{a\sqrt{k_z^2 + \kappa_D^2}}{2}(1+\cos\eta_2(\tau-\tau'))\right)J_{n'-2m'-l'}(ak_z\sin\eta_2(\tau-\tau'))$$

$$\zeta_{surf0}^{(1)}(l, k_z\cos\eta_1(\tau))(\tilde{\zeta}_{surf1,0}^{(2)}(l', j', -k_z\cos\eta_2(\tau)) + \sin(\eta_1(\tau)+\eta_2(\tau))\tilde{\zeta}_{surf1,1}^{(2)}(l', j', -k_z\cos\eta_2(\tau)))$$

$$\exp(ij'\xi_1(\tau-\tau'))\exp(-in(\xi_1(\tau+\tau')-\gamma_A(\tau+\tau';\tau)))\exp(-in'\gamma_A(\tau-\tau';\tau)))\exp(-i(l+n)\xi_1(\tau+\tau')),$$

(6.10)

$$\tilde{E}_{img1,1}(\tau,\tau') = \frac{2\pi e^2}{\varepsilon_w l_c^2} \frac{1}{(2\pi)^2} \int_{-\infty}^{\infty} dk_z \sum_{l,l',n,n',m,m',j'} (-1)^{n'} \frac{(n-n')\sin\eta_1(\tau)}{R}\exp(2i\tau'k_z)K_{n'-n}\left(\sqrt{k_z^2+\kappa_D^2}R\right)$$

$$I_{n-m}\left(\frac{a\sqrt{k_z^2 + \kappa_D^2}}{2}(1-\cos\eta_1(\tau+\tau'))\right)I_m\left(\frac{a\sqrt{k_z^2 + \kappa_D^2}}{2}(1+\cos\eta_1(\tau+\tau'))\right)J_{2m-n-l}(ak_z\sin\eta_1(\tau+\tau'))$$

$$I_{n'-m'}\left(\frac{a\sqrt{k_z^2 + \kappa_D^2}}{2}(1-\cos\eta_2(\tau-\tau'))\right)I_{m'}\left(\frac{a\sqrt{k_z^2 + \kappa_D^2}}{2}(1+\cos\eta_2(\tau-\tau'))\right)J_{n'-2m'-l'}(ak_z\sin\eta_2(\tau-\tau'))$$

$$\zeta_{surf0}^{\prime(1)}(l, k_z\cos\eta_1(\tau))\tilde{\zeta}_{surf1,0}^{(2)}(l', j', -k_z\cos\eta_2(\tau))\exp(ij'\xi_1(\tau-\tau'))\exp(-in'\gamma_A(\tau-\tau';\tau)))$$

$$\exp(-in(\xi_1(\tau+\tau')-\gamma_A(\tau+\tau';\tau)))\exp(i(n+l)\xi_1(\tau+\tau')),$$

(6.11)

$$\tilde{E}_{img1,2}(\tau,\tau') = \frac{2\pi e^2}{\varepsilon_w l_c^2} \frac{1}{(2\pi)^2} \int_{-\infty}^{\infty} dk_z \sum_{l,l',n,n',m,m',j'} (-1)^{n'} \frac{(n-n')\sin\eta_2(\tau)}{R} K_{n'-n}\left(\sqrt{k_z^2 + \kappa_D^2} R\right) \exp(2i\tau' k_z)$$

$$I_{n-m}\left(\frac{a\sqrt{k_z^2+\kappa_D^2}}{2}(1-\cos\eta_1(\tau+\tau'))\right) I_m\left(\frac{a\sqrt{k_z^2+\kappa_D^2}}{2}(1+\cos\eta_1(\tau+\tau'))\right) J_{2m-n-l}(ak_z \sin\eta_1(\tau+\tau'))$$

$$I_{n'-m'}\left(\frac{a\sqrt{k_z^2+\kappa_D^2}}{2}(1-\cos\eta_2(\tau-\tau'))\right) I_{m'}\left(\frac{a\sqrt{k_z^2+\kappa_D^2}}{2}(1+\cos\eta_2(\tau-\tau'))\right) J_{n'-2m'-l'}(ak_z \sin\eta_2(\tau-\tau'))$$

$$\zeta^{(1)}_{surf\,0}(l, k_z \cos\eta_1(\tau)) \tilde{\zeta}'^{(2)}_{surf\,1,0}(l', j', -k_z \cos\eta_2(\tau)) \exp(ij'\xi_1(\tau-\tau')) \exp(-in(\xi_1(\tau+\tau')-\gamma_A(\tau+\tau';\tau)))$$

$$\exp(-in'\gamma_A(\tau-\tau';\tau)) \exp(i(n+l)\xi_1(\tau+\tau')).$$

(6.12)

Now we can make the final Taylor expansions using Eqs. (2.28)-(2.30), giving us on performing the $\tau'$ and $k_z$ integrations

$$E_{img1} = E_{img1,0} + E_{img1,1} + E_{img1,2} + E_{img1,3} \tag{6.13}$$

$$E_{img1,0} = \frac{e^2}{\varepsilon_w l_c^2} \int_0^{L_B} d\tau \sigma_1(\tau)^2 \sum_{l,l',n,n',m,m',j'} (-1)^{n'}$$

$$K_{n'-n}\left(\tilde{\kappa}^{(1)}_{l,n,n',j'}(\tau)R\right) I_{n-m}\left(\frac{a\tilde{\kappa}^{(1)}_{l,n,n',j'}(\tau)}{2}(1-\cos\eta_1(\tau))\right) I_m\left(\frac{a\tilde{\kappa}^{(1)}_{l,n,n',j'}(\tau)}{2}(1+\cos\eta_1(\tau))\right)$$

$$I_{n'-m'}\left(\frac{a\tilde{\kappa}^{(1)}_{l,n,n',j'}(\tau)}{2}(1-\cos\eta_2(\tau))\right) I_{m'}\left(\frac{a\tilde{\kappa}^{(1)}_{l,n,n',j'}(\tau)}{2}(1+\cos\eta_2(\tau))\right)$$

$$\zeta^{(1)}_{surf\,0}(l, -\tilde{k}^{(1)}_{l,n,n',j'}(\tau)\cos\eta_1(\tau)) \tilde{\zeta}^{(2)}_{surf\,1,0}(l', j', -\tilde{k}^{(1)}_{l,n,n',j'}(\tau)\cos\eta_2(\tau))$$

$$\exp(i(j'+l)\xi_1(\tau)) J_{2m-n-l}(a\tilde{k}^{(1)}_{l,n,n',j'}(\tau)\sin\eta_1(\tau)) J_{n'-2m'-l'}(a\tilde{k}^{(1)}_{l,n,n',j'}(\tau)\sin\eta_2(\tau)),$$

(6.14)

$$E_{img1,1} = -\frac{e^2}{\varepsilon_w l_c^2} \int_0^{L_B} d\tau \sigma_1(\tau)^2 \sum_{l,l',n,n',m,m',j'} (-1)^{n'} \frac{(n-n')\sin\eta_1(\tau)}{R}$$

$$K_{n'-n}\left(\tilde{\kappa}^{(1)}_{l,n,n',j'}(\tau)R\right) I_{n-m}\left(\frac{a\tilde{\kappa}^{(1)}_{l,n,n',j'}(\tau)}{2}(1-\cos\eta_1(\tau))\right) I_m\left(\frac{a\tilde{\kappa}^{(1)}_{l,n,n',j'}(\tau)}{2}(1+\cos\eta_1(\tau))\right)$$

$$I_{n'-m'}\left(\frac{a\tilde{\kappa}^{(1)}_{l,n,n',j'}(\tau)}{2}(1-\cos\eta_2(\tau))\right) I_{m'}\left(\frac{a\tilde{\kappa}^{(1)}_{l,n,n',j'}(\tau)}{2}(1+\cos\eta_2(\tau))\right) \tag{6.15}$$

$$\zeta'^{(1)}_{surf\,0}(l, -\tilde{k}^{(1)}_{l,n,n',j'}(\tau)\cos\eta_1(\tau)) \tilde{\zeta}^{(2)}_{surf\,1,0}(l', j', -\tilde{k}^{(1)}_{l,n,n',j'}(\tau)\cos\eta_2(\tau))$$

$$\exp(i(j'+l)\xi_1(\tau)) J_{2m-n-l}(a\tilde{k}^{(1)}_{l,n,n',j'}(\tau)\sin\eta_1(\tau)) J_{n'-2m'-l'}(a\tilde{k}^{(1)}_{l,n,n',j'}(\tau)\sin\eta_2(\tau)),$$

$$E_{img1,2} = \frac{e^2}{\varepsilon_w l_c^2} \int_0^{L_B} d\tau \sigma_1(\tau)^2 \sum_{l,l',n,n',m,m',j'} (-1)^{n'} \frac{(n-n')\sin\eta_2(\tau)}{R}$$

$$K_{n'-n}\left(\tilde{\kappa}^{(1)}_{l,n,n',j'}(\tau)R\right) I_{n-m}\left(\frac{a\tilde{\kappa}^{(1)}_{l,n,n',j'}(\tau)}{2}(1-\cos\eta_1(\tau))\right) I_m\left(\frac{a\tilde{\kappa}^{(1)}_{l,n,n',j'}(\tau)}{2}(1+\cos\eta_1(\tau))\right)$$

$$I_{n'-m'}\left(\frac{a\tilde{\kappa}^{(1)}_{l,n,n',j'}(\tau)}{2}(1-\cos\eta_2(\tau))\right) I_{m'}\left(\frac{a\tilde{\kappa}^{(1)}_{l,n,n',j'}(\tau)}{2}(1+\cos\eta_2(\tau))\right) \qquad (6.16)$$

$$\tilde{\zeta}^{(1)}_{surf0}(l, -\tilde{k}^{(1)}_{l,n,n',j'}(\tau)\cos\eta_1(\tau)) \tilde{\zeta}'^{(2)}_{surf1,0}(l', j', -\tilde{k}^{(1)}_{l,n,n',j'}(\tau)\cos\eta_2(\tau))$$

$$\exp(i(j'+l)\xi_1(\tau)) J_{2m-n-l}(a\tilde{k}^{(1)}_{l,n,n',j'}(\tau)\sin\eta_1(\tau)) J_{n'-2m'-l'}(a\tilde{k}^{(1)}_{l,n,n',j'}(\tau)\sin\eta_2(\tau)),$$

$$E_{img1,3} = \frac{e^2}{\varepsilon_w l_c^2} \int_0^{L_B} d\tau \sigma_1(\tau)^2 \sum_{l,l',n,n',m,m',j'} (-1)^{n'} \sin(\eta_1(\tau)+\eta_2(\tau))$$

$$K_{n'-n}\left(\tilde{\kappa}^{(1)}_{l,n,n',j'}(\tau)R\right) I_{n-m}\left(\frac{a\tilde{\kappa}^{(1)}_{l,n,n',j'}(\tau)}{2}(1-\cos\eta_1(\tau))\right) I_m\left(\frac{a\tilde{\kappa}^{(1)}_{l,n,n',j'}(\tau)}{2}(1+\cos\eta_1(\tau))\right)$$

$$I_{n'-m'}\left(\frac{a\tilde{\kappa}^{(1)}_{l,n,n',j'}(\tau)}{2}(1-\cos\eta_2(\tau))\right) I_{m'}\left(\frac{a\tilde{\kappa}^{(1)}_{l,n,n',j'}(\tau)}{2}(1+\cos\eta_2(\tau))\right)$$

$$\tilde{\zeta}^{(1)}_{surf0}(l, -\tilde{k}^{(1)}_{l,n,n',j'}(\tau)\cos\eta_1(\tau)) \tilde{\zeta}'^{(2)}_{surf1,1}(l', j', -\tilde{k}^{(1)}_{l,n,n',j'}(\tau)\cos\eta_2(\tau))$$

$$\exp(i(j'+l)\xi_1(\tau)) J_{2m-n-l}(a\tilde{k}^{(1)}_{l,n,n',j'}(\tau)\sin\eta_1(\tau)) J_{n'-2m'-l'}(a\tilde{k}^{(1)}_{l,n,n',j'}(\tau)\sin\eta_2(\tau)),$$

$$(6.17)$$

where

$$\tilde{\kappa}^{(1)}_{l,n,n',j'}(\tau) = \sqrt{\left(\tilde{k}^{(1)}_{l,n,n',j'}(\tau)\right)^2 + \kappa_D^2}, \qquad (6.18)$$

and $\tilde{k}^{(1)}_{l,n,n',j'}(\tau) = -\frac{(l-j')}{2}\frac{d\xi_1(\tau)}{d\tau} - \frac{n'+n}{2}\omega_{A,1}(\tau),$ (6.19)

where we have used the fact that $\tilde{\zeta}^{(1)}_{surf0}(l', j', -k_z) = \tilde{\zeta}^{(1)}_{surf0}(l', j', k_z)$ and $\tilde{\zeta}'^{(2)}_{surf1,1}(l', j', -k_z) = \tilde{\zeta}'^{(2)}_{surf1,1}(l', j', k_z)$.

Now we look at $E_{img2}$. We set $t_0 = \xi_2(s')$ and $s_0 = s'$ in the expression for the effective charges on rod 1. In the expressions describing the image charges on rod 2 induced by rod 1 we set $t_0 = \xi_1(s)$ and $s_0 = s$.

$$E_{img2} = 2\int_0^{L_B} d\tau \int_{-\infty}^{\infty} d\tau' \sigma_2(\tau+\tau')\sigma_2(\tau-\tau')\tilde{E}_{img2}(\tau,\tau'), \qquad (6.20)$$

$$\tilde{E}_{img\,2}(\tau,\tau') = \frac{4\pi e^2}{\varepsilon_w l_c^2} \frac{1}{(2\pi)^5} \sum_{l,l'} \int_0^{2\pi} d\phi \int_0^{2\pi} d\phi' \int d^3k \exp(-il\phi)\exp(-il'\phi') \frac{\exp\left(i\mathbf{k}.(\mathbf{r}_1(\tau+\tau') - \mathbf{r}_2(\tau-\tau'))\right)}{\mathbf{k}^2 + \kappa_D^2}$$
$$\exp(ia\mathbf{k}.\left(\hat{\mathbf{N}}_1^0(\tau+\tau',\phi) - \hat{\mathbf{N}}_2^0(\tau-\tau',\phi' + \xi_2(s'))\right)\zeta_{surf,1}^{(1)}(l,\mathbf{k}.\hat{\mathbf{t}}_1(s))\zeta_{surf,0}^{(2)}(l',-\hat{\mathbf{k}}.\hat{\mathbf{t}}_2(s)).$$

(6.21)

Again, we Taylor expand through Eqs. (2.4) and (5.4). We again choose $\hat{\mathbf{t}}_A(\tau).\mathbf{k} = k_z$ $\hat{\mathbf{d}}(\tau).\mathbf{k} = k_x = K\cos\phi_K$ so that Eq. (5.5) is satisfied. Then with similar steps to those of Eq. (5.5)-(5.11) we can write Eq. (6.21) as

$$\tilde{E}_{img\,2}(\tau,\tau') \approx \tilde{E}_{img\,2,0}(\tau,\tau') + \tilde{E}_{img\,2,1}(\tau,\tau') + \tilde{E}_{img\,2,2}(\tau,\tau'),$$

(6.22)

$$\tilde{E}_{img\,2,0}(\tau,\tau') = \frac{2\pi e^2}{\varepsilon_w l_c^2} \frac{1}{(2\pi)^5} \int_0^{2\pi} d\phi_K \int_{-\infty}^{\infty} dk_z \int_0^{\infty} KdK \int_0^{2\pi} d\phi \int_0^{2\pi} d\phi' \sum_{l,l'} \frac{\exp(-iKR\cos\phi_K)\exp(2i\tau'k_z)}{\mathbf{k}^2 + \kappa_D^2}$$
$$\zeta_{surf\,1}^{(1)}(l,k_z \cos\eta_1(\tau);\xi_2(\tau+\tau'))\zeta_{surf\,0}^{(2)}(l',k_z\cos\eta_2(\tau))\exp(-il'\phi')\exp(-il\phi)$$
$$\exp\left(iaK\overline{R}_{H1}(\tau+\tau',\phi)\cos\left(\tilde{\Phi}_1(\tau+\tau',\phi) + \gamma_A(\tau+\tau';\tau) - \phi_K\right)\right)\exp\left(iak_z\overline{Z}_{H1}(\tau+\tau',\phi)\right)$$
$$\exp\left(-iaK\tilde{R}_{H2}(\tau-\tau',\phi')\cos\left(\tilde{\xi}_2(\tau-\tau',\phi') + \gamma_A(\tau-\tau';\tau) - \phi_K\right)\right)\exp\left(-iak_z\tilde{Z}_{H2}(\tau-\tau',\phi')\right),$$

(6.23)

$$\tilde{E}_{img\,2,1}(\tau,\tau') = \frac{2\pi e^2}{\varepsilon_w l_c^2} \frac{1}{(2\pi)^5} \int_0^{2\pi} d\phi_K \int_{-\infty}^{\infty} dk_z \int_0^{\infty} KdK \int_0^{2\pi} d\phi \int_0^{2\pi} d\phi' \sum_{l,l'} \frac{\exp(-iKR\cos\phi_K)\exp(2i\tau'k_z)}{\mathbf{k}^2 + \kappa_D^2}$$
$$\zeta_{surf\,1}'^{(1)}(l,k_z \cos\eta_1(\tau);\xi_2(\tau+\tau'))\zeta_{surf\,0}^{(2)}(l',k_z\cos\eta_2(\tau))K\sin\phi_K \sin\eta_1(\tau)\exp(-il'\phi')\exp(-il\phi)$$
$$\exp\left(iaK\overline{R}_{H1}(\tau+\tau',\phi)\cos\left(\tilde{\Phi}_1(\tau+\tau',\phi) + \gamma_A(\tau+\tau';\tau) - \phi_K\right)\right)\exp\left(iak_z\overline{Z}_{H1}(\tau+\tau',\phi)\right)$$
$$\exp\left(-iaK\tilde{R}_{H2}(\tau-\tau',\phi')\cos\left(\tilde{\xi}_2(\tau-\tau',\phi') + \gamma_A(\tau-\tau';\tau) - \phi_K\right)\right)\exp\left(-iak_z\tilde{Z}_{H2}(\tau-\tau',\phi')\right),$$

(6.24)

$$\tilde{E}_{img\,2,2}(\tau,\tau') = -\frac{2\pi e^2}{\varepsilon_w l_c^2} \frac{1}{(2\pi)^5} \int_0^{2\pi} d\phi_K \int_{-\infty}^{\infty} dk_z \int_0^{\infty} KdK \int_0^{2\pi} d\phi \int_0^{2\pi} d\phi' \sum_{l,l'} \frac{\exp(-iKR\cos\phi_K)\exp(2i\tau'k_z)}{\mathbf{k}^2 + \kappa_D^2}$$
$$\zeta_{surf\,1}^{(1)}(l,k_z \cos\eta_1(\tau);\xi_2(\tau+\tau'))\zeta_{surf\,0}'^{(2)}(l',k_z\cos\eta_2(\tau))K\sin\phi_K \sin\eta_2(\tau)\exp(-il'\phi')\exp(-il\phi)$$
$$\exp\left(iaK\overline{R}_{H1}(\tau+\tau',\phi)\cos\left(\tilde{\Phi}_1(\tau+\tau',\phi) + \gamma_A(\tau+\tau';\tau) - \phi_K\right)\right)\exp\left(iak_z\overline{Z}_{H1}(\tau+\tau',\phi)\right)$$
$$\exp\left(-iaK\tilde{R}_{H2}(\tau-\tau',\phi')\cos\left(\tilde{\xi}_2(\tau-\tau',\phi') + \gamma_A(\tau-\tau';\tau) - \phi_K\right)\right)\exp\left(-iak_z\tilde{Z}_{H2}(\tau-\tau',\phi')\right).$$

(6.25)

Again, following similar steps as for $E_{img\,1}$ with similar intermediate expressions Eqs. (5.19)-(5.27) and Eqs. (6.10)-(6.12), we may write

$$E_{img\,2} = E_{img\,2,0} + E_{img\,2,1} + E_{img\,2,2} + E_{img\,2,3},$$

(6.26)

$$E_{img2,0} = \frac{e^2}{\varepsilon_w l_c^2} \int_0^{L_B} d\tau \sigma_2(\tau)^2 \sum_{l,l',n,n',m,m',j'} (-1)^{n'} J_{2m-l-n}(a\tilde{k}^{(2)}_{l',n,n',j'}(\tau)\sin\eta_1(\tau)) J_{n'-l'-2m'}(a\tilde{k}^{(2)}_{l',n,n',j'}(\tau)\sin\eta_2(\tau))$$

$$K_{n'-n}\left(\tilde{\kappa}^{(2)}_{l',n,n',j'}(\tau)R\right) I_{n-m}\left(\frac{a\tilde{\kappa}^{(2)}_{l',n,n',j'}(\tau)}{2}(1-\cos\eta_1(\tau))\right) I_m\left(\frac{a\tilde{\kappa}^{(2)}_{l',n,n',j'}(\tau)}{2}(1+\cos\eta_1(\tau))\right)$$

$$I_{n'-m'}\left(\frac{a\tilde{\kappa}^{(2)}_{l',n,n',j'}(\tau)}{2}(1-\cos\eta_2(\tau))\right) I_{m'}\left(\frac{a\tilde{\kappa}^{(2)}_{l',n,n',j'}(\tau)}{2}(1+\cos\eta_2(\tau))\right)$$

$$\tilde{\zeta}^{(1)}_{surf1,0}(l,j',\tilde{k}^{(2)}_{l',n,n',j'}(\tau)\cos\eta_1(\tau)) \zeta^{(2)}_{surf0}(l',\tilde{k}^{(2)}_{l',n,n',j'}(\tau)\cos\eta_2(\tau)) \exp(i(j'+l')\xi_2(\tau)),$$

(6.27)

$$E_{img2,1} = \frac{e^2}{\varepsilon_w l_c^2} \int_0^{L_B} d\tau \sigma_2(\tau)^2 \sum_{l,l',n,n',m,m',j'} (-1)^{n'} \frac{(n-n')\sin\eta_1(\tau)}{R} K_{n'-n}\left(\tilde{\kappa}^{(2)}_{l',n,n',j'}(\tau)R\right)$$

$$I_{n-m}\left(\frac{a\tilde{\kappa}^{(2)}_{l',n,n',j'}(\tau)}{2}(1-\cos\eta_1(\tau))\right) I_m\left(\frac{a\tilde{\kappa}^{(2)}_{l',n,n',j'}(\tau)}{2}(1+\cos\eta_1(\tau))\right) J_{2m-l-n}(a\tilde{k}^{(2)}_{l',n,n',j'}(\tau)\sin\eta_1(\tau))$$

$$I_{n'-m'}\left(\frac{a\tilde{\kappa}^{(2)}_{l',n,n',j'}(\tau)}{2}(1-\cos\eta_2(\tau))\right) I_{m'}\left(\frac{a\tilde{\kappa}^{(2)}_{l',n,n',j'}(\tau)}{2}(1+\cos\eta_2(\tau))\right) J_{n'-l'-2m'}(a\tilde{k}^{(2)}_{l',n,n',j'}(\tau)\sin\eta_2(\tau))$$

$$\tilde{\zeta}'^{(1)}_{surf1,0}(l,j',\tilde{k}^{(2)}_{l',n,n',j'}(\tau)\cos\eta_1(\tau)) \zeta^{(2)}_{surf0}(l',\tilde{k}^{(2)}_{l',n,n',j'}(\tau)\cos\eta_2(\tau)) \exp(i(j'+l')\xi_2(\tau)),$$

(6.28)

$$E_{img2,2} = -\frac{e^2}{\varepsilon_w l_c^2} \int_0^{L_B} d\tau \sigma_2(\tau)^2 \sum_{l,l',n,n',m,m',j'} (-1)^{n'} \frac{(n-n')\sin\eta_2(\tau)}{R} K_{n'-n}\left(\tilde{\kappa}^{(2)}_{l',n,n',j'}(\tau)R\right)$$

$$I_{n-m}\left(\frac{a\tilde{\kappa}^{(2)}_{l',n,n',j'}(\tau)}{2}(1-\cos\eta_1(\tau))\right) I_m\left(\frac{a\tilde{\kappa}^{(2)}_{l',n,n',j'}(\tau)}{2}(1+\cos\eta_1(\tau))\right) J_{2m-l-n}(a\tilde{k}^{(2)}_{l',n,n',j'}(\tau)\sin\eta_1(\tau))$$

$$I_{n'-m'}\left(\frac{a\tilde{\kappa}^{(2)}_{l',n,n',j'}(\tau)}{2}(1-\cos\eta_2(\tau))\right) I_{m'}\left(\frac{a\tilde{\kappa}^{(2)}_{l',n,n',j'}(\tau)}{2}(1+\cos\eta_2(\tau))\right) J_{n'-l'-2m'}(a\tilde{k}^{(2)}_{l',n,n',j'}(\tau)\sin\eta_2(\tau))$$

$$\tilde{\zeta}^{(1)}_{surf1,0}(l,j',\tilde{k}^{(2)}_{l',n,n',j'}(\tau)\cos\eta_1(\tau)) \zeta'^{(2)}_{surf0}(l',\tilde{k}^{(2)}_{l',n,n',j'}(\tau)\cos\eta_2(\tau)) \exp(i(j'+l')\xi_2(\tau)),$$

(6.29)

$$E_{img2,3} = \frac{e^2}{\varepsilon_w l_c^2} \int_0^{L_B} d\tau \sigma_2(\tau)^2 \sum_{l,l',n,n',m,m',j'} (-1)^{n'} \sin\eta(\tau) K_{n'-n}\left(\tilde{\kappa}^{(2)}_{l',n,n',j'}(\tau)R\right)$$

$$I_{n-m}\left(\frac{a\tilde{\kappa}^{(2)}_{l',n,n',j'}(\tau)}{2}(1-\cos\eta_1(\tau+\tau'))\right) I_m\left(\frac{a\tilde{\kappa}^{(2)}_{l',n,n',j'}(\tau)}{2}(1+\cos\eta_1(\tau+\tau'))\right) J_{2m-l-n}(a\tilde{k}^{(2)}_{l',n,n',j'}(\tau)\sin\eta_1(\tau))$$

$$I_{n'-m'}\left(\frac{a\tilde{\kappa}^{(2)}_{l',n,n',j'}(\tau)}{2}(1-\cos\eta_2(\tau-\tau'))\right) I_{m'}\left(\frac{a\tilde{\kappa}^{(2)}_{l',n,n',j'}(\tau)}{2}(1+\cos\eta_2(\tau-\tau'))\right) J_{n'-l'-2m'}(a\tilde{k}^{(2)}_{l',n,n',j'}(\tau)\sin\eta_2(\tau))$$

$$\tilde{\zeta}^{(1)}_{surf1,1}(l,j',\tilde{k}^{(2)}_{l',n,n',j'}(\tau)\cos\eta_1(\tau)) \zeta^{(2)}_{surf0}(l',\tilde{k}^{(2)}_{l',n,n',j'}(\tau)\cos\eta_2(\tau)) \exp(i(j'+l')\xi_2(\tau)),$$

(6.30)

where

$$\tilde{\kappa}^{(2)}_{l',n,n',j'}(\tau) = \sqrt{\left(\tilde{k}^{(2)}_{l',n,n',j'}(\tau)\right)^2 + \kappa_D^2}, \tag{6.31}$$

and $\quad \tilde{k}^{(2)}_{l',n,n',j'}(\tau) = -\dfrac{(j'-l')}{2}\dfrac{d\xi_2(\tau)}{d\tau} - \dfrac{n'+n}{2}\omega_{A,1}(\tau). \tag{6.32}$

## 7. Diagonal Mode Approximation

We can write

$$\xi_1(\tau) = \omega_\xi \tau + \Delta\xi_1(\tau), \qquad \xi_2(\tau) = \omega_\xi \tau + \Delta\xi_2(\tau). \tag{7.1}$$

Both $\Delta\xi_1(\tau)$ and $\Delta\xi_2(\tau)$ can be considered deviations away from the symmetric, straight, regular braided configuration of ideal helices.

If

$$\dfrac{d\Delta\xi_1(\tau)}{d\tau} \ll \omega_\xi \quad \text{and} \quad \dfrac{d\Delta\xi_2(\tau)}{d\tau} \ll \omega_\xi. \tag{7.2}$$

Then, for the direct electrostatic interaction the modes with $l = -l'$ dominate so we can approximate Eqs. (5.29)-(5.31), by expanding $\sin\eta_\mu(\tau)$ and $\sigma_\mu(\tau)$ out to $O(R\omega_{3,A})$, as

$$\begin{aligned}
E_{dir,0} &\approx \dfrac{2e^2}{\varepsilon_w l_c^2}\int_0^{L_B} d\tau \sum_{l,n,n',m,m'} (-1)^{n'} K_{n'-n}\left(R\hat{\kappa}_{l,n,n'}(\tau)\right)\exp(il(\xi_1(\tau)-\xi_2(\tau))) \\
&\zeta^{(1)}_{surf}\left(l,\hat{k}_{l,n,n'}(\tau)\left(\cos\left(\dfrac{\eta(s)}{2}\right) + \dfrac{R\omega_{3,A}(s)\sin\eta(s)}{4}\sin\left(\dfrac{\eta(s)}{2}\right)\right)\right) \\
&\zeta^{(2)}_{surf}\left(l,\hat{k}_{l,n,n'}(\tau)\left(\cos\left(\dfrac{\eta(s)}{2}\right) - \dfrac{R\omega_{3,A}(s)\sin\eta(s)}{4}\sin\left(\dfrac{\eta(s)}{2}\right)\right)\right) \\
&\chi_{n,n',m,m',l,-l}\left(\hat{k}_{l,n,n'}(\tau),\eta(s),R\omega_{3,A}(s)\right)
\end{aligned} \tag{7.3}$$

$$\begin{aligned}
E_{dir,1} + E_{dir,2} &\approx -\dfrac{e^2}{\varepsilon_w l_c^2}\int_0^{L_B} d\tau \sum_{l,n,n',m,m'} (-1)^{n'}\omega_{3,A}(s)(n-n')\cos\left(\dfrac{\eta(s)}{2}\right)\sin\eta(s) K_{n'-n}\left(R\hat{\kappa}_{l,n,n'}(\tau)\right) \\
&\chi_{n,n',m,m',l,-l}\left(\hat{k}_{l,n,n'}(\tau),\eta(s),R\omega_{3,A}(s)\right)\zeta^{(1)}_{surf}\left(l,\hat{k}_{l,n,n'}(\tau)\left(\cos\left(\dfrac{\eta(s)}{2}\right) + \dfrac{R\omega_{3,A}(s)\sin\eta(s)}{4}\sin\left(\dfrac{\eta(s)}{2}\right)\right)\right) \\
&\zeta'^{(2)}_{surf}\left(l,\hat{k}_{l,n,n'}(\tau)\left(\cos\left(\dfrac{\eta(s)}{2}\right) - \dfrac{R\omega_{3,A}(s)\sin\eta(s)}{4}\sin\left(\dfrac{\eta(s)}{2}\right)\right)\right)\exp(il(\xi_1(\tau)-\xi_2(\tau))),
\end{aligned} \tag{7.4}$$

where

$$\chi_{n,n',m,m',l,l'}\left(\hat{k}_{l,n,n'}(\tau),\eta(s),R\omega_{3,A}(s)\right)=$$

$$I_{n-m}\left(\frac{a\hat{\kappa}_{l,n,n'}(\tau)}{2}\left(1-\cos\left(\frac{\eta(s)}{2}\right)-\frac{R\omega_{3,A}(s)\sin\eta(s)}{4}\sin\left(\frac{\eta(s)}{2}\right)\right)\right)$$

$$I_{m}\left(\frac{a\hat{\kappa}_{l,n,n'}(\tau)}{2}\left(1+\cos\left(\frac{\eta(s)}{2}\right)+\frac{R\omega_{3,A}(s)\sin\eta(s)}{4}\sin\left(\frac{\eta(s)}{2}\right)\right)\right)$$

$$I_{n'-m'}\left(\frac{a\hat{\kappa}_{l,n,n'}(\tau)}{2}\left(1-\cos\left(\frac{\eta(s)}{2}\right)+\frac{R\omega_{3,A}(s)\sin\eta(s)}{4}\sin\left(\frac{\eta(s)}{2}\right)\right)\right)$$

$$I_{m'}\left(\frac{a\hat{\kappa}_{l,n,n'}(\tau)}{2}\left(1+\cos\left(\frac{\eta(s)}{2}\right)-\frac{R\omega_{3,A}(s)\sin\eta(s)}{4}\sin\left(\frac{\eta(s)}{2}\right)\right)\right)$$

$$J_{2m-l-n}\left(a\hat{k}_{l,n,n'}(\tau)\left(\sin\left(\frac{\eta(s)}{2}\right)-\frac{R\omega_{3,A}(s)}{4}\cos\left(\frac{\eta(s)}{2}\right)\sin(\eta(s))\right)\right)$$

$$J_{n'-2m'-l'}\left(a\hat{k}_{l,n,n'}(\tau)\left(\sin\left(\frac{\eta(s)}{2}\right)+\frac{R\omega_{3,A}(s)}{4}\cos\left(\frac{\eta(s)}{2}\right)\sin(\eta(s))\right)\right), \tag{7.5}$$

$$\hat{\kappa}_{l,n,n'}(\tau)=\sqrt{\hat{k}_{l,n,n'}(\tau)^2+\kappa_D^2}, \tag{7.6}$$

$$\tilde{k}_{l,n,n'}(\tau)=\left(\frac{(n+n')}{2}\left(\frac{2(1-\cos\eta(s))}{R\sin\eta(s)}\right)-\frac{l}{2}\left(\frac{d\xi_1(\tau)}{d\tau}+\frac{d\xi_2(\tau)}{d\tau}\right)\right). \tag{7.7}$$

For the image charge interactions for $E_{img1}$ and $E_{img2}$ the modes $l=-j'$ and $l'=-j'$, repectively dominate if Eq. (7.2) is satisfied. Therefore from Eqs. (6.14)-(6.17) and (6.27)-(6.30), we obtain

$$E_{img1,0}=\frac{e^2}{\varepsilon_w l_c^2}\int_0^{L_B}d\tau\left(\cos\left(\frac{\eta(s)}{2}\right)^{-1}+R\omega_{A,3}(s)\cos\left(\frac{\eta(s)}{2}\right)\right)\sum_{l,l',n,n',m,m'}(-1)^{n'}K_{n'-n}\left(\hat{\kappa}_{l,n,n'}^{(1)}(\tau)R\right)$$

$$\chi_{n,n',m,m',l,l'}\left(\hat{k}_{l,n,n'}^{(1)}(\tau),\eta(s),R\omega_{3,A}(s)\right)\zeta_{surf0}^{(1)}\left(l,\hat{k}_{l,n,n'}(\tau)\left(\cos\left(\frac{\eta(s)}{2}\right)+\frac{R\omega_{3,A}(s)\sin\eta(s)}{4}\sin\left(\frac{\eta(s)}{2}\right)\right)\right)$$

$$\tilde{\zeta}_{surf1,0}^{(2)}\left(l',-l,-\hat{k}_{l,n,n'}^{(1)}(\tau)\left(\cos\left(\frac{\eta(s)}{2}\right)-\frac{R\omega_{3,A}(s)\sin\eta(s)}{4}\sin\left(\frac{\eta(s)}{2}\right)\right)\right), \tag{7.8}$$

$$E_{img1,1}\approx-\frac{e^2}{2\varepsilon_w l_c^2}\int_0^{L_B}d\tau\left(\cos\left(\frac{\eta(s)}{2}\right)^{-1}+\frac{R\omega_{A,3}(s)}{2}\cos\left(\frac{\eta(s)}{2}\right)\right)\sum_{l,l',n,n',m,m'}(-1)^{n'}\frac{(n-n')\sin\eta(\tau)}{R}$$

$$\chi_{n,n',m,m',l,l'}\left(\hat{k}_{l,n,n'}^{(1)}(\tau),\eta(s),R\omega_{3,A}(s)\right)\zeta_{surf0}^{\prime(1)}\left(l,-\hat{k}_{l,n,n'}(\tau)\left(\cos\left(\frac{\eta(s)}{2}\right)+\frac{R\omega_{3,A}(s)\sin\eta(s)}{4}\sin\left(\frac{\eta(s)}{2}\right)\right)\right)$$

$$\tilde{\zeta}_{surf1,0}^{(2)}\left(l',-l,-\hat{k}_{l,n,n'}^{(1)}(\tau)\left(\cos\left(\frac{\eta(s)}{2}\right)-\frac{R\omega_{3,A}(s)\sin\eta(s)}{4}\sin\left(\frac{\eta(s)}{2}\right)\right)\right)K_{n'-n}\left(\hat{\kappa}_{l,n,n'}^{(1)}(\tau)R\right), \tag{7.9}$$

$$E_{img1,2} \approx \frac{e^2}{2\varepsilon_w l_c^2} \int_0^{L_B} d\tau \left( \cos\left(\frac{\eta(s)}{2}\right)^{-1} + \frac{3R\omega_{A,3}(s)}{2}\cos\left(\frac{\eta(s)}{2}\right) \right) \sum_{l,l',n,n',m,m'} (-1)^{n'} \frac{(n-n')\sin\eta(\tau)}{R}$$

$$\chi_{n,n',m,m',l,l'}\left(\hat{k}_{l,n,n'}^{(1)}(\tau),\eta(s),R\omega_{3,A}(s)\right) \zeta_{surf\,0}^{(1)}\left(l,-\hat{k}_{l,n,n'}^{(1)}(\tau)\left(\cos\left(\frac{\eta(s)}{2}\right) + \frac{R\omega_{3,A}(s)\sin\eta(s)}{4}\sin\left(\frac{\eta(s)}{2}\right)\right)\right)$$

$$\tilde{\zeta}_{surf\,1,0}^{\prime(2)}\left(l',-l,-\hat{k}_{l,n,n'}^{(1)}(\tau)\left(\cos\left(\frac{\eta(s)}{2}\right) - \frac{R\omega_{3,A}(s)\sin\eta(s)}{4}\sin\left(\frac{\eta(s)}{2}\right)\right)\right) K_{n'-n}\left(\hat{k}_{l,n,n'}^{(1)}(\tau)R\right),$$

(7.10)

$$E_{img1,3} = \frac{e^2}{\varepsilon_w l_c^2} \int_0^{L_B} d\tau \left( \cos\left(\frac{\eta(s)}{2}\right)^{-1} + R\omega_{A,3}(s)\cos\left(\frac{\eta(s)}{2}\right) \right) \sum_{l,l',n,n',m,m'} (-1)^{n'} \sin\eta(\tau) K_{n'-n}\left(\hat{k}_{l,n,n'}^{(1)}(\tau)R\right)$$

$$\chi_{n,n',m,m',l,l'}\left(\hat{k}_{l,n,n'}^{(1)}(\tau),\eta(s),R\omega_{3,A}(s)\right) \zeta_{surf\,0}^{(1)}\left(l,-\hat{k}_{l,n,n'}^{(1)}(\tau)\left(\cos\left(\frac{\eta(s)}{2}\right) + \frac{R\omega_{3,A}(s)\sin\eta(s)}{4}\sin\left(\frac{\eta(s)}{2}\right)\right)\right)$$

$$\tilde{\zeta}_{surf\,1,1}^{(2)}\left(l',-l,-\hat{k}_{l,n,n'}^{(1)}(\tau)\left(\cos\left(\frac{\eta(s)}{2}\right) - \frac{R\omega_{3,A}(s)\sin\eta(s)}{4}\sin\left(\frac{\eta(s)}{2}\right)\right)\right),$$

(7.11)

$$E_{img2,0} \approx \frac{e^2}{\varepsilon_w l_c^2} \int_0^{L_B} d\tau \left( \cos\left(\frac{\eta(s)}{2}\right)^{-1} - R\omega_{A,3}(s)\cos\left(\frac{\eta(s)}{2}\right) \right) \sum_{l,l',n,n',m,m'} (-1)^{n'} K_{n'-n}\left(\hat{\kappa}_{l',n,n'}^{(2)}(\tau)R\right)$$

$$\chi_{n,n',m,m',l,l'}\left(\hat{k}_{l,n,n'}^{(2)}(\tau),\eta(s),R\omega_{3,A}(s)\right) \tilde{\zeta}_{surf\,1,0}^{(1)}\left(l,-l',\hat{k}_{l',n,n'}^{(2)}(\tau)\left(\cos\left(\frac{\eta(s)}{2}\right) + \frac{R\omega_{3,A}(s)\sin\eta(s)}{4}\sin\left(\frac{\eta(s)}{2}\right)\right)\right)$$

$$\zeta_{surf\,0}^{(2)}\left(l',\hat{k}_{l',n,n'}^{(2)}(\tau)\left(\cos\left(\frac{\eta(s)}{2}\right) - \frac{R\omega_{3,A}(s)\sin\eta(s)}{4}\sin\left(\frac{\eta(s)}{2}\right)\right)\right),$$

(7.12)

$$E_{img2,1} \approx \frac{e^2}{2\varepsilon_w l_c^2} \int_0^{L_B} d\tau \left( \cos\left(\frac{\eta(s)}{2}\right)^{-1} - \frac{3R\omega_{A,3}(s)}{2}\cos\left(\frac{\eta(s)}{2}\right) \right) \sum_{l,l',n,n',m,m'} (-1)^{n'} \frac{(n-n')\sin\eta(\tau)}{R}$$

$$\chi_{n,n',m,m',l,l'}\left(\hat{k}_{l,n,n'}^{(2)}(\tau),\eta(s),R\omega_{3,A}(s)\right) \tilde{\zeta}_{surf\,1,0}^{\prime(1)}\left(l,-l',\hat{k}_{l',n,n'}^{(2)}(\tau)\left(\cos\left(\frac{\eta(s)}{2}\right) + \frac{R\omega_{3,A}(s)\sin\eta(s)}{4}\sin\left(\frac{\eta(s)}{2}\right)\right)\right)$$

$$\zeta_{surf\,0}^{(2)}\left(l',\hat{k}_{l',n,n'}^{(2)}(\tau))\left(\cos\left(\frac{\eta(s)}{2}\right) - \frac{R\omega_{3,A}(s)\sin\eta(s)}{4}\sin\left(\frac{\eta(s)}{2}\right)\right)\right) K_{n'-n}\left(\hat{\kappa}_{l',n,n'}^{(2)}(\tau)R\right),$$

(7.13)

$$E_{img\,2,2} = -\frac{e^2}{2\varepsilon_w l_c^2} \int_0^{L_B} d\tau \left( \cos\left(\frac{\eta(s)}{2}\right)^{-1} - \frac{R\omega_{A,3}(s)}{2} \cos\left(\frac{\eta(s)}{2}\right) \right) \sum_{l,l',n,n',m,m'} (-1)^{n'} \frac{(n-n')\sin\eta(\tau)}{R}$$

$$\chi_{n,n',m,m',l,l'}\left(\hat{k}^{(2)}_{l,n,n'}(\tau), \eta(s), R\omega_{3,A}(s)\right) \tilde{\zeta}^{(1)}_{surf\,1,0}\left( l, -l', \hat{k}^{(2)}_{l',n,n'}(\tau) \left( \cos\left(\frac{\eta(s)}{2}\right) + \frac{R\omega_{3,A}(s)\sin\eta(s)}{4} \sin\left(\frac{\eta(s)}{2}\right) \right) \right)$$

$$\zeta'^{(2)}_{surf\,0}\left( l', \hat{k}^{(2)}_{l',n,n'}(\tau) \left( \cos\left(\frac{\eta(s)}{2}\right) - \frac{R\omega_{3,A}(s)\sin\eta(s)}{4} \sin\left(\frac{\eta(s)}{2}\right) \right) \right) K_{n'-n}\left( \hat{\kappa}^{(2)}_{l',n,n'}(\tau) R \right),$$

(7.14)

$$E_{img\,2,3}(\tau,\tau') = \frac{e^2}{\varepsilon_w l_c^2} \int_0^{L_B} d\tau \left( \cos\left(\frac{\eta(s)}{2}\right)^{-1} - R\omega_{A,3}(s)\cos\left(\frac{\eta(s)}{2}\right) \right) \sum_{l,l',n,n',m,m'} (-1)^{n'} \sin\eta(\tau)$$

$$\chi_{n,n',m,m',l,l'}\left(\hat{k}^{(2)}_{l,n,n'}(\tau), \eta(s), R\omega_{3,A}(s)\right) \tilde{\zeta}^{(1)}_{surf\,1,1}\left( l, -l', \hat{k}^{(2)}_{l',n,n'}(\tau) \left( \cos\left(\frac{\eta(s)}{2}\right) + \frac{R\omega_{3,A}(s)\sin\eta(s)}{4} \sin\left(\frac{\eta(s)}{2}\right) \right) \right)$$

$$\zeta^{(2)}_{surf\,0}\left( l', \hat{k}^{(2)}_{l',n,n'}(\tau) \left( \cos\left(\frac{\eta(s)}{2}\right) - \frac{R\omega_{3,A}(s)\sin\eta(s)}{4} \sin\left(\frac{\eta(s)}{2}\right) \right) \right) K_{n'-n}\left( \hat{\kappa}^{(2)}_{l',n,n'}(\tau) R \right),$$

(7.15)

where

$$\hat{\kappa}^{(\mu)}_{l',n,n'}(\tau) = \sqrt{\left(\hat{k}^{(\mu)}_{l',n,n'}(\tau)\right)^2 + \kappa_D^2},$$
(7.16)

$$\hat{k}^{(\mu)}_{l,n,n'}(\tau) = -l\delta_\mu \frac{d\xi_1}{d\tau} + (n'+n)\frac{(1-\cos\eta(s))}{R\sin\eta(s)}.$$
(7.17)

## 8. Small angle approximation for next to leading order images

Next we consider the small angle approximations of Eqs (4.90), (4.91), (4.102) and (4.103) for the image charges. We calculate the leading and next to leading orders in a power series in $\sin\eta(s)$. First of all we can make the approximations

$$I_{n'-m'}\left( \frac{a\sqrt{k_z^2 + \kappa_D^2}}{2}(1-\cos\eta(s_0)) \right) \approx \delta_{n'-m'} + O(\sin\eta(s_0)),$$
(8.1)

$$J_{n'-l}(ak_z \sin\eta(s_0)) \approx \delta_{n'-l,0} + \frac{ak_z \sin\eta(s_0)}{2}\left( \delta_{n'-l,1} - \delta_{n'-l,-1} \right).$$
(8.2)

Then we can write Eq. (4.90) in the form

$$\zeta^{(2)}_{surf\,1,0}(n,k_z) \approx \zeta^{(2)}_{surf\,1,0,0}(n,k_z) + \sin\eta(s_0)\zeta^{(2)}_{surf\,1,0,1}(n,k_z),$$
(8.3)

where

$$\zeta^{(2)}_{surf,1,0,0}(n,k_z) \approx -\frac{I'_n\left(a\sqrt{k_z^2+\kappa_D^2}\right)}{K'_n\left(a\sqrt{k_z^2+\kappa_D^2}\right)I_n\left(a\sqrt{k_z^2+\kappa_D^2}\right)}\sum_l (-1)^n$$

$$I_l\left(a\sqrt{k_z^2+\kappa_D^2}\right)K_{n-l}\left(R\sqrt{k_z^2+\kappa_D^2}\right)\zeta^{(1)}_{surf,0}(l,k_z\cos\eta(s_0))\exp(ilt_0) \equiv \sum_l \tilde{\zeta}^{(2)}_{surf,1,0,0}(n,l,k_z)\exp(ilt_0),$$

(8.4)

$$\zeta^{(2)}_{surf,1,0,1}(n,k_z) = \frac{ak_z I'_n\left(a\sqrt{k_z^2+\kappa_D^2}\right)}{2K'_n\left(a\sqrt{k_z^2+\kappa_D^2}\right)I_n\left(a\sqrt{k_z^2+\kappa_D^2}\right)}\sum_{l,n'} (-1)^n$$

$$\left(I_{l-1}\left(a\sqrt{k_z^2+\kappa_D^2}\right)K_{n-l+1}\left(R\sqrt{k_z^2+\kappa_D^2}\right) - I_{l+1}\left(a\sqrt{k_z^2+\kappa_D^2}\right)K_{n-l-1}\left(R\sqrt{k_z^2+\kappa_D^2}\right)\right)$$

(8.5)

$$\zeta^{(1)}_{surf,0}(l,k_z\cos\eta(s_0))\exp(ilt_0) \equiv \sum_l \tilde{\zeta}^{(2)}_{surf,1,0,1}(n,l,k_z)\exp(ilt_0),$$

and Eq. (4.102) in the form

$$\zeta^{(1)}_{surf,1,0}(n,k_z) \approx \zeta^{(1)}_{surf,1,0,0}(n,k_z) + \sin\eta(s_0)\zeta^{(1)}_{surf,1,0,1}(n,k_z) \tag{8.6}$$

$$\zeta^{(1)}_{surf\,1,0,0}(n,k_z) = -\frac{I'_n\left(a\sqrt{k_z^2+\kappa_D^2}\right)}{I_n\left(a\sqrt{k_z^2+\kappa_D^2}\right)K'_n\left(a\sqrt{k_z^2+\kappa_D^2}\right)}\sum_l (-1)^l K_{n-l}\left(R\sqrt{k_z^2+\kappa_D^2}\right)I_l\left(a\sqrt{k_z^2+\kappa_D^2}\right)$$

$$\zeta^{(2)}_{surf,0}(l,k_z\cos\eta(s_0))\exp(ilt_0) \equiv \sum_l \tilde{\zeta}^{(1)}_{surf,1,0,0}(n,l,k_z)\exp(ilt_0),$$

(8.7)

$$\zeta^{(1)}_{surf\,1,0,1}(n,k_z) = -\frac{I'_n\left(a\sqrt{k_z^2+\kappa_D^2}\right)ak_z}{I_n\left(a\sqrt{k_z^2+\kappa_D^2}\right)K'_n\left(a\sqrt{k_z^2+\kappa_D^2}\right)}\sum_l (-1)^l$$

$$\left(K_{n-l-1}\left(R\sqrt{k_z^2+\kappa_D^2}\right)I_{l+1}\left(a\sqrt{k_z^2+\kappa_D^2}\right) - K_{n-l+1}\left(R\sqrt{k_z^2+\kappa_D^2}\right)I_{l-1}\left(a\sqrt{k_z^2+\kappa_D^2}\right)\right)$$

(8.8)

$$\zeta^{(2)}_{surf,0}(l,k_z\cos\eta(s_0))\exp(ilt_0) \equiv \sum_l \tilde{\zeta}^{(1)}_{surf,1,0,0}(n,l,k_z)\exp(ilt_0).$$

Also, we can write for Eq. (4.91)

$$\zeta^{(2)}_{surf,1,1}(n,k_z) \approx \frac{I'_n\left(a\sqrt{k_z^2+\kappa_D^2}\right)}{I_n\left(a\sqrt{k_z^2+\kappa_D^2}\right)K'_n\left(a\sqrt{k_z^2+\kappa_D^2}\right)} \sum_l (-1)^n(n-l)I_l\left(a\sqrt{k_z^2+\kappa_D^2}\right)K_{n-l}\left(R\sqrt{k_z^2+\kappa_D^2}\right)$$

$$\frac{\zeta'^{(1)}_{surf,0}(l,k_z\cos\eta(s_0))}{R}\exp(ilt_0) \equiv \sum_l \tilde{\zeta}^{(2)}_{surf,1,1}(n,l,k_z)\exp(ilt_0),$$

(8.9)

Finally, we can write for Eq. (4.103)

$$\zeta^{(1)}_{surf,1,1}(n,k_z) \approx \frac{I'_n\left(a\sqrt{k_z^2+\kappa_D^2}\right)}{I_n\left(a\sqrt{k_z^2+\kappa_D^2}\right)K'_n\left(a\sqrt{k_z^2+\kappa_D^2}\right)} \sum_l (-1)^l(n-l)I_l\left(a\sqrt{k_z^2+\kappa_D^2}\right)K_{n-l}\left(R\sqrt{k_z^2+\kappa_D^2}\right)$$

$$\frac{\zeta'^{(2)}_{surf,0}(l,k_z\cos\eta(s_0))}{R}\exp(ilt_0) \equiv \sum_l \tilde{\zeta}^{(1)}_{surf,1,1}(n,l,k_z)\exp(ilt_0),$$

(8.10)

## 9. Small tilt angle expansion for direct interaction in diagonal mode approximation

First of all to $O(\sin\eta(s))$ we may approximate Eq. (7.5) with

$$\chi_{n,n',m,m',l,l'}\left(\hat{k}_{l,n,n'}(\tau),\eta(s),R\omega_{3,A}(s)\right) = \delta_{m,n}\delta_{m',n'}I_n\left(a\hat{k}_{l,n,n'}(\tau)\right)I_{n'}\left(a\hat{k}_{l,n,n'}(\tau)\right)$$

$$\left(\delta_{n-l,0}\delta_{n'+l',0} + \delta_{n'+l',0}\frac{a\hat{k}_{l,n,n'}(\tau)\sin\eta(s)}{4}\left(1-\frac{R\omega_{3,A}(s)}{2}\right)(\delta_{n-l,1}-\delta_{n-l,-1})\right.$$

$$\left.\delta_{n-l,0}\frac{a\hat{k}_{l,n,n'}(\tau)\sin\eta(s)}{4}\left(1+\frac{R\omega_{3,A}(s)}{2}\right)(\delta_{n'+l',-1}-\delta_{n'+l',1})\right).$$

(9.1)

Therefore, by substituting in Eq. (9.1), we can write for Eq. (7.3)

$$E_{dir,0} = E_{dir,0,0} + E_{dir,0,1},$$

(9.2)

$$E_{dir,0} \approx \frac{2e^2}{\varepsilon_w l_c^2}\int_0^{L_B} d\tau \sum_{l,n,n'} (-1)^{n'}K_0\left(R\bar{\kappa}_n(\tau)\right)\exp(in(\xi_1(\tau)-\xi_2(\tau)))\zeta^{(1)}_{surf}\left(l,\bar{k}_n(\tau)\right)\zeta^{(2)}_{surf}\left(l,\bar{k}_n(\tau)\right)$$

(9.3)

$$E_{dir,0,1} \approx -\frac{2e^2}{\varepsilon_w l_c^2} \int_0^{L_B} d\tau \sum_n (-1)^n n \bar{k}_n(\tau) K_1(R\bar{\kappa}_n(\tau)) \exp(in(\xi_1(\tau)-\xi_2(\tau))) I_{n'}(R\bar{\kappa}_n(\tau)) I_n(R\bar{\kappa}_n(\tau))$$
$$\zeta_{surf}^{(1)}(n,\bar{k}_n(\tau)) \zeta_{surf}^{(2)}(n,\bar{k}_n(\tau))$$
(9.4)

In deriving Eqs.(9.3) and (9.4) we have also made the approximation, valid to $O(\sin\eta(s))$

$$\hat{k}_{l,n,n-1}(\tau) \approx \hat{k}_{l,n,n+1}(\tau) \approx \hat{k}_{n,n,n}(\tau) \approx \bar{k}_n(\tau) = -\frac{n}{2}(\tilde{\omega}_{1,1}(\tau)+\tilde{\omega}_{2,1}(\tau)) - \frac{nR\omega_{3,A}(s)}{4}\left(\frac{d\xi_1(\tau)}{d\tau} - \frac{d\xi_2(\tau)}{d\tau}\right),$$
(9.5)

and

$$\bar{\kappa}_n(\tau) = \sqrt{\kappa_D^2 + \bar{k}_n(\tau)^2}.$$
(9.6)

To this order of approximation we neglect $E_{dir,1}+E_{dir,2}$. More explicitly, using Eqs. (4.67) and (4.94), as well as expanding out for small $R\omega_{3,A}(s)$, we can write:

$$E_{dir} \approx \bar{E}_{dir,0} + \bar{E}_{dir,\omega} + \bar{E}_{dir,\eta},$$
(9.7)

where

$$\bar{E}_{dir,0} \approx \frac{2e^2}{\varepsilon_w l_c^2} \int_0^{L_B} d\tau \sum_n (-1)^n \frac{K_0(R\bar{\kappa}_n(\tau))}{a^2 \bar{\kappa}_n(\tau)^2 K_n'(a\bar{\kappa}_n(\tau))^2} \cos(n(\xi_1(\tau)-\xi_2(\tau))),$$
(9.8)

$$\bar{E}_{dir,\omega} \approx -\frac{e^2 R}{4\varepsilon_w l_c^2} \int_0^{L_B} d\tau \omega_{A,3}(s) \sum_n (-1)^n \left(\frac{d\xi_1(\tau)}{d\tau} - \frac{d\xi_2(\tau)}{d\tau}\right) \cos(n(\xi_1(\tau)-\xi_2(\tau)))$$
$$\frac{n^2(\tilde{\omega}_{1,1}(\tau)+\tilde{\omega}_{2,1}(\tau))}{\tilde{\kappa}_n(\tau)} \left(\frac{RK_1(R\tilde{\kappa}_n(\tau))}{a^2\tilde{\kappa}_n(\tau)^2 K_n'(a\tilde{\kappa}_n(\tau))^2} + \frac{2K_0(R\tilde{\kappa}_n(\tau))}{a^2\tilde{\kappa}_n(\tau)^2 K_n'(a\tilde{\kappa}_n(\tau))^2}\left(\frac{1}{\tilde{\kappa}_n(\tau)} + \frac{aK_n''(a\tilde{\kappa}_n(\tau))}{K_n'(a\tilde{\kappa}_n(\tau))}\right)\right)$$
(9.9)

$$E_{dir,\eta} \approx \frac{e^2}{\varepsilon_w l_c^2} \int_0^{L_B} d\tau \sin\eta(s) \sum_n (-1)^n \frac{n^2(\tilde{\omega}_{1,1}(\tau)+\tilde{\omega}_{2,1}(\tau))K_1(R\tilde{\kappa}_n(\tau))}{a^2\tilde{\kappa}_n(\tau)^2 K_n'(a\tilde{\kappa}_n(\tau))^2} \cos(n(\xi_1(\tau)-\xi_2(\tau)))$$ (9.10)

## 10. Small tilt angle expansion for image charge interaction in diagonal mode approximation

### *10.1 Expanding out the energy*

Subsitution of Eq. (9.1) into Eqs. (7.8) and (7.12) allows us to write for small tilt angle

$$E_{img1,0} \approx \frac{e^2}{\varepsilon_w l_c^2} \int_0^{L_B} d\tau \left(1 + R\omega_{A,3}(s)\right) \sum_{l,l',n,n'} (-1)^{n'} K_{n'-n}\left(\hat{k}_{l,n,n'}^{(1)}(\tau)R\right) I_n\left(a\hat{k}_{l,n,n'}^{(1)}(\tau)\right) I_{n'}\left(a\hat{k}_{l,n,n'}^{(1)}(\tau)\right)$$

$$\left(\delta_{n,l}\delta_{-n',l'} + \frac{a\hat{k}_{l,n,n'}^{(1)}(\tau)\sin\eta(\tau)}{4}\left(\delta_{-n',l'}\left(\delta_{n,l+1} - \delta_{n,l-1}\right) + \delta_{n,l}\left(\delta_{-n',l'+1} - \delta_{-n',l'-1}\right)\right)\right)$$

$$\zeta_{surf\,0}^{(1)}(l, -\hat{k}_{l,n,n'}^{(1)}(\tau))\tilde{\zeta}_{surf\,1,0}^{(2)}(l', -l, -\hat{k}_{l,n,n'}^{(1)}(\tau)),$$

(10.1)

$$E_{img2,0} \approx \frac{e^2}{\varepsilon_w l_c^2} \int_0^{L_B} d\tau \left(1 - R\omega_{A,3}(s)\right) \sum_{l,l',n,n'} (-1)^{n'} K_{n'-n}\left(\hat{k}_{l,n,n'}^{(2)}(\tau)R\right) I_n\left(a\hat{k}_{l,n,n'}^{(2)}(\tau)\right) I_{n'}\left(a\hat{k}_{l,n,n'}^{(2)}(\tau)\right)$$

$$\left(\delta_{n,l}\delta_{-n',l'} + \frac{a\hat{k}_{l',n,n'}^{(2)}(\tau)\sin\eta(\tau)}{4}\left(\delta_{-n',l'}\left(\delta_{n,l+1} - \delta_{n,l-1}\right) + \delta_{n,l}\left(\delta_{-n',l'+1} - \delta_{-n',l'-1}\right)\right)\right)$$

$$\tilde{\zeta}_{surf\,1,0}^{(1)}(l, -l', \hat{k}_{l',n,n'}^{(2)}(\tau))\zeta_{surf\,0}^{(2)}(l', \hat{k}_{l,n,n'}^{(2)}(\tau)).$$

(10.2)

Also, for small tilt angle we may approximate

$$\hat{k}_{l,n,n'}^{(\mu)}(\tau) = -l\delta_\mu \tilde{\omega}_{\mu,1}(\tau) - l\frac{R\omega_{A,3}(\tau)}{2}\frac{d\xi_1(\tau)}{d\tau} - (2\delta_\mu l - n' - n)\frac{\sin\eta(\tau)}{2R}.$$  (10.3)

Using Eqs. (8.4),(8.5), (10.1) and (10.3), as well as expanding out in powers of $\sin\eta(s)$, we obtain for $E_{img1,0}$

$$E_{img1,0} = E_{img1,0,0} + E_{img1,0,1} + E_{img1,0,2} + E_{img1,0,3} + E_{img1,0,4} + E_{img1,0,5} + E_{img1,0,6},$$  (10.4)

$$E_{img1,0,0} = -\frac{e^2}{\varepsilon_w l_c^2} \int_0^{L_B} d\tau \left(1 + R\omega_{A,3}(s)\right) \sum_{n,n'} \frac{I_{n'}'\left(a\bar{\kappa}_n^{(1)}(\tau)\right)}{K_{n'}'\left(a\bar{\kappa}_n^{(1)}(\tau)\right)} K_{n'-n}\left(\bar{\kappa}_n^{(1)}(\tau)R\right) K_{n'-n}\left(\bar{\kappa}_n^{(1)}(\tau)R\right)$$

$$\zeta_{surf,0}^{(1)}(n, n\tilde{\omega}_{1,1}(\tau))\zeta_{surf,0}^{(1)}(n, n\tilde{\omega}_{1,1}(\tau)) I_n\left(a\bar{\kappa}_n^{(1)}(\tau)\right) I_n\left(a\bar{\kappa}_n^{(1)}(\tau)\right),$$

(10.5)

$$E_{img1,0,1} = -\frac{ae^2}{2\varepsilon_w l_c^2} \int_0^{L_B} d\tau \sum_{n,n'} n\tilde{\omega}_{1,1}(\tau)\sin\eta(\tau) K_{n'-n}\left(\bar{\kappa}_n^{(1)}(\tau)R\right) I_n\left(a\bar{\kappa}_n^{(1)}(\tau)\right)\frac{I_{n'}'\left(a\bar{\kappa}_n^{(1)}(\tau)\right)}{K_{n'}'\left(a\bar{\kappa}_n^{(1)}(\tau)\right)}$$

$$\left(I_{n-1}\left(a\bar{\kappa}_n^{(1)}(\tau)\right)K_{n'-n+1}\left(R\bar{\kappa}_n^{(1)}(\tau)\right) - I_{n+1}\left(a\bar{\kappa}_n^{(1)}(\tau)\right)K_{n'-n-1}\left(R\bar{\kappa}_n^{(1)}(\tau)\right)\right)$$

$$\zeta_{surf\,0}^{(1)}(n, n\tilde{\omega}_{1,1}(\tau))\zeta_{surf\,0}^{(1)}(n, n\tilde{\omega}_{1,1}(\tau)),$$

(10.6)

$$E_{img1,0,2} = \frac{ae^2}{4\varepsilon_w l_c^2} \int_0^{L_B} d\tau \sum_{n,n'} n\tilde{\omega}_{1,1}(\tau) \sin\eta(\tau) K_{n'-n}\left(R\bar{\kappa}_n^{(1)}(\tau)\right) I_n\left(a\bar{\kappa}_n^{(1)}(\tau)\right) \frac{I'_{n'}\left(a\bar{\kappa}_n^{(1)}(\tau)\right)}{K'_{n'}\left(a\bar{\kappa}_n^{(1)}(\tau)\right)}$$

$$\left(K_{n'-n-1}\left(\bar{\kappa}_n^{(1)}(\tau)R\right) I_{n+1}\left(a\bar{\kappa}_n^{(1)}(\tau)\right) - K_{n'-n+1}\left(\bar{\kappa}_n^{(1)}(\tau)R\right) I_{n-1}\left(a\bar{\kappa}_n^{(1)}(\tau)\right)\right) \quad (10.7)$$

$$\zeta_{surf\,0}^{(1)}(n, n\tilde{\omega}_{1,1}(\tau)) \zeta_{surf,0}^{(1)}(n, n\tilde{\omega}_{1,1}(\tau)),$$

$$E_{img1,0,3} = \frac{e^2}{4\varepsilon_w l_c^2} \int_0^{L_B} d\tau \sum_{n,n'} an\tilde{\omega}_{1,1}(\tau) \sin\eta(\tau) K_{n'-n}\left(R\bar{\kappa}_n^{(1)}(\tau)\right) \frac{I'_{n'}\left(a\bar{\kappa}_n^{(1)}(\tau)\right)}{K'_{n'}\left(a\bar{\kappa}_n^{(1)}(\tau)\right) I_{n'}\left(a\bar{\kappa}_n^{(1)}(\tau)\right)}$$

$$I_n\left(a\bar{\kappa}_n^{(1)}(\tau)\right) I_n\left(a\bar{\kappa}_n^{(1)}(\tau)\right) \left(K_{n'+1-n}\left(\bar{\kappa}_n^{(1)}(\tau)R\right) I_{n'+1}\left(a\bar{\kappa}_n^{(1)}(\tau)\right) - K_{n'-1-n}\left(\bar{\kappa}_n^{(1)}(\tau)R\right) I_{n'-1}\left(a\bar{\kappa}_n^{(1)}(\tau)\right)\right)$$

$$\zeta_{surf\,0}^{(1)}(n, n\tilde{\omega}_{1,1}(\tau)) \zeta_{surf,0}^{(1)}(n, n\tilde{\omega}_{1,1}(\tau)),$$

$$(10.8)$$

$$E_{img1,0,4} = -\frac{e^2}{4\varepsilon_w l_c^2} \int_0^{L_B} d\tau \sum_{n,n'} an\tilde{\omega}_{1,1} \sin\eta(\tau) K_{n'-n}\left(\bar{\kappa}_n^{(1)}(\tau)R\right)$$

$$\frac{I'_{n'}\left(a\bar{\kappa}_n^{(1)}(\tau)\right)}{K'_{n'}\left(a\bar{\kappa}_n^{(1)}(\tau)\right) I_{n'}\left(a\bar{\kappa}_n^{(1)}(\tau)\right)} I_n\left(a\bar{\kappa}_n^{(1)}(\tau)\right) \zeta_{surf\,0}^{(1)}(n, n\tilde{\omega}_{1,1}(\tau)) \zeta_{surf\,0}^{(1)}(n, n\tilde{\omega}_{1,1}(\tau)) \quad (10.9)$$

$$\left(I_n(a\bar{\kappa}_n^{(1)}(\tau))\left[I_{n'-1}(a\bar{\kappa}_n^{(1)}(\tau)) K_{n'-n-1}(a\bar{\kappa}_n^{(1)}(\tau)) - I_{n'+1}(a\bar{\kappa}_n^{(1)}(\tau)) K_{n'-n+1}(a\bar{\kappa}_n^{(1)}(\tau))\right]\right.$$

$$\left. + I_{n'}(a\bar{\kappa}_n^{(1)}(\tau))\left[I_{n+1}(a\bar{\kappa}_n^{(1)}(\tau)) K_{n'-n-1}(a\bar{\kappa}_n^{(1)}(\tau)) - I_{n-1}(a\bar{\kappa}_n^{(1)}(\tau)) K_{n'-n+1}(a\bar{\kappa}_n^{(1)}(\tau))\right]\right),$$

$$E_{img1,0,5} \approx -\frac{e^2}{\varepsilon_w l_c^2} \int_0^{L_B} d\tau \sum_{n,n'} \frac{(n-n')\sin\eta(\tau)}{R} K_{n'-n}\left(\bar{\kappa}_n^{(1)}(\tau)R\right) K_{n'-n}\left(\bar{\kappa}_n^{(1)}(\tau)R\right)$$

$$\frac{I'_{n'}\left(a\bar{\kappa}_n^{(1)}(\tau)\right)}{K'_{n'}\left(a\bar{\kappa}_n^{(1)}(\tau)\right)} I_n\left(a\bar{\kappa}_{n'}^{(1)}(\tau)\right) I_n\left(a\bar{\kappa}_{n'}^{(1)}(\tau)\right) \zeta_{surf\,0}^{\prime(1)}(n, n\tilde{\omega}_{1,1}(\tau)) \zeta_{surf\,0}^{(1)}(n, n\tilde{\omega}_{1,1}(\tau))$$

$$-\frac{e^2}{2\varepsilon_w l_c^2} \int_0^{L_B} d\tau \sum_{n,n'} \frac{(n-n')\sin\eta(\tau)}{R} K_{n'-n}\left(\bar{\kappa}_n^{(1)}(\tau)R\right) I_n\left(a\bar{\kappa}_n^{(1)}(\tau)\right) \frac{n\tilde{\omega}_{1,1}(\tau)}{\bar{\kappa}_n^{(1)}(\tau)}$$

$$\frac{d}{dx}\left[\frac{I'_{n'}(ax) I_n(ax) K_{n'-n}(Rx)}{K'_{n'}(ax) I_{n'}(ax)}\right]_{x=\bar{\kappa}_n^{(1)}(\tau)} I_{n'}\left(a\bar{\kappa}_n^{(1)}(\tau)\right) \zeta_{surf\,0}^{(1)}(n, n\tilde{\omega}_{1,1}(\tau)) \zeta_{surf\,0}^{(1)}(n, n\tilde{\omega}_{1,1}(\tau)),$$

$$(10.11)$$

$$E_{img1,0,6} = -\frac{e^2}{\varepsilon_w l_c^2} \int_0^{L_B} d\tau \frac{\omega_{A,3}(\tau)R}{2} \frac{d\xi_1(\tau)}{d\tau} \sum_{n,n'} \left( \frac{n^2 \tilde{\omega}_{1,1}(\tau)}{\bar{\kappa}_n^{(1)}(\tau)} 2R K_{n'-n}\left(\bar{\kappa}_n^{(1)}(\tau)R\right) K'_{n'-n}\left(\bar{\kappa}_n^{(1)}(\tau)R\right) \frac{I'_{n'}\left(a\bar{\kappa}_n^{(1)}(\tau)\right)}{K'_{n'}\left(a\bar{\kappa}_n^{(1)}(\tau)\right)} \right.$$

$$\zeta_{surf,0}^{(1)}(n, n\tilde{\omega}_{1,1}(\tau))\zeta_{surf,0}^{(1)}(n, n\tilde{\omega}_{1,1}(\tau)) I_n\left(a\bar{\kappa}_n^{(1)}(\tau)\right) I_n\left(a\bar{\kappa}_n^{(1)}(\tau)\right) + n K_{n'-n}\left(\bar{\kappa}_n^{(1)}(\tau)R\right) K_{n'-n}\left(\bar{\kappa}_n^{(1)}(\tau)R\right)$$

$$\frac{d}{d(n\tilde{\omega}_{1,1}(\tau))} \left[ \frac{I'_{n'}\left(a\bar{\kappa}_n^{(1)}(\tau)\right)}{K'_{n'}\left(a\bar{\kappa}_n^{(1)}(\tau)\right)} \zeta_{surf,0}^{(1)}(n, n\tilde{\omega}_{1,1}(\tau))\zeta_{surf,0}^{(1)}(n, n\tilde{\omega}_{1,1}(\tau)) I_n\left(a\bar{\kappa}_n^{(1)}(\tau)\right) I_n\left(a\bar{\kappa}_n^{(1)}(\tau)\right) \right] \right),$$
(10.12)

where $\bar{\kappa}_n^{(1)}(\tau) = \sqrt{n^2 \tilde{\omega}_{1,1}(\tau)^2 + \kappa_D^2}$. In obtaining Eq. (10.9) we have used the identity

$$\frac{n-n'}{R\bar{\kappa}_n^{(1)}(\tau)} \left( a \left( I'_n\left(a\bar{\kappa}_n^{(1)}(\tau)\right) I_{n'}\left(a\bar{\kappa}_n^{(1)}(\tau)\right) + I_n\left(a\bar{\kappa}_n^{(1)}(\tau)\right) I'_{n'}\left(a\bar{\kappa}_n^{(1)}(\tau)\right) \right) K_{n'-n}\left(\bar{\kappa}_n^{(1)}(\tau)R\right) \right.$$

$$+ R I_{n'}\left(a\bar{\kappa}_n^{(1)}(\tau)\right) I_n\left(a\bar{\kappa}_n^{(1)}(\tau)\right) K'_{n'-n}\left(\bar{\kappa}_n^{(1)}(\tau)R\right) \right)$$
(10.13)

$$\equiv \frac{a}{2} \left( I_n(a\bar{\kappa}_n^{(1)}(\tau)) \left[ I_{n'-1}(a\bar{\kappa}_n^{(1)}(\tau)) K_{n'-n-1}(a\bar{\kappa}_n^{(1)}(\tau)) - I_{n'+1}(a\bar{\kappa}_n^{(1)}(\tau)) K_{n'-n+1}(a\bar{\kappa}_n^{(1)}(\tau)) \right] \right.$$

$$+ I_{n'}(a\bar{\kappa}_n^{(1)}(\tau)) \left[ I_{n+1}(a\bar{\kappa}_n^{(1)}(\tau)) K_{n'-n-1}(a\bar{\kappa}_n^{(1)}(\tau)) - I_{n-1}(a\bar{\kappa}_n^{(1)}(\tau)) K_{n'-n+1}(a\bar{\kappa}_n^{(1)}(\tau)) \right] \right).$$

We can also obtain, using Eqs. (8.7), (8.8), (10.2) and (10.3),

$$E_{img2,0} = E_{img2,0,0} + E_{img2,0,1} + E_{img2,0,2} + E_{img2,0,3} + E_{img2,0,4} + E_{img2,0,5} + E_{img2,0,6},$$
(10.14)

$$E_{img2,0,0} = -\frac{e^2}{\varepsilon_w l_c^2} \int_0^{L_B} d\tau \left(1 - R\omega_{A,3}(s)\right) \sum_{n,n'} \frac{I'_n\left(a\bar{\kappa}_{n'}^{(2)}(\tau)\right)}{K'_n\left(a\bar{\kappa}_{n'}^{(2)}(\tau)\right)} K_{n'-n}\left(\bar{\kappa}_{n'}^{(2)}(\tau)R\right) K_{n'-n}\left(\bar{\kappa}_{n'}^{(2)}(\tau)R\right)$$
(10.15)

$$\zeta_{surf,0}^{(2)}(n, n'\tilde{\omega}_{2,1}(\tau))\zeta_{surf,0}^{(2)}(n, n'\tilde{\omega}_{2,1}(\tau)) I_n\left(a\bar{\kappa}_{n'}^{(2)}(\tau)\right) I_n\left(a\bar{\kappa}_{n'}^{(2)}(\tau)\right),$$

$$E_{img2,0,1} = \frac{e^2 a}{2\varepsilon_w l_c^2} \int_0^{L_B} d\tau \sum_{n,n'} n'\tilde{\omega}_{2,1}(\tau) \sin\eta(\tau) K_{n'-n}\left(\bar{\kappa}_{n'}^{(2)}(\tau)R\right) \frac{I'_n\left(a\bar{\kappa}_{n'}^{(2)}(\tau)\right)}{K'_n\left(a\bar{\kappa}_{n'}^{(2)}(\tau)\right)}$$

$$I_{n'}\left(a\bar{\kappa}_{n'}^{(2)}(\tau)\right)\left(K_{n-n'-1}\left(R\bar{\kappa}_{n'}^{(2)}(\tau)\right) I_{n'+1}\left(a\bar{\kappa}_{n'}^{(2)}(\tau)\right) - K_{n-n'+1}\left(R\bar{\kappa}_{n'}^{(2)}(\tau)\right) I_{n'-1}\left(a\bar{\kappa}_{n'}^{(2)}(\tau)\right)\right)$$
(10.16)

$$\zeta_{surf0}^{(2)}(n', n'\tilde{\omega}_{2,1}(\tau))\zeta_{surf0}^{(2)}(n', n'\tilde{\omega}_{2,1}(\tau)),$$

$$E_{img2,0,2} = \frac{e^2}{4\varepsilon_w l_c^2} \int_0^{L_B} d\tau \sum_{n,n'} an'\tilde{\omega}_{2,1}(\tau) \sin\eta(\tau) K_{n-n'}\left(R\bar{\kappa}_{n'}^{(2)}(\tau)\right) \frac{I'_n\left(a\bar{\kappa}_{n'}^{(2)}(\tau)\right)}{I_n\left(a\bar{\kappa}_{n'}^{(2)}(\tau)\right) K'_n\left(a\bar{\kappa}_{n'}^{(2)}(\tau)\right)}$$

$$\left(K_{n'-n-1}\left(\bar{\kappa}_{n'}^{(2)}(\tau)R\right) I_{n+1}\left(a\bar{\kappa}_{n'}^{(2)}(\tau)\right) - K_{n'-n+1}\left(\bar{\kappa}_{n'}^{(2)}(\tau)R\right) I_{n-1}\left(a\bar{\kappa}_{n'}^{(2)}(\tau)\right)\right) I_{n'}\left(a\bar{\kappa}_{n'}^{(2)}(\tau)\right) I_{n'}\left(a\bar{\kappa}_{n'}^{(2)}(\tau)\right)$$

$$\zeta_{surf,0}^{(2)}(n', n'\tilde{\omega}_{2,1}(\tau))\zeta_{surf0}^{(2)}(-n', n'\tilde{\omega}_{2,1}(\tau)),$$

(10.17)

$$E_{img\,2,0,3} = \frac{e^2}{4\varepsilon_w l_c^2} \int_0^{L_B} d\tau \sum_{n,n'} an'\tilde{\omega}_{2,1}(\tau)\sin\eta(\tau) K_{n-n'}\left(R\bar{\kappa}_{n'}^{(2)}(\tau)\right) I_{n'}\left(a\bar{\kappa}_{n'}^{(2)}(\tau)\right) \frac{I_n'\left(a\bar{\kappa}_{n'}^{(2)}(\tau)\right)}{K_n'\left(a\bar{\kappa}_{n'}^{(2)}(\tau)\right)}$$

$$\left(K_{n'-n+1}\left(\bar{\kappa}_{n'}^{(2)}(\tau)R\right)I_{n'+1}\left(a\bar{\kappa}_{n'}^{(2)}(\tau)\right) - K_{n'-n-1}\left(\bar{\kappa}_{n'}^{(2)}(\tau)R\right)I_{n'-1}\left(a\bar{\kappa}_{n'}^{(2)}(\tau)\right)\right)$$

$$\zeta_{surf\,0}^{(2)}(n',n'\tilde{\omega}_{2,1}(\tau))\zeta_{surf\,0}^{(2)}(n',n'\tilde{\omega}_{2,1}(\tau)),$$

(10.18)

$$E_{img\,2,0,4} = \frac{e^2}{4\varepsilon_w l_c^2} \int_0^{L_B} d\tau \sum_{n,n'} \sin\eta(\tau) an'\tilde{\omega}_{2,1}(\tau) K_{n-n'}\left(R\bar{\kappa}_{n'}^{(2)}(\tau)\right)$$

$$\frac{I_n'\left(a\bar{\kappa}_{n'}^{(2)}(\tau)\right)}{I_n\left(a\bar{\kappa}_{n'}^{(2)}(\tau)\right) K_n'\left(a\bar{\kappa}_{n'}^{(2)}(\tau)\right)} I_{n'}\left(a\bar{\kappa}_{n'}^{(2)}(\tau)\right) \zeta_{surf\,0}^{(2)}(n',n'\tilde{\omega}_{2,1}(\tau))\zeta_{surf,0}^{(2)}(n',n'\tilde{\omega}_{2,1}(\tau))$$

$$\left(I_n(a\bar{\kappa}_{n'}^{(2)}(\tau))\left[I_{n'-1}(a\bar{\kappa}_{n'}^{(2)}(\tau))K_{n'-n-1}(a\bar{\kappa}_{n'}^{(2)}(\tau)) - I_{n'+1}(a\bar{\kappa}_{n'}^{(2)}(\tau))K_{n'-n+1}(a\bar{\kappa}_{n'}^{(2)}(\tau))\right]\right.$$

$$\left. + I_{n'}(a\bar{\kappa}_n^{(2)}(\tau))\left[I_{n+1}(a\bar{\kappa}_{n'}^{(2)}(\tau))K_{n'-n-1}(a\bar{\kappa}_{n'}^{(2)}(\tau)) - I_{n-1}(a\bar{\kappa}_{n'}^{(2)}(\tau))K_{n'-n+1}(a\bar{\kappa}_{n'}^{(2)}(\tau))\right]\right),$$

(10.19)

$$E_{img\,2,0,5} = \frac{e^2}{\varepsilon_w l_c^2} \int_0^{L_B} d\tau \sum_{n,n'} \frac{(n-n')\sin\eta(\tau)}{R} K_{n'-n}\left(\bar{\kappa}_{n'}^{(2)}(\tau)R\right) K_{n'-n}\left(\bar{\kappa}_n^{(2)}(\tau)R\right)$$

$$\frac{I_n'\left(a\bar{\kappa}_{n'}^{(2)}(\tau)\right)}{K_n'\left(a\bar{\kappa}_{n'}^{(2)}(\tau)\right)} I_{n'}\left(a\bar{\kappa}_{n'}^{(2)}(\tau)\right) I_{n'}\left(a\bar{\kappa}_{n'}^{(2)}(\tau)\right) \zeta_{surf\,0}^{(2)}(n',n'\tilde{\omega}_{2,1}(\tau))\zeta_{surf,0}'^{(2)}(n',n'\tilde{\omega}_{2,1}(\tau))$$

$$+ \frac{e^2}{2\varepsilon_w l_c^2} \int_0^{L_B} d\tau \sigma_1(\tau)^2 \sum_{n,n'} (n-n')\frac{\sin\eta(\tau)}{R} K_{n'-n}\left(\bar{\kappa}_{n'}^{(2)}(\tau)R\right) I_{n'}\left(a\bar{\kappa}_{n'}^{(2)}(\tau)\right)$$

$$\frac{n\tilde{\omega}_{2,1}(\tau)}{\bar{\kappa}_{n'}^{(2)}(\tau)} \frac{d}{dx}\left[\frac{I_n'(ax)K_{n-n'}(Rx)I_{n'}(ax)}{I_n(ax)K_n'(ax)}\right]_{x=\bar{\kappa}_{n'}^{(2)}(\tau)} I_n\left(a\bar{\kappa}_{n'}^{(2)}(\tau)\right) \zeta_{surf\,0}^{(2)}(n',n'\tilde{\omega}_{2,1}(\tau))\zeta_{surf\,0}^{(2)}(n',n'\tilde{\omega}_{2,1}(\tau)),$$

(10.20)

$$E_{img\,2,0,6} = \frac{e^2}{\varepsilon_w l_c^2} \int_0^{L_B} d\tau \frac{\omega_{A,3}(\tau)R}{2} \frac{d\xi_2(\tau)}{d\tau} \sum_{n,n'} \left(\frac{n^2\tilde{\omega}_{1,1}(\tau)}{\bar{\kappa}_n^{(1)}(\tau)} 2RK_{n'-n}\left(\bar{\kappa}_{n'}^{(2)}(\tau)R\right) K_{n'-n}'\left(\bar{\kappa}_{n'}^{(2)}(\tau)R\right) \frac{I_n'\left(a\bar{\kappa}_{n'}^{(2)}(\tau)\right)}{K_n'\left(a\bar{\kappa}_{n'}^{(2)}(\tau)\right)}\right.$$

$$\zeta_{surf,0}^{(1)}(n,n'\tilde{\omega}_{1,2}(\tau))\zeta_{surf,0}^{(1)}(n,n'\tilde{\omega}_{1,2}(\tau)) I_n\left(a\bar{\kappa}_{n'}^{(2)}(\tau)\right) I_n\left(a\bar{\kappa}_{n'}^{(2)}(\tau)\right) + nK_{n'-n}\left(\bar{\kappa}_{n'}^{(2)}(\tau)R\right) K_{n'-n}\left(\bar{\kappa}_{n'}^{(2)}(\tau)R\right)$$

$$\left.\frac{d}{d(n'\tilde{\omega}_{1,2}(\tau))}\left[\frac{I_{n'}'\left(a\bar{\kappa}_n^{(2)}(\tau)\right)}{K_{n'}'\left(a\bar{\kappa}_n^{(2)}(\tau)\right)}\zeta_{surf,0}^{(1)}(n,n'\tilde{\omega}_{1,2}(\tau))\zeta_{surf,0}^{(1)}(n,n'\tilde{\omega}_{1,2}(\tau))I_n\left(a\bar{\kappa}_{n'}^{(2)}(\tau)\right) I_n\left(a\bar{\kappa}_{n'}^{(2)}(\tau)\right)\right]\right),$$

(10.21)

where $\bar{\kappa}_{n'}^{(2)}(\tau) = \sqrt{n'^2\tilde{\omega}_{2,1}(\tau)^2 + \kappa_D^2}$. Also, for small $\sin\eta(s)$, Eqs (7.9)-(7.11) simplify so that

$$E_{img\,1,1} \approx \frac{e^2}{2\varepsilon_w l_c^2} \int_0^{L_B} d\tau \sum_{n,n'} \frac{(n-n')\sin\eta(\tau)}{R} K_{n'-n}\left(\bar{\kappa}_n^{(1)}(\tau)R\right) K_{n'-n}\left(\bar{\kappa}_n^{(1)}(\tau)R\right) \frac{I_{n'}'\left(a\bar{\kappa}_n^{(1)}(\tau)\right)}{K_{n'}'\left(a\bar{\kappa}_n^{(1)}(\tau)\right)}$$

$$I_n\left(a\bar{\kappa}_{n'}^{(1)}(\tau)\right) I_n\left(a\bar{\kappa}_{n'}^{(1)}(\tau)\right) \zeta_{surf\,0}'^{(1)}(n,n\tilde{\omega}_{1,1}(\tau))\zeta_{surf\,0}^{(1)}(n,n\tilde{\omega}_{1,1}(\tau)),$$

(10.22)

$$E_{img1,2} \approx -\frac{e^2}{2\varepsilon_w l_c^2} \int_0^{L_B} d\tau \sum_{n,n'} \frac{(n-n')\sin\eta(\tau)}{R} K_{n'-n}\left(\bar{\kappa}_n^{(1)}(\tau)R\right) K_{n'-n}\left(\bar{\kappa}_n^{(1)}(\tau)R\right) \frac{I_{n'}'\left(a\bar{\kappa}_n^{(1)}(\tau)\right)}{K_{n'}'\left(a\bar{\kappa}_n^{(1)}(\tau)\right)}$$

$$I_n\left(a\bar{\kappa}_{n'}^{(1)}(\tau)\right) I_n\left(a\bar{\kappa}_{n'}^{(1)}(\tau)\right) \zeta_{surf\,0}^{\prime(1)}(n,n\tilde{\omega}_{1,1}(\tau))\zeta_{surf\,0}^{(1)}(n,n\tilde{\omega}_{1,1}(\tau))$$

$$-\frac{e^2}{2\varepsilon_w l_c^2} \int_0^{L_B} d\tau \sum_{n,n'} \frac{(n-n')\sin\eta_2(\tau)}{R} K_{n'-n}\left(\bar{\kappa}_n^{(1)}(\tau)R\right) I_n\left(a\bar{\kappa}_n^{(1)}(\tau)\right) \frac{n\tilde{\omega}_{1,1}(\tau)}{\bar{\kappa}_n^{(1)}(\tau)}$$

$$\frac{d}{dx}\left[\frac{I_{n'}'(ax)I_n(ax)K_{n'-n}(Rx)}{K_{n'}'(ax)I_{n'}(ax)}\right]_{x=\bar{\kappa}_n^{(1)}(\tau)} I_{n'}\left(a\bar{\kappa}_n^{(1)}(\tau)\right) \zeta_{surf\,0}^{(1)}(n,n\tilde{\omega}_{1,1}(\tau))\zeta_{surf\,0}^{(1)}(n,n\tilde{\omega}_{1,1}(\tau)),$$

(10.23)

$$E_{img1,3} \approx \frac{e^2}{\varepsilon_w l_c^2} \int_0^{L_B} d\tau \sum_{n,n'} \frac{(n-n')\sin(\eta(\tau))}{R} K_{n'-n}\left(\bar{\kappa}_n^{(1)}(\tau)R\right) K_{n'-n}\left(\bar{\kappa}_n^{(1)}(\tau)R\right)$$

$$\frac{I_{n'}'\left(a\bar{\kappa}_n^{(1)}(\tau)\right)}{K_{n'}'\left(a\bar{\kappa}_n^{(1)}(\tau)\right)} I_n\left(a\bar{\kappa}_n^{(1)}(\tau)\right) I_n\left(a\bar{\kappa}_n^{(1)}(\tau)\right) \zeta_{surf\,0}^{(1)}(n,n\tilde{\omega}_{1,1}(\tau))\zeta_{surf,0}^{\prime(1)}(n,n\tilde{\omega}_{1,1}(\tau)),$$

(10.24)

and Eqs. (7.13)-(7.15) simplify so that

$$E_{img2,1} \approx \frac{e^2}{2\varepsilon_w l_c^2} \int_0^{L_B} d\tau \sum_{n,n'} \frac{(n-n')\sin\eta(\tau)}{R} K_{n'-n}\left(\bar{\kappa}_{n'}^{(2)}(\tau)R\right) K_{n'-n}\left(\bar{\kappa}_{n'}^{(2)}(\tau)R\right) \frac{I_n'\left(a\bar{\kappa}_{n'}^{(2)}(\tau)\right)}{K_n'\left(a\bar{\kappa}_{n'}^{(2)}(\tau)\right)}$$

$$I_{n'}\left(a\bar{\kappa}_{n'}^{(2)}(\tau)\right) I_{n'}\left(a\bar{\kappa}_{n'}^{(2)}(\tau)\right) \zeta_{surf,0}^{\prime(2)}(n',n'\tilde{\omega}_{2,1}(\tau))\zeta_{surf\,0}^{(2)}(n',n'\tilde{\omega}_{2,1}(\tau))$$

$$+\frac{e^2}{2\varepsilon_w l_c^2} \int_0^{L_B} d\tau \sum_{n,n'} \frac{(n-n')\sin\eta(\tau)}{R} K_{n'-n}\left(\bar{\kappa}_{n'}^{(2)}(\tau)R\right) I_n\left(a\bar{\kappa}_{n'}^{(2)}(\tau)\right) \frac{n'\tilde{\omega}_{2,1}(\tau)}{\bar{\kappa}_{n'}^{(2)}(\tau)}$$

$$\frac{d}{dx}\left[\frac{I_n'(ax)K_{n'-n}(Rx)I_{n'}(ax)}{I_n(ax)K_n'(ax)}\right]_{x=\bar{\kappa}_{n'}^{(2)}(\tau)} I_{n'}\left(a\bar{\kappa}_{n'}^{(2)}(\tau)\right) \zeta_{surf,0}^{(2)}(n',n'\tilde{\omega}_{2,1}(\tau))\zeta_{surf\,0}^{(2)}(n',n'\tilde{\omega}_{2,1}(\tau)),$$

(10.25)

$$E_{img2,2} \approx -\frac{e^2}{2\varepsilon_w l_c^2} \int_0^{L_B} d\tau \sum_{n,n'} \frac{(n-n')\sin\eta(\tau)}{R} K_{n'-n}\left(\bar{\kappa}_{n'}^{(2)}(\tau)R\right) K_{n'-n}\left(\bar{\kappa}_{n'}^{(2)}(\tau)R\right) \frac{I_n'\left(a\bar{\kappa}_{n'}^{(2)}(\tau)\right)}{K_n'\left(a\bar{\kappa}_{n'}^{(2)}(\tau)\right)}$$

$$I_{n'}\left(a\bar{\kappa}_{n'}^{(2)}(\tau)\right) I_{n'}\left(a\bar{\kappa}_{n'}^{(2)}(\tau)\right) \zeta_{surf,0}^{(2)}(n',n'\tilde{\omega}_{2,1}(\tau))\zeta_{surf}^{\prime(2)}(n',n'\tilde{\omega}_{2,1}(\tau))),$$

(10.26)

$$E_{img2,3} \approx -\frac{e^2}{\varepsilon_w l_c^2} \int_0^{L_B} d\tau \sum_{n,n'} \frac{(n-n')\sin(\eta_1(\tau)+\eta_2(\tau))}{R} K_{n'-n}\left(\bar{\kappa}_{n'}^{(2)}(\tau)R\right) K_{n'-n}\left(R\bar{\kappa}_{n'}^{(2)}(\tau)\right) \frac{I_n'\left(a\bar{\kappa}_{n'}^{(2)}(\tau)\right)}{K_n'\left(a\bar{\kappa}_{n'}^{(2)}(\tau)\right)}$$

$$I_{n'}\left(a\bar{\kappa}_{n'}^{(2)}(\tau)\right) I_{n'}\left(a\bar{\kappa}_{n'}^{(2)}(\tau)\right) \zeta_{surf,0}^{\prime(2)}(n',n'\tilde{\omega}_{2,1}(\tau))\zeta_{surf}^{(2)}(n',n'\tilde{\omega}_{2,1}(\tau))).$$

(10.27)

## 10.2 Grouping and simplification of terms.

We now want to group the terms so that we can write

$$E_{img1} = \bar{E}_{img1,0} + \bar{E}_{img1,\eta} + \bar{E}_{img1,\omega}, \quad E_{img2} = \bar{E}_{img2,0} + \bar{E}_{img2,\eta} + \bar{E}_{img2,\omega} \tag{10.28}$$

where

$$\bar{E}_{img\mu,0} = E_{img\mu,0,0} \quad \bar{E}_{img2,\omega} = E_{img\mu,0,6}$$
$$\bar{E}_{img\mu,\eta} = E_{img\mu,0,1} + E_{img\mu,0,2} + E_{img\mu,0,3} + E_{img\mu,0,4} + E_{img\mu,0,5} + E_{img\mu,1} + E_{img\mu,2} + E_{img\mu,3} \tag{10.29}$$

Substituting in Eqs. (4.67) and (4.94) into Eqs. (10.5) and (10.15) we obtain for $\bar{E}_{img\mu,0}$

$$\bar{E}_{img\mu,0} = -\frac{e^2}{\varepsilon_w l_c^2} \int_0^{L_B} d\tau \left(1 + (-1)^\mu R\omega_{A,3}(s)\right) \sum_{n,n'} \frac{I'_{n'}\left(a\bar{\kappa}_n^{(\mu)}(\tau)\right) K_{n'-n}\left(\bar{\kappa}_n^{(\mu)}(\tau)R\right) K_{n'-n}\left(\bar{\kappa}_n^{(\mu)}(\tau)R\right)}{K'_{n'}\left(a\bar{\kappa}_n^{(\mu)}(\tau)\right) a^2 \bar{\kappa}_n^{(\mu)}(\tau)^2 K'_n\left(a\bar{\kappa}_n^{(\mu)}(\tau)\right)^2}.$$

$$\tag{10.30}$$

Now in obtaining $\bar{E}_{img\mu,\eta}$ we will first deal with the image charge terms due to the charges on rod 1. We first write:

$$E_{1T1} = E_{img1,0,1} = -\frac{ae^2}{\varepsilon_w l_c^2} \int_0^{L_B} d\tau \sum_{n,n'} n\tilde{\omega}_{1,1}(\tau) \sin\eta(\tau) K_{n'+n}\left(\bar{\kappa}_n^{(1)}(\tau)R\right) I_n\left(a\bar{\kappa}_n^{(1)}(\tau)\right) \frac{I'_{n'}\left(a\bar{\kappa}_n^{(1)}(\tau)\right)}{K'_{n'}\left(a\bar{\kappa}_n^{(1)}(\tau)\right)}$$
$$I_{n-1}\left(a\bar{\kappa}_n^{(1)}(\tau)\right) K_{n'+n-1}\left(R\bar{\kappa}_n^{(1)}(\tau)\right) \zeta_{surf,0}^{(1)}(n, n\tilde{\omega}_{1,1}(\tau)) \zeta_{surf,0}^{(1)}(n, n\tilde{\omega}_{1,1}(\tau)).$$

$$\tag{10.31}$$

Then, we combine

$$E_{1T2} = E_{img1,0,4} + E_{img1,0,3} + E_{img1,0,2} = \frac{e^2 a}{\varepsilon_w l_c^2} \int_0^{L_B} d\tau \sum_{n,n'} n\tilde{\omega}_{1,1}(\tau) K_{n'-n}\left(\bar{\kappa}_n^{(1)}(\tau)R\right) \sin\eta(\tau)$$
$$\frac{I'_{n'}\left(a\bar{\kappa}_n^{(1)}(\tau)\right)}{K'_{n'}\left(a\bar{\kappa}_n^{(1)}(\tau)\right) I_{n'}\left(a\bar{\kappa}_n^{(1)}(\tau)\right)} I_n\left(a\bar{\kappa}_n^{(1)}(\tau)\right) \zeta_{surf,0}^{(1)}(n, n\tilde{\omega}_{1,1}(\tau)) \zeta_{surf,0}^{(1)}(n, n\tilde{\omega}_{1,1}(\tau)) K_{n'+n-1}(a\bar{\kappa}_n^{(1)}(\tau))$$
$$I_n(a\bar{\kappa}_n^{(1)}(\tau)) I_{n'-1}(a\bar{\kappa}_n^{(1)}(\tau)).$$

$$\tag{10.32}$$

Next, we group

$$E_{img1,0,5} + E_{img1,1} + E_{img1,2} + E_{img1,3} = E_{1T3} + E_{1T4} + E_{1T5}, \tag{10.33}$$

where

$$E_{1T3} = \frac{e^2}{\varepsilon_w l_c^2} \int_0^{L_B} d\tau \sum_{n,n'} a \sin\eta(\tau) K_{n'+n}\left(\bar{\kappa}_n^{(1)}(\tau)R\right) I_n\left(a\bar{\kappa}_n^{(1)}(\tau)\right) \frac{n\tilde{\omega}_{1,1}(\tau) I_{n'}'(ax)}{I_{n'}(ax) K_{n'}'(ax)}$$

$$\left(I_{n'}\left(a\bar{\kappa}_n^{(1)}(\tau)\right) I_n'\left(a\bar{\kappa}_n^{(1)}(\tau)\right) - I_n\left(a\bar{\kappa}_n^{(1)}(\tau)\right) I_{n'}'\left(a\bar{\kappa}_n^{(1)}(\tau)\right)\right) \quad (10.34)$$

$$K_{n'+n-1}\left(R\bar{\kappa}_n^{(1)}(\tau)\right) \zeta_{surf\,0}^{(1)}(n, n\tilde{\omega}_{1,1}(\tau)) \zeta_{surf\,0}^{(1)}(n, n\tilde{\omega}_{1,1}(\tau)),$$

$$E_{1T4} = \frac{e^2}{\varepsilon_w l_c^2} \int_0^{L_B} d\tau \sum_{n,n'} \sin\eta(\tau) I_n\left(a\bar{\kappa}_n^{(1)}(\tau)\right) K_{n'+n-1}\left(\bar{\kappa}_n^{(1)}(\tau)R\right)$$

$$\frac{n\tilde{\omega}_{1,1}(\tau) I_{n'}'\left(a\bar{\kappa}_n^{(1)}(\tau)\right) I_n\left(a\bar{\kappa}_n^{(1)}(\tau)\right)}{\bar{\kappa}_n^{(1)}(\tau) K_{n'}'\left(a\bar{\kappa}_n^{(1)}(\tau)\right)} \zeta_{surf\,0}^{(1)}(n, n\tilde{\omega}_{1,1}(\tau)) \zeta_{surf\,0}^{(1)}(n, n\tilde{\omega}_{1,1}(\tau)) \quad (10.35)$$

$$\left((n+n') K_{n'+n}\left(R\bar{\kappa}_n^{(1)}(\tau)\right) - \bar{\kappa}_n^{(1)}(\tau) R K_{n'+n+1}\left(R\bar{\kappa}_n^{(1)}(\tau)\right)\right),$$

$$E_{1T5} = \frac{e^2}{\varepsilon_w l_c^2} \int_0^{L_B} d\tau \sum_{n,n'} a n \tilde{\omega}_{1,1}(\tau) \sin\eta(\tau) K_{n'+n}\left(\bar{\kappa}_n^{(1)}(\tau)R\right) I_n\left(a\bar{\kappa}_n^{(1)}(\tau)\right)$$

$$\frac{d}{dx}\left[\frac{I_{n'}'(ax)}{K_{n'}'(ax)}\right]_{x=a\bar{\kappa}_n^{(1)}(\tau)} \frac{I_n\left(a\bar{\kappa}_n^{(1)}(\tau)\right)}{I_{n'}\left(a\bar{\kappa}_n^{(1)}(\tau)\right)} K_{n'+n-1}\left(R\bar{\kappa}_n^{(1)}(\tau)\right) I_{n'}\left(a\bar{\kappa}_n^{(1)}(\tau)\right) \zeta_{surf\,0}^{(1)}(n, n\tilde{\omega}_{1,1}(\tau)) \zeta_{surf\,0}^{(1)}(n, n\tilde{\omega}_{1,1}(\tau)).$$

$$(10.36)$$

We next, can consider the sum

$$E_{1,T1} + E_{1,T2} + E_{1,T3} = -\frac{e^2}{\varepsilon_w l_c^2} \int_0^{L_B} d\tau \sigma_1(\tau)^2 \sum_{n,n'} n(n-n') \tilde{\omega}_{1,1}(\tau) \sin(\eta_1(\tau) + \eta_2(\tau)) K_{n'+n}\left(\bar{\kappa}_n^{(1)}(\tau)R\right)$$

$$I_n\left(a\bar{\kappa}_n^{(1)}(\tau)\right) I_n\left(a\bar{\kappa}_n^{(1)}(\tau)\right) \frac{I_{n'}'\left(a\bar{\kappa}_n^{(1)}(\tau)\right)}{\bar{\kappa}_n^{(1)}(\tau) K_{n'}'\left(a\bar{\kappa}_n^{(1)}(\tau)\right)} K_{n'+n-1}\left(R\bar{\kappa}_n^{(1)}(\tau)\right) \zeta_{surf\,0}^{(1)}(n, n\tilde{\omega}_{1,1}(\tau)) \zeta_{surf\,0}^{(1)}(n, n\tilde{\omega}_{1,1}(\tau)).$$

$$(10.37)$$

Finally, we consider the sum of all the terms (substituting in Eq. (4.67))

$$\bar{E}_{img\,1,\eta} = E_{1T1} + E_{1T2} + E_{1T3} + E_{1T4} + E_{1T5}, \quad (10.38)$$

$$\bar{E}_{img\,1,\eta} = \frac{e^2}{\varepsilon_w l_c^2} \int_0^{L_B} d\tau \sum_{n=-\infty}^{\infty} \frac{a\tilde{\omega}_{1,1}(\tau) \sin\eta(\tau)}{(\bar{\kappa}_n^{(1)}(\tau)a)^2 \left[K_n'(a\bar{\kappa}_n^{(1)}(\tau))\right]^2} \left[n\tilde{\Omega}_{n,n}(R\bar{\kappa}_n^{(1)}(\tau), a\bar{\kappa}_n^{(1)}(\tau))\right], \quad (10.39)$$

$$\tilde{\Omega}_{n,n}(x, y) = \sum_{n'=-\infty}^{\infty} K_{n+n'+1}(x) K_{n+n'}(x) \left[2\frac{n'}{y} \frac{I_{n'}'(y)}{K_{n'}'(y)} - \frac{d}{dz}\left(\frac{I_{n'}'(z)}{K_{n'}'(z)}\right)\bigg|_{z=y}\right]. \quad (10.40)$$

For Rod 2, following exactly the same analysis as rod 1 we obtain

$$\bar{E}_{img\,2,\eta} = \frac{e^2}{\varepsilon_w l_c^2} \int_0^{L_B} d\tau \sum_{n=-\infty}^{\infty} \frac{a\tilde{\omega}_{2,1}(\tau) \sin\eta(\tau)}{(\bar{\kappa}_n^{(2)}(\tau)a)^2 \left[K_n'(a\bar{\kappa}_n^{(2)}(\tau))\right]^2} n\tilde{\Omega}_{n,n}(R\bar{\kappa}_n^{(2)}(\tau), a\bar{\kappa}_n^{(2)}(\tau)). \quad (10.41)$$

Last of all we have $\bar{E}_{img\mu,\omega}$

$$E_{img\mu,\omega} = -\frac{e^2}{\varepsilon_w l_c^2} \int_0^{L_B} d\tau \frac{\omega_{A,3}(\tau)R}{2} (-1)^\mu \frac{d\xi_\mu(\tau)}{d\tau} \sum_{n,n'} \frac{n^2 \tilde{\omega}_{\mu,1}(\tau)}{\bar{\kappa}_n^{(1)}(\tau)} \left( 2R \frac{I'_{n'}\left(a\bar{\kappa}_n^{(\mu)}(\tau)\right)}{K'_{n'}\left(a\bar{\kappa}_n^{(\mu)}(\tau)\right)} \right.$$

$$\frac{K_{n'-n}\left(\bar{\kappa}_n^{(\mu)}(\tau)R\right) K'_{n'-n}\left(\bar{\kappa}_n^{(\mu)}(\tau)R\right)}{a^2 \bar{\kappa}_n^{(\mu)}(\tau)^2 K'_n\left(a\bar{\kappa}_n^{(\mu)}(\tau)\right)^2} + na K_{n'-n}\left(\bar{\kappa}_n^{(\mu)}(\tau)R\right) K_{n'-n}\left(\bar{\kappa}_n^{(\mu)}(\tau)R\right) \frac{d}{dx}\left[\frac{I'_{n'}(x)}{K'_{n'}(x)} \frac{1}{x^2 K'_n(x)^2}\right]\Bigg|_{x=a\bar{\kappa}_n^{(\mu)}(\tau)} \Bigg).$$

(10.42)

After further algebraic manipulation [1] of $\tilde{\Omega}_{n,n}(x,y)$

$$\tilde{\Omega}_{n,n}(x,y) = \sum_{j=-\infty}^{\infty} \frac{K_{n+j+1}(x) K_{n+j}(x)}{y\left[K'_j(y)\right]^2} \left[1 + 2j I'_j(y) K'_j(y) + \frac{j^2}{y^2}\right].$$

(10.43)

## 11. Extending the results to more general helical charge distributions

We can modify the effective charge density so that, so that for instance for Eqs. (4.55) and (4.58) we have

$$\rho_{eff,0}^{(\mu)}(\mathbf{k};s_0) = \frac{1}{2\pi} \int_0^{L_B} ds' \int_0^{2\pi} dt' \int_0^{2\pi} dt_0 \exp\left(i\mathbf{k}.\mathbf{r}_\mu(s_0) + i\sigma_\mu(s_0)\mathbf{k}.\hat{\mathbf{t}}_\mu(s_0)(s'-s_0) + ia\cos t'\mathbf{k}.\hat{\mathbf{d}}(s_0) + ia\sin t'\mathbf{k}.\hat{\mathbf{n}}_\mu(s_0)\right)$$

$$\left(\sigma_{ind,0,0}^{(\mu)}(t',s') + \delta(t'-t_0, s'-s_0)\right) \sigma_{rad}(t_0 - \xi_\mu(s_0)),$$

(11.1)

$$\rho_{img,0}^{(\nu,\mu)}(\mathbf{k};s_0) = \frac{1}{2\pi} \int_0^{L_B} ds' \int_0^{2\pi} dt' \int_0^{2\pi} dt_0 \exp\left(i\mathbf{k}.\mathbf{r}_\nu(s_0) + i\sigma_1(s_0)\mathbf{k}.\hat{\mathbf{t}}_\nu(s_0)(s'-s_0) + ia\cos t'\mathbf{k}.\hat{\mathbf{d}}(s_0) + ia\sin t'\mathbf{k}.\hat{\mathbf{n}}_\nu(s_0)\right)$$

$$\sigma_{ind,2,0}^{(\nu)}(t',s';t_0) \sigma_{rad}(t_0 - \xi_\mu(s_0)).$$

(11.2)

Here $\sigma_{rad}(t_0 - \xi_\mu(s_0))$ is a radial charge distribution measured respect to a helical line with phase $\xi_\mu(s_0)$. The addition of this generates up a more general set of charge distributions with helical symmetry.

We can express

$$\sigma_{rad}(t_0 - \xi_\mu(s_0)) = \sum_n \zeta_n \exp(-in(t_0 - \xi_\mu(s_0)))$$

(11.3)

This allows us to express (using Eqs. (4.65), (4.71), (4.93), (4.96), (11.1), (11.2) and (11.3))

$$\rho_{eff,0}^{(\mu)}(\mathbf{k};s_0) = \frac{1}{(2\pi)} \sum_n \int_0^{2\pi} d\phi \exp(-in\phi) \exp\left(i\mathbf{k}.\mathbf{r}_\mu(s_0) + ia\cos(\phi+\xi_\mu(s_0))\mathbf{k}.\hat{\mathbf{d}}(s_0) + ia\sin(\phi+\xi_\mu(s_0))\mathbf{k}.\hat{\mathbf{n}}_\mu(s_0)\right)$$
$$\zeta_n \zeta_{surf,0}^{(\mu)}(n,\mathbf{k}.\hat{\mathbf{t}}_2),$$
(11.4)

$$\rho_{img,0}^{(\nu,\mu)}(\mathbf{k};s_0) = \frac{1}{2\pi} \sum_n \int_0^{L_B} ds' \int_0^{2\pi} d\phi \exp\left(i\mathbf{k}.\mathbf{r}_\nu(s_0) + ia\mathbf{k}.\hat{\mathbf{d}}(s_0)\cos\phi + ia\mathbf{k}.\hat{\mathbf{n}}_\nu(s_0)\sin\phi\right)$$
$$\exp(in\phi)\zeta_n \zeta_{surf,1}^{(1)}(n,\mathbf{k}.\hat{\mathbf{t}}_\nu;\xi_\mu(s_0)).$$
(11.5)

Therefore, Eqs. (9.8), (9.9), (9.10), (10.30), (10.39), (10.41) and (10.42) are simply modified to be

$$\bar{E}_{dir,0} \approx \frac{2e^2}{\varepsilon_w l_c^2} \int_0^{L_B} d\tau \sum_n (-1)^n \frac{\zeta_n \zeta_{-n} K_0(R\bar{\kappa}_n(\tau))}{a^2 \bar{\kappa}_n(\tau)^2 K_n'(a\bar{\kappa}_n(\tau)) K_n'(a\bar{\kappa}_n(\tau))} \cos(n(\xi_1(\tau) - \xi_2(\tau))),$$
(11.6)

$$\bar{E}_{dir,\omega} \approx -\frac{e^2 R}{4\varepsilon_w l_c^2} \int_0^{L_B} d\tau \omega_{A,3}(s) \sum_n (-1)^n \zeta_n \zeta_{-n} \left(\frac{d\xi_1(\tau)}{d\tau} - \frac{d\xi_2(\tau)}{d\tau}\right) \frac{n^2(\tilde{\omega}_{1,1}(\tau) + \tilde{\omega}_{2,1}(\tau))}{\tilde{\kappa}_n(\tau)} \cos(n(\xi_1(\tau) - \xi_2(\tau)))$$
$$\left(\frac{RK_1(R\tilde{\kappa}_n(\tau))}{a^2 \tilde{\kappa}_n(\tau)^2 K_n'(a\tilde{\kappa}_n(\tau)) K_n'(a\tilde{\kappa}_n(\tau))} + \frac{2K_0(R\tilde{\kappa}_n(\tau))}{a^2 \tilde{\kappa}_n(\tau)^2 K_n'(a\tilde{\kappa}_n(\tau)) K_n'(a\tilde{\kappa}_n(\tau))} \right)\left(\frac{1}{\tilde{\kappa}_n(\tau)} + \frac{aK_n''(a\tilde{\kappa}_n(\tau))}{K_n'(a\tilde{\kappa}_n(\tau))}\right),$$
(11.7)

$$\bar{E}_{dir,\eta} \approx \frac{e^2}{\varepsilon_w l_c^2} \int_0^{L_B} d\tau \sin\eta(s) \sum_n (-1)^n \frac{\zeta_n \zeta_{-n} n^2 (\tilde{\omega}_{1,1}(\tau) + \tilde{\omega}_{2,1}(\tau)) K_1(R\tilde{\kappa}_n(\tau))}{a^2 \tilde{\kappa}_n(\tau)\tilde{\kappa}_n(\tau) K_n'(a\tilde{\kappa}_n(\tau)) K_n'(a\tilde{\kappa}_n(\tau))} \cos(n(\xi_1(\tau) - \xi_2(\tau))),$$
(11.8)

$$\bar{E}_{img\mu,0} = -\frac{e^2}{\varepsilon_w l_c^2} \int_0^{L_B} d\tau \left(1 + (-1)^\mu R\omega_{A,3}(s)\right) \sum_{n,n'} \frac{\zeta_n \zeta_{-n} I_{n'}'(a\bar{\kappa}_n^{(\mu)}(\tau)) K_{n'-n}(\bar{\kappa}_n^{(\mu)}(\tau)R) K_{n'-n}(\bar{\kappa}_n^{(\mu)}(\tau)R)}{K_{n'}'(a\bar{\kappa}_n^{(\mu)}(\tau)) a^2 \bar{\kappa}_n^{(\mu)}(\tau)^2 K_n'(a\bar{\kappa}_n^{(\mu)}(\tau))^2},$$
(11.9)

$$\bar{E}_{img\mu,\eta} = \frac{e^2}{\varepsilon_w l_c^2} \int_0^{L_B} d\tau \sum_{n=-\infty}^{\infty} \frac{a\tilde{\omega}_{\mu,1}(\tau)\sin\eta(\tau)\zeta_n \zeta_{-n}}{(\bar{\kappa}_n^{(\mu)}(\tau)a)^2 \left[K_n'(a\bar{\kappa}_n^{(\mu)}(\tau))\right]^2} \left[n\tilde{\Omega}_{n,n}(R\bar{\kappa}_n^{(\mu)}(\tau), a\bar{\kappa}_n^{(\mu)}(\tau))\right],$$
(11.10)

$$\bar{E}_{img\mu,\omega} = -\frac{e^2}{\varepsilon_w l_c^2} \int_0^{L_B} d\tau \frac{\omega_{A,3}(\tau)R}{2} (-1)^\mu \frac{d\xi_\mu(\tau)}{d\tau} \sum_{n,n'} \frac{n^2 \tilde{\omega}_{\mu,1}(\tau)\zeta_n \zeta_{-n}}{\bar{\kappa}_n^{(1)}(\tau)} \left(2R \frac{I_{n'}'(a\bar{\kappa}_n^{(\mu)}(\tau))}{K_{n'}'(a\bar{\kappa}_n^{(\mu)}(\tau))}\right.$$
$$\left.\frac{K_{n'-n}(\bar{\kappa}_n^{(\mu)}(\tau)R) K_{n'-n}'(\bar{\kappa}_n^{(\mu)}(\tau)R)}{a^2 \bar{\kappa}_n^{(\mu)}(\tau)^2 K_n'(a\bar{\kappa}_n^{(\mu)}(\tau))^2} + naK_{n'-n}(\bar{\kappa}_n^{(\mu)}(\tau)R) K_{n'-n}(\bar{\kappa}_n^{(\mu)}(\tau)R) \frac{d}{dx}\left[\frac{I_{n'}'(x)}{K_{n'}'(x)} \frac{1}{x^2 K_n'(x)^2}\right]\bigg|_{x=a\bar{\kappa}_n^{(\mu)}(\tau)}\right).$$
(11.11)

For a DNA like helical charge distribution [3] we may write

$$\sigma_{rad}(t_0 - \xi_\mu(s_0)) = -\frac{1}{2}\left(\delta(t_0 - \xi_\mu(s_0) - \tilde{\phi}_s) + \delta(t_0 - \xi_\mu(s_0) + \tilde{\phi}_s)\right)$$
$$+ \theta(1 - f_1 - f_2) + \theta f_1 \delta(t_0 - \xi_\mu(s_0)) + \theta f_2 \delta(t_0 - \xi_\mu(s_0) - \pi).$$
(11.12)

Here, the centre of the minor groove of the DNA like charge distributions are chosen to lie at $\hat{\mathbf{v}}_\mu(s)$. The charge distribution given by Eq. (11.12) traces out four helical lines of charge. The first two delta functions correspond to the negative phosphate charges. Positive ions lie at the surface that neutralize these phosphate charges by a faction $\theta$. These ions are either at the centres of the major and minor grooves or uniformly smeared over the molecule. The proportion of positive ions that lie at the centre of the minor groove and major groove are given by $f_1$ and $f_2$, respectively, and the charge distributions of ions in the minor and major grooves are given by the third and fourth delta function, respectively. The term $\theta(1 - f_1 - f_2)$ corresponds to a uniformly smeared contribution. From Eq. (11.12) we get

$$\zeta_n = \delta_{n,0}\theta(1 - f_1 - f_2) + \theta f_1 + (-1)^n \theta f_2 - \cos n\tilde{\phi}_s.$$
(11.13)

## 12. Discussion and outlook

Some care should be taken with the results given in Eqs. (11.6)-(11.11) that explicitly depend on $\omega_{A,3}(s)$ as we have not systematically expanded the energy out to linear order in $\omega_{A,3}(s)$; we may have neglected terms arising from the curvature expansion for the calculation image charges (section 4) as well as terms that arise from higher order corrections from the Taylor expansions considered in sections 2, 5 and 6 that are also linear in $\omega_{A,3}(s)$. However, from a physical point of view, the most important terms in $\omega_{A,3}(s)$ may be to do with the rescaling of the charge densities to the braid axis in the image charges seen in Eq. (11.9) (as well as those encountered within expressions for the twisting elastic energy [2]). The terms given by Eq. (11.7) and (11.11) are probably not that important when compared to terms in the twisting elastic energy that also depend on the derivatives of $\xi_\mu(s)$, and so probably Eq. (11.7) and (11.11) can be neglected provided that the elastic torsional modulus is large enough. On the other hand, this still has yet to be seen, and it would be useful to have all the terms that linearly depend on $\sin \eta(s)$, $\omega_{A,3}(s)$ and $\omega_{A,2}(s)$, and perhaps some higher order terms in the expansion. A full systematic expansion is a task yet to be done. This could form the basis of some variational procedure to minimize the electrostatic free energy for the system [11].

Another place where the calculation should be extended to is the case where the inter-axial separation $R$ depends on $s$ in a slowly varying way. Indeed, in a combined electrostatic elastic rod theory, $R(s)$ should also be determined self consistently by minimization of a combined energy functional of elastic and electrostatic energies, as well as the other geometric parameters. In develop the statistical mechanics of charged braids, as applied to biopolymers; we also need such energy functionals to describe the Boltzmann weight in a path integral over all the degrees of geometric freedom for the braid. We hope in future work to develop these calculations further to take account of some of these issues.

## Acknowledgements

D.J. Lee would like to acknowledge useful discussions with T. Liverpool, A. A. Korynshev and R. Cortini. He would also like to acknowledge the support of the Human Frontiers Science Program (grant RGP0049/2010-C102).

[1] R. Cortini, A. A. Kornyshev, D. J. Lee, S. Leikin; Biophys. J. **101,** 875 (2011) and supplemental material
[2] http://www.ucl.ac.uk/~ucesgvd/braids.html
[3] A.A. Kornyshev, D. J. Lee, S. Leikin; Rev. Mod. Phys. **79,** 943 (2007)
[4] G. Charvin, A. Vologodskii, D. Bensimon, V. Croquette; Biophys. J. **88,** 4124 (2005)
[5] F. Mosconi, J. Allemand, D. Bensimon, V. Croquette, Phys. Rev. Lett. **102** (2009), 7: 78301
[6] Q. Shao, S. Goyal, L. Finzi, D. Dunlap; Macromolecules **45**, 3188 (2012)
[7] C. Maffeao, R. Schöpflin, H. Brutzer, R. Stehr, A. Aksimentiev, G. Wedemann, R. Seidel; Phys. Rev. Lett. **105**, 158101 (2010)
[8] D. Argudo, P. K. Purohit; Acta Biomat. **8,** 2133 (2012)
[9] S. Neukirch, J. F. Marko; Phys. Rev. Lett. **106**, 138104 (2011)
[10] S. Leikin; Unpuplished
[11] R.R. Netz and H. Orland; Eur. Phys. J. E **11**, 301 (2003)